%% file: article.tex
\input{macro2e.tex}

\documentclass[12pt,twoside]{article}
\usepackage[eqsecnum]{cmp2e}
\usepackage{amsbsy,amsfonts,amssymb,amstext,rsfs,epsf}

\renewcommand{\d}{{\rm d}}
\newcommand{\e}{{\rm e}}
\newcommand{\xnt}{\lp x^N;t\rp}
\newcommand{\vt}{\vartriangle\!\!}
\newcommand{\bmv}[1]{\boldsymbol{#1}}
\newcommand{\xntp}{\left(x^N;t'\right)}
\newcommand{\tensor}[1]{\stackrel{\leftrightarrow}{#1}}

\title[A consistent description of kinetics and hydrodynamics]
{A consistent description of kinetics and hydrodynamics of systems of
interacting particles by means of the nonequilibrium statistical operator
method}

\author[M.V.Tokarchuk, I.P.Omelyan, A.E.Kobryn]
{M.V.Tokarchuk, I.P.Omelyan, A.E.Kobryn}

\address{Institute for Condensed Matter Physics of
\newline the National Academy of Sciences of Ukraine
\newline 1 Svientsitskii St., 290011 Lviv--11, Ukraine}

\date{\today}

\begin{document}

\maketitle

\begin{abstract}
A statistical approach to a self-consistent description of kinetic
and hydrodynamic processes in systems of interacting particles is
formulated on the basis of the nonequilibrium statistical operator
method by D.N.Zubarev. It is shown how to obtain the kinetic equation of
the revised Enskog theory for a hard sphere model, the kinetic equations for
multistep potentials of interaction and the Enskog-Landau kinetic equation
for a system of charged hard spheres. The BBGKY hierarchy is analyzed on the
basis of modified group expansions. Generalized transport equations are
obtained in view of a self-consistent description of kinetics and
hydrodynamics. Time correlation functions, spectra of collective excitations
and generalized transport coefficients are investigated in the case of
weakly nonequilibrium systems of interacting particles.
\pacs 05.20.Dd, 05.60.+w, 52.25.Fi, 71.45.G, 82.20.M
%
\end{abstract}

\section{Introduction}
\input{s_01.tex}
\section{Conception of a consistent description of kinetics and
hydrodynamics for dense gases and liquids}
\input{s_02_1.tex}
\input{s_02_2.tex}
\input{s_02_3.tex}
\input{s_02_31.tex}
\input{s_02_32.tex}
\input{s_02_33.tex}
\input{s_02_41.tex}
\input{s_02_42.tex}
\input{s_02_43.tex}
\section{Generalized transport equations and time correlation
\protect\newline functions}
\input{s_03_1.tex}
\input{s_03_2.tex}
\input{s_03_3.tex}
\input{s_03_4.tex}
\section{Conclusion}
\input{s_04.tex}
\newpage

\input{bibliogr.tex}
\label{last@page}
\end{document}

%% file: macro2e.tex
\newcommand{\be}{\begin{equation}}
\newcommand{\ee}{\end{equation}}
\newcommand{\ba}{\begin{array}}
\newcommand{\ea}{\end{array}}
\newcommand{\bea}{\begin{eqnarray}}
\newcommand{\eea}{\end{eqnarray}}
\newcommand{\bean}{\begin{eqnarray*}}
\newcommand{\eean}{\end{eqnarray*}}

\newcommand{\lp}{\left(}
\newcommand{\rp}{\right)}
\newcommand{\ls}{\left[}
\newcommand{\rs}{\right]}
\newcommand{\lc}{\left\{}
\newcommand{\rc}{\right\}}

\newcommand{\la}{\langle}
\newcommand{\La}{\left\la}
\newcommand{\ra}{\rangle}
\newcommand{\Ra}{\right\ra}

\renewcommand{\d}{\mbox{\rm d}}
\newcommand{\g}{\text{\slshape g}}
\newcommand{\gb}%
	{\text{\bf\slshape g%
	}}

\newcommand{\intl}{\int\limits}

\newcommand{\Imm}{\,\Im{\mathfrak{m}}\,}

\newcommand{\im}{{\rm i}}

\newcommand{\veps}{\varepsilon}
\newcommand{\vphi}{\varphi}
\newcommand{\vrho}{\varrho}

\newcommand{\mGamma}{{\mit\Gamma}}

\newcommand{\mTheta}{{\mit\Theta}}

\newcommand{\mPhi}{{\mit\Phi}}

\newcommand{\Chi}{\mbox{\parbox[c]{0.8em}{\large$\chi$}}}

\newcommand{\El}{{\scr L}}
\newcommand{\SP}{{\scr P}}

\newcommand{\ddt}{\frac{\partial}{\partial t}}
\newcommand{\dd}[1]{\frac{\partial}{\partial #1}}
\newcommand{\refp}[1]{(\ref{#1})}

\newcommand{\ds}{\displaystyle}

%% file: s_01.tex
The problem of a self-consistent description of fast and slow processes
which are connected with both linear and non-linear fluctuations of observed
quantities for various physical systems, such as dense gases, liquids,
their mixtures and plasma, remains an actual problem in nonequilibrium
statistical mechanics. An important task in this direction is the
construction of kinetic equations for dense systems with taking into
account collective effects (hydrodynamic contribution) into the collision
integrals. On the other hand, there is a problem of the calculation of time
correlation functions and collective excitations spectra in the range
of intermediate values of wavevectors and frequencies, since low- and
high-frequencies regions are described sufficiently well within the
frameworks of molecular hydrodynamics and kinetic equations, respectively.
In the intermediate region, the kinetic and hydrodynamic processes are
connected  and must be considered simultaneously.

A significant interest in these problems was exhibited by Prof. D.N.Zubarev
in his investigations. One of the approaches to the unification of 
kinetics and hydrodynamics in the theory of transport phenomena for
systems of interacting particles was proposed by Zubarev and Morozov in
\cite{1} and developed in further papers \cite{2,3,4,5,6,7,8,9,9a,10,11}. 
This statistical approach is based on a modification of the boundary 
conditions for the weakening of correlations to the BBGKY hierarchy for 
many-particle distribution functions and to the Liouville equation for a
full nonequilibrium distribution function (nonequilibrium statistical 
operator -- NSO). The modification of the boundary conditions consists in 
the fact that both the one-particle nonequilibrium distribution function 
and the local conservation laws for mass, momentum and total energy are 
included into the parameters for an abbreviated description of a 
nonequilibrium state of the system. The concept of a self-consistent 
description of kinetics and hydrodynamics was applied to plasma in an 
electro-magnetic field \cite{12,13,14}, to quantum systems with taking into 
account coupled states \cite{9,9a,15}, quantum Bose systems \cite{16} and 
the nonequilibrium thermo field dynamics \cite{17}. The questions as to the 
necessity and possibility of a unified description of kinetic and 
hydrodynamic processes were discussed in papers by Klimontovich 
\cite{18,19,20}.

In this paper we present some actual results obtained recently within the 
concept of a self-consistent description of kinetics and hydrodynamics for 
dense gases and simple liquids. In section 2 we present the results for 
kinetic equations of RET \cite{3,7}, GDRS \cite{8} and Enskog-Landau 
theories \cite{3,7}, obtained in the polarization approximation for a 
system of hard spheres and a Lenard-Balescu kinetic equation \cite{10,11} 
derived from the BBGKY hierarchy with modified boundary conditions. 
Generalized transfer equations, time correlation functions and transport 
coefficients are presented in section 3 on the basis of a self-consistent 
description of kinetics and hydrodynamics within the 
framework of the nonequilibrium statistical operator method.

%% file: s_02_1.tex
\subsection{Overview}

Bogolubov \cite{21} proposed a consequent approach to the construction of
kinetic equations which is based on the chain of equations for $s$-particle
distribution functions and a weakening correlation principle. There are
a lot of theories \cite{22,23,24,25} which differ in forms of presentation, 
but all these approaches use the same weakening correlation condition and 
are most effective when a small parameter (interaction, density, etc.) is 
presented.

In the kinetic theory of classical gases we can point out two principal
problems. The first problem is connected with the fact that collision
integrals can depend nonanalytically on density. Because of this, even
in the case of a small-density gas, in order to calculate corrections
to transport coefficients it is necessary to perform a partial resummation
of the BBGKY hierarchy \cite{26,27}. The second problem is the construction
of kinetic equations for the gases of high densities. In this case we cannot
restrict ourselves by several terms of the expansion of collision integrals
on densities and the analysis of the BBGKY hierarchy becomes very
complicated (generally speaking, the density is not a small parameter
here). That is why the construction of kinetic equations for dense
gases and liquids with model interparticle potentials of interaction
has a great importance in the problem under investigation.

The first theory in this direction is a semi-phenomenological standard 
Enskog theory (SET) \cite{28,29} of dense gases. Ideas similar to those 
used at the derivation of the Boltzmann equation were also used in this 
approach. The Enskog equation plays an essential role in the kinetic 
theory, next to the Boltzmann one \cite{29}. This equation was obtained at 
the modelling of molecules by hard spheres for dense gases. As a result, 
the collision integral was presented in an analytical form. It is obtained 
by means of a hard sphere model where collisions can be considered as 
momentary and by the fact that the multiparticle contact at the same time 
is reputed to be infinitely small. Density correction introduced by 
Enskog proved to be considerable, as far as transport, due to collisions 
in a dense system, is the main mechanism of transport. Each molecule is 
almost localized at one point of space by the surrounding neighbour 
molecules and, therefore, the flow transport is suppressed. Though this 
theory properly describes density dependence of the kinetic coefficients, at 
the same time the suppositions about the structure of the collision integral 
remain sufficiently rough and phenomenological. Despite approximate 
assumptions on the collision integral in the kinetic equation for a 
one-particle distribution function of hard spheres, the Enskog theory 
very well describes a set of properties for real dense gases \cite{24,30}.

Davis, Rise and Sengers \cite{31} proposed a kinetic DRS theory. Within the 
framework of this theory, the interparticle potential of an interaction is 
chosen in a square well form. An attractive part of the real potential is 
approximated here by a finite height wall.

From the point of view of the statistical theory of nonequilibrium 
processes, SET and DRS contain two essential drawbacks. The first one is 
that their kinetic equations are not obtained within the framework of some 
consequential theoretical scheme and one does not know how to improve these 
theories. And the second one is that the $H$-theorem has not been proved. 
Nevertheless, not long ago a way of constructing the SET entropy functional 
was given \cite{32}. In order to overcome these drawbacks the authors of 
\cite{33} obtained a kinetic equation for a revised version of the Enskog 
theory (RET) using the diagram method. R\'esibois \cite{34,35} proved the 
$H$-theorem for it. In 1985, a revised version of the DRS theory (RDRS) was 
proposed \cite{39}. The kinetic equation of RDRS satisfies the $H$-theorem 
as well. A generalized version of RDRS --- GDRS was considered in 
\cite{2,8}, being based on the approaches of \cite{1,3}. Here, in order to 
treat a more realistic model, a kinetic equation for dense classical 
systems is offered with an interparticle interaction potential in the shape 
of a multistep function, where its sequential derivation and the normal 
solution are obtained.

In the paper by Stell et al. \cite{36,37} the authors completed the 
construction of the kinetic variational theory (KVT) introducing a whole 
family of theories, namely, KVT I, KVT II and KVT III. The KVT III refers 
to the local energy constraint in maximizing entropy as distinguished from 
the global energy constraint (KVT II) or the hard-core constraint (KVT I).
According to this classification, the kinetic equation of RDRS \cite{39} is
obtained by applying KVT III to the square well potential and KMFT \cite{36}
is derived from KVT I as its application to the potential with a smooth 
tail. In \cite{37}, the KVT III version of KMFT was proposed. In this case 
the quasiequilibrium binary distribution function (QBDF) of hard spheres is 
to be substituted by the full QBDF, which takes into account the full
potential of the interaction. The main conclusion of the KVT I version of 
KMFT \cite{37,38} is that the smooth part of the potential does not 
contribute explicitly to transport coefficients. This fact is caused by 
the approximations of the theory. When the full QBDF is used, as it is in 
the KVT III version of KMFT, transport coefficients are determined by 
the soft part of the potential. The case when the potential between the 
hard sphere and the attractive walls is considered as a smooth tail 
(instead of the constant potential in DRS) has also been investigated 
\cite{40}. The kinetic equation was obtained for this potential applying a 
mean field approximation for a smooth tail.

At this stage of investigations in the kinetic theory of dense gases
and liquids, paper \cite{1} by Zubarev and Morozov played a fundamental 
role. In this paper a new formulation of boundary conditions for the BBGKY 
hierarchy is presented. Such a formulation took into account correlations 
connected with local conservation laws. In the binary collision 
approximation, this modification of the weakening correlation condition
by Bogolubov led to a kinetic equation for a system of hard spheres which 
is close in structure to the usual Enskog kinetic equation. Similar ideas 
were proposed independently by the authors of paper \cite{39} at the 
derivation of a kinetic equation when an interaction between the particles 
is modelled by a square well potential. It is necessary to point out that a 
somewhat different modification of the Bogolubov approach \cite{21} was 
presented in the papers by Rudyak \cite{41,42,43}. This allows one to 
obtain an Enskog-like kinetic equation and extend it to the systems with a 
soft potential of interaction between the particles.

The modification of weakening correlation conditions to a chain of 
equations by Bogolubov was developed in papers 
\cite{2,3,4,5,6,7,8,9,9a,10,11}. An important achievement of the given 
approach is the fact that the modified boundary condition gives the 
possibility \cite{4,7,13} to derive consequently the kinetic equation of 
the RET theory \cite{33} for the first time. On the basis of result 
\cite{1}, a kinetic equation was obtained \cite{2,8} for systems with 
a multistep potential of interaction (in particular, the $H$-theorem was 
proved for this equation in paper \cite{8}), whereas an Enskog-Landau 
kinetic equation was derived \cite{3,7} for a system of charged hard 
spheres.

Normal solutions for the obtained kinetic equations were found using the
Chap\-man-Enskog method. With the help of these solutions, the numerical
calculations of transport coefficients (bulk and shear viscosities, thermal 
conductivity) were carried out for Argon-like systems \cite{2,44}, ionized 
Argon \cite{7} and mixtures \cite{45}. It is important to stress that the 
RET kinetic equation, the kinetic equation for systems with multistep 
potentials of interaction and the Enskog-Landau kinetic equation for 
charged hard spheres were derived from the BBGKY hierarchy with a modified 
boundary condition in a binary collision approximation. Obviously, the 
presented equations have a restricted region of application. They cannot be 
applied to the description of systems with significantly collective effects 
caused by Coulomb, dipole or other long-range forces of interaction between 
the particles. To describe the collective effects in systems with a 
long-range character of interaction it is necessary to consider higher-order
approximations for collision integrals. An example of such an equation
is the Lenard-Balescu kinetic equation \cite{46,47,48} for Coulomb plasma.

To analyze solutions to the BBGKY hierarchy with a modified boundary 
condition in higher-order approximations on interparticle correlations we 
have applied a concept of group expansions \cite{10,11}.

%% file: s_02_2.tex
\subsection{The Liouville equation and the BBGKY hierarchy with a modified 
\protect\linebreak boundary condition}

Let us consider a system of $N$ identical classical particles which are
enclosed in volume $V$, with the Hamiltonian:
%
%
\be
H=\sum\limits_{j=1}^N\frac {p_{j}^{2}}{2m}+\frac{1}{2}
\mathop{\sum_{j=1}^N\sum_{k=1}^N}\limits_{j\ne
k}\mPhi\left(|\bmv{r}_{jk}|\right),\label{e2.1}
\ee
where $\mPhi\left(|\bmv{r}_{jk}|\right)$ is the interaction energy
between two particles $j$ and $k$;
$|\bmv{r}_{jk}|$=$|\bmv{r}_{j}-\bmv{r}_{k}|$ is the distance between a
pair of interacting particles;
$\bmv{p}_{j}$ is the momentum of $j$th particle and $m$ denotes its mass.
A nonequilibrium state of such a system is described by the $N$-particle
nonequilibrium distribution function
$\vrho\lp x^N;t\rp$ = $\vrho\lp x_{1},\ldots,x_{N};t\rp$
which satisfies the Liouville equation:
%
%
\be
\lp\ddt+\im L_N\rp\vrho\lp x^N,t\rp=0,\label{e2.2}
\ee
where $\im=\sqrt{-1}$, $x_j=\{\bmv{r}_{j},\bmv{p}_{j}\}$ is a set of
phase variables (coordinates and momenta), $L_N$ is
the Liouville operator:
%
%
\be
L_N=\ds\sum\limits_{j=1}^NL(j)+
        \frac 12\mathop{\sum_{j=1}^N\sum_{k=1}^N}\limits_{j\ne k}L(j,k),
	\label{e2.3}
\ee
\newpage\noindent
where
\[
\begin{array}{lll}
L(j)&=&\ds-\im\frac{\bmv{p}_j}{m}\dd{\bmv{r}_j},\\
L(j,k)&=&\ds\phantom{-}\im\frac{\partial\mPhi\left(|\bmv{r}_{jk}|\right)}
        {\partial\bmv{r}_{jk}}\left(\dd{\bmv{p}_j}-\dd{\bmv{p}_k}\right).
\end{array}
\]

The function $\vrho\lp x^N;t\rp$ is symmetrical with respect to permutations
$x_l\leftrightarrows x_j$ of phase variables for an arbitrary pair of 
particles. This function satisfies the normalization condition
%
%
\be
\int\d\mGamma_N\vrho\xnt=1,
\quad\d\mGamma_N=(\d x)^N/N!\,=\d\bmv{r}\d\bmv{p}.\label{e2.4}
\ee

In order to solve the Liouville equation \refp{e2.2}, it is necessary to 
introduce a boundary condition. It must be chosen in such a way that the 
solution to the Liouville equation will correspond to a physical state
of the system under consideration. Using the nonequilibrium statistical 
operator method by D.N.Zubarev \cite{9,9a,49,50} (NSO method), we shall 
search for such solutions to the Liouville equation depending on time 
explicitly via the values for some set of observable variables, which is 
sufficient for describing a nonequilibrium state without depending on the 
initial time $t_0$.

The solution to the Liouville equation, which satisfies the initial 
condition
\[
\vrho\xnt|_{t=t_{0}}=\vrho_{\rm q}\lp x^N;t_{0}\rp,
\]
has the form:
%
%
\be
\vrho\lp x^N;t,t_0\rp=\e^{-\im L_N (t-t_0)}\vrho_{\rm q}\lp x^N;t_0\rp.
\label{e2.5}
\ee

We shall consider such times $t \gg t_0$ when the details of the initial 
state of the system can be neglected. Then, to avoid the dependence on 
$t_0$, let us average \refp{e2.5} with respect to the initial times from 
$t_0$ t® $t$ and perform the boundary transition $\lp t-t_0\rp\to\infty$. 
As a result, one obtains \cite{9,9a,48}:
%
%
\be
\vrho\lp x^N;t\rp = \veps \int \limits^{0}_{-\infty}\d t'\;
	\e^{\veps t'}\e^{\im L_N t'}\vrho_{\rm q}\lp x^N;t+t'\rp,\qquad
	t'=t_0-t,\label{e2.6}
\ee
where $\veps$ tends to $+0$ after the thermodynamic limit transition.
It can be shown by straightforward differentiations that solution 
\refp{e2.6} satisfies the Liouville equation with an infinitesimal source 
in the right-hand side:
%
%
\be
\lp\ddt+\im L_N\rp\vrho\lp x^N,t\rp=-\varepsilon\Big(\vrho\lp x^N,t\rp-
        \vrho_{\rm q}\lp x^N,t\rp\Big).\label{e2.7}
\ee
The source breaks the symmetry of the Liouville equation with respect to 
time inversion and selects retarded solutions which correspond to an 
abbreviated description of a nonequilibrium state of the system. The 
auxiliary function $\vrho_{\rm q}\lp x^N;t\rp$ denotes a quasiequilibrium 
distribution function that is defined from an extreme condition for the 
informational entropy of the system, provided the normalization condition 
is preserved and the average values for variables of the abbreviated 
description are fixed.

The choice of $\vrho_{\rm q}\xnt$ depends mainly on a nonequilibrium state
of the system under consideration. In the case of low density gases, where
times of free motion are essentially larger than collision times,
higher-order distribution functions of particles become dependent on
time only via one-particle distribution functions \cite{21,51,52}. It does 
mean that an abbreviated description of nonequilibrium states is available, 
and the total nonequilibrium distribution function depends on time via
$f_1(x;t)$. In such a case, the quasiequilibrium distribution function
$\vrho_{\rm q}\xnt$ reads \cite{51,52}:
%
%
\be
\vrho_{\rm q}\xnt=\prod_{j=1}^N\frac{f_1(x_j;t)}{\e},\label{e2.8}
\ee
where $\e$ is the natural logarithm base. Then, the Liouville equation with 
a small source \refp{e2.7} in view of \refp{e2.8} corresponds to the 
abbreviated description of time evolution of the system on a kinetic stage, 
when only a one-particle distribution function is considered as a slow 
variable. However, there are always additional quantities which vary in time 
slowly because they are locally conserved. In the case of a one-component 
system, the mass density $\rho(\bmv{r};t)$, momentum $\bmv{j}(\bmv{r};t)$ 
and total energy ${\cal E}(\bmv{r};t)$ belong to such quantities. At long 
times they satisfy the generalized hydrodynamics equations. Generally 
speaking, the equation for $f_1(x;t)$ must be conjugated with these 
equations. For low density gases, such a conjugation can be done, in 
principle, with an arbitrary precision in each order on density. In high 
density gases and liquids, when a small parameter is absent, the correlation 
times corresponding to hydrodynamic quantities become commensurable with 
the characteristic times of varying one-particle distribution functions. 
Therefore, in dense gases and liquids, the kinetics and hydrodynamics are 
closely connected and should be considered 
simultaneously. That is why many-particle correlations, related to the 
local conservation laws of mass, momentum and total energy, cannot be 
neglected \cite{1,5,9,9a}. The local conservation laws affect the kinetic 
processes due to an interaction of the selected particles group with 
other particles of the system. This interaction is especially important in 
the case of high densities, and it must be taken into consideration. 
Then, at the construction of kinetic equations for high densities, it is 
necessary to choose the abbreviated description of a nonequilibrium system 
in the form to satisfy the true dynamics of conserved quantities 
automatically. To this end, the densities of hydrodynamic variables should 
be included together with the one-particle distribution function $f_1(x;t)$ 
into the initial set of parameters of the abbreviated description 
\cite{1,5,9,9a}. The next phase functions correspond to the densities of 
the hydrodynamic variables $\rho(\bmv{r};t)$, $\bmv{j}(\bmv{r};t)$ and 
${\cal E}(\bmv{r};t)$:\newpage
%
%
\be
\begin{array}{lcl}
\hat{\rho}(\bmv{r})&=&\ds\int\d\bmv{p}\;\hat{n}_1(x)m,\\
\hat{\bmv{j}}(\bmv{r})&=&\ds\int\d\bmv{p}\;\hat{n}_1(x)\bmv{p},\\
\hat{\cal E}(\bmv{r})&=&\ds\int\d\bmv{p}\;\hat{n}_1(x)\frac{p^2}{2m}+
	\frac 12\int\d\bmv{r}'\;\d\bmv{p}\;\d\bmv{p}'\;
        \hat{n}_2(x,x')\mPhi(|\bmv{r}-\bmv{r}'|),\\
\end{array}
\label{e2.9}
\ee
where $\hat{n}_1(x)$ and $\hat{n}_2(x,x')$ are one- and two-particle
microscopic phase densities by Klimontovich \cite{23}:
%
%
\bea
\hat{n}_1(x)&=&\sum_{j=1}^N\delta(x-x_j)=
        \sum_{j=1}^N\delta(\bmv{r}-\bmv{r}_j)\delta(\bmv{p}-\bmv{p}_j),
        \label{e2.10}\\
\hat{n}_2(x,x')&=&\sum_{j\ne k=1}^N\delta(x-x_j)\delta(x'-x_k).
        \label{e2.11}
\eea
Relations \refp{e2.9} show
a distinctive role of the potential interaction energy. Contrary to
$\rho(\bmv{r};t)=\la\hat{\rho}(\bmv{r})\ra^t$ and
$\bmv{j}(\bmv{r};t)=\la\hat{\bmv{j}}(\bmv{r})\ra^t$, the nonequilibrium 
values of the total energy 
${\cal E}(\bmv{r};t)=\la\hat{\cal E}(\bmv{r})\ra^t$
cannot be expressed via the one-particle distribution function
$f_1(x;t)=\la\hat{n}_1(x)\ra^t$ only, because in order to evaluate
the potential part of ${\cal E}_{\rm int}(\bmv{r};t)=\la\hat{\cal
E}_{\rm int}(\bmv{r})\ra^t$ it is necessary to involve
the two-particle distribution function 
$f_2(x,x';t)=\la\hat{n}_2(x,x')\ra^t$. Here
%
%
\be
\hat{\cal E}_{\rm int}(\bmv{r})=\frac 12
        \int\d\bmv{r}'\;\d\bmv{p}\;\d\bmv{p}'\;
        \hat{n}_2(x,x')\mPhi(|\bmv{r}-\bmv{r}'|)\label{e2.12}
\ee
is the density of the potential energy of interaction. The next conclusion
can be formulated as follows. If the one-particle distribution function
$f_1(x;t)$ is chosen as a parameter of an abbreviated description, then
the density of the interaction energy (\ref{e2.12}) can be considered as
an additional independent parameter. One can find the quasiequilibrium 
distribution function $\vrho_{\rm q}\xnt$ from the condition of extremum 
for a functional of the informational entropy 
$S_{\rm inf}(t)=-\int\d\mGamma_N\;\vrho\xnt\ln\vrho\xnt$ at fixed average 
values of $\la\hat{n}_1(x)\ra^t=f_1(x;t)$, 
$\la\hat{\cal E}_{\rm int}(\bmv{r})\ra^t={\cal E}_{\rm int}(\bmv{r};t)$, 
including the normalization condition for $\vrho_{\rm q}\xnt$ 
\cite{5,9,9a,48}: 
$\int\d\mGamma_N\vrho_{\rm q}\xnt=\int\d\mGamma_N\vrho\xnt$ $=1$. 
This is equivalent to finding an unconditional extreme for the functional
{
\small
\[
L(\vrho)=\!\!\int\d\mGamma_N\vrho\xnt\lc 1-\ln\vrho\xnt-\Phi(t)-
        \!\!\int\d\bmv{r}\beta(\bmv{r};t)\hat{\cal E}_{\rm int}(\bmv{r})-
        \!\!\int\d x\,a(x;t)\hat{n}_1(x)\rc,
\]
}

\noindent
where $\Phi(t)$, $\beta(\bmv{r};t)$, $a(x;t)$ are the Lagrange multipliers.

Taking the variation $\ds\frac{\delta}{\delta\vrho}L(\vrho)$, after some 
simple manipulations one obtains the quasiequilibrium distribution function
%
%
\bea
\vrho_{\rm q}\xnt&=&\exp\lc-\Phi(t)-\!\!\int\d\bmv{r}\;\beta(\bmv{r};t)
        \hat{\cal E}_{\rm int}(\bmv{r})-\!\!\int\d x\;a(x;t)\hat{n}_1(x)\rc,
        \label{e2.13}\\
\Phi(t)&=&\ln\int\d\mGamma_N\;\exp\lc-\!\!\int\d\bmv{r}\;\beta(\bmv{r};t)
        \hat{\cal E}_{\rm int}(\bmv{r})-
        \!\!\int\d x\;a(x;t)\hat{n}_1(x)\rc\!.\label{e2.14}
\eea
Here, $\Phi(t)$ is the Massieu-Planck functional. It is determined from the
condition of normalization for the distribution $\vrho_{\rm q}\xnt$. 
Relation \refp{e2.13} was obtained for the first time in \cite{1}. To 
determine the physical meaning of parameters $\beta(\bmv{r};t)$ and 
$a(x;t)$ let us rewrite $\vrho_{\rm q}\xnt$ (\ref{e2.13}) in the form:
%
%
\bea
\vrho_{\rm q}\xnt&=&\exp\lc-\Phi(t)-\int\d\bmv{r}\;\beta(\bmv{r};t)
        \hat{\cal E}'(\bmv{r})-\int\d x\;a'(x;t)\hat{n}_1(x)\rc,
        \label{e2.15}\\
\Phi(t)&=&\ln\int\d\mGamma_N\;
        \exp\lc-\int\d\bmv{r}\;\beta(\bmv{r};t)\hat{\cal E}'(\bmv{r})-
        \int\d x\;a'(x;t)\hat{n}_1(x)\rc,\nonumber
\eea
where $\hat{\cal E}'(\bmv{r})$ is the density of the total energy in a 
reference frame which moves together with a system element of the mass 
velocity $\bmv{V}(\bmv{r};t)$ \cite{5}
%
%
\be
\hat{\cal E}'(\bmv{r})=\hat{\cal E}(\bmv{r})-
        \bmv{V}(\bmv{r};t)\hat{\bmv{j}}(\bmv{r})+\frac m2V^2(\bmv{r};t)
        \hat{n}(\bmv{r}).\label{e2.16}
\ee
Here $\hat{n}(\bmv{r})=\int\d\bmv{p}\;\hat{n}_1(x)$ is the density of the 
particles number. Parameters $\beta(\bmv{r};t)$ and $a'(x;t)$ in
(\ref{e2.15}) are determined from the conditions of self-consistency, namely,
the equality of quasi-average values $\la\hat{n}_1(x)\ra^t_{\rm q}$ and
$\la\hat{\cal E}'(\bmv{r})\ra^t_{\rm q}$ with their real averages 
$\la\hat{n}_1(x)\ra^t$, $\la\hat{\cal E}'(\bmv{r})\ra^t$:
%
%
\be
\begin{array}{rcl}
\la\hat{n}_1(x)\ra^t_{\rm q}&=&\la\hat{n}_1(x)\ra^t=f_1(x;t),\\
\la\hat{\cal E}'(\bmv{r})\ra^t_{\rm q}&=&
        \la\hat{\cal E}'(\bmv{r})\ra^t,\\
\end{array}\quad\mbox{here}\quad
\la\ldots\ra^t_{\rm q}=\int\d\mGamma_N\;\ldots\vrho_{\rm q}\xnt.
\label{e2.17}
\ee
In these transformations the parameters $a'(x;t)$ and
$a(x;t)$ are connected by the relation
\[
a'(x;t)=a(x;t)-\beta(\bmv{r};t)\lc\frac{p^2}{2m}-
        \bmv{V}(\bmv{r};t)\bmv{p}+\frac m2V^2(\bmv{r};t)\rc.
\]
In the case, when the conditions (\ref{e2.17}) take place, one can obtain 
some relations taking into account self-consistency conditions and varying 
the modified Massieu-Planck functional $\Phi(t)$ after \refp{e2.15}
with respect to parameters $\beta(\bmv{r};t)$ and $a'(x;t)$
%
%
\be
\begin{array}{lcl}
\ds\frac{\delta\Phi(t)}{\delta\beta(\bmv{r};t)}&=&
        -\la\hat{\cal E}'(\bmv{r})\ra^t_{\rm q}=
        -\la\hat{\cal E}'(\bmv{r})\ra^t,\\
\ds\frac{\delta\Phi(t)}{\delta a(x;t)}&=&
        -\la\hat{n}_1(x)\ra^t_{\rm q}=-\la\hat{n}_1(x)\ra^t=-f_1(x;t).\\
\end{array}
\label{e2.18}
\ee
It means that parameter $\beta(\bmv{r};t)$ is conjugated to the average 
energy in an accompanying reference frame, and $a'(x;t)$ is conjugated
to the nonequilibrium one-particle distribution function $f_1(x;t)$. To
determine the physical meaning of these parameters let us define the 
entropy of a system taking into account the self-con\-sis\-ten\-cy 
conditions (\ref{e2.17}):
%
%
\be
S(t)=-\la\ln\vrho_{\rm q}\xnt\ra^t_{\rm q}=\Phi(t)+\int\d\bmv{r}\;
        \beta(\bmv{r};t)\la\hat{\cal E}'(\bmv{r})\ra^t+
        \int\d x\;a'(x;t)\la\hat{n}_1(x)\ra^t.\label{e2.19}
\ee
Taking functional derivatives of $S(t)$ (\ref{e2.19}) with respect to
$\la\hat{\cal E}'(\bmv{r})\ra^t$ and $\la\hat{n}_1(x)\ra^t$ at fixed
corresponding averaged values gives the following thermodynamic relations:
%
%
\be
\frac{\delta S(t)}{\delta\la\hat{\cal E}'(\bmv{r})\ra^t}=\beta(\bmv{r};t),
        \quad\frac{\delta S(t)}{\delta f_1(x;t)}=a'(x;t).\label{e2.20}
\ee
Hence, $\beta(\bmv{r};t)$ is an analogue of the  inverse local temperature.

Two limiting cases follow from the structure of expression (\ref{e2.19})
for entropy. At $a'(x;t)=-\beta(\bmv{r};t)\mu(\bmv{r};t)$, (\ref{e2.19})
transforms into an expression for entropy which corresponds to the 
hydrodynamic description of the nonequilibrium state of the system 
\cite{5,9,9a}:
%
%
\be
S(t)=\Phi(t)+\int\d\bmv{r}\;\beta(\bmv{r};t)\lp\la\hat{\cal E}'(\bmv{r})\ra^t-
        \mu(\bmv{r};t)\la\hat{n}(\bmv{r})\ra^t\rp,\label{e2.21}
\ee
with the following quasiequilibrium distribution function \cite{5,9,9a}:
%
%
\bea
\vrho_{\rm q}\xnt&=&\exp\lc-\Phi(t)-
        \int\d\bmv{r}\;\beta(\bmv{r};t)\Big(\hat{\cal E}'(\bmv{r})-
        \mu(\bmv{r};t)\hat{n}(\bmv{r})\Big)\rc,\label{e2.22}\\
\Phi(t)&=&\ln\int\d\mGamma_N\;
        \exp\lc\int\d\bmv{r}\;\beta(\bmv{r};t)\Big(\hat{\cal E}'(\bmv{r})-
        \mu(\bmv{r};t)\hat{n}(\bmv{r})\Big)\rc\nonumber
\eea
and the corresponding self-consistency conditions for the definition
of thermodynamic parameters $\beta(\bmv{r};t)$,
$\mu(\bmv{r};t)$ (local value of the chemical potential):
\[
\la\hat{n}(\bmv{r})\ra^t_{\rm q}=\la\hat{n}(\bmv{r})\ra^t,\quad
\la\hat{\cal E}'(\bmv{r})\ra^t_{\rm q}=\la\hat{\cal E}'(\bmv{r})\ra^t.
\]

In the case when the contribution of the interaction energy between the 
particles can be neglected (a dilute gas), the quasiequilibrium 
distribution function (\ref{e2.15}) has the form:
\[
\vrho_{\rm q}\xnt=\exp\lc-\Phi(t)-\int\d\bmv{r}\;\beta(\bmv{r};t)
        \hat{\cal E}'_{\rm kin}(\bmv{r})-\int\d x\;a'(x;t)\hat{n}_1(x)\rc,
\]
or, taking into account \refp{e2.16} and the relation between $a'(x;t)$
and $a(x;t)$, one obtains:
%
%
\bea
\vrho_{\rm q}\xnt&=&\exp\lc-\Phi(t)-\int\d x\;a(x;t)\hat{n}_1(x)\rc,
        \label{e2.23}\\
\Phi(t)&=&\ln\int\d\mGamma_N\;\exp\lc-\int\d x\;a(x;t)\hat{n}_1(x)\rc,
        \nonumber
\eea
where
\[
\hat{\cal E}'_{\rm kin}(\bmv{r})=
        \hat{\cal E}_{\rm kin}(\bmv{r})-\bmv{V}(\bmv{r};t)
        \hat{\bmv{j}}(\bmv{r})+\frac m2V^2(\bmv{r};t)\hat{n}(\bmv{r}),
\]
$\hat{\cal E}_{\rm kin}(\bmv{r})=\int\d\bmv{p}\;\frac{p^2}{2m}\hat{n}_1(x)$
is the density of the kinetic energy. Now, determining in \refp{e2.23}
parameter $a(x;t)$ with the help of the self-consistency condition
$\la\hat{n}(x)\ra^t_{\rm q}=\la\hat{n}(x)\ra^t$, one can show \cite{5} that
\refp{e2.23} for $\vrho_{\rm q}\xnt$ transforms into distribution
\refp{e2.8} when it is assumed that the only parameter of the abbreviated 
description for a nonequilibrium state of the system is a one-particle 
distribution function. As it is known \cite{50,52}, the quasiequilibrium 
distribution function \refp{e2.23} corresponds to the Boltzmann entropy of 
a dilute gas:
%
%
\be
S_{\rm B}(t)=-\int\d x\;f_1(x;t)\ln\frac{f_1(x;t)}{\e}.\label{e2.24}
\ee

In a general case, when kinetic and hydrodynamic processes are considered
simultaneously, the quasi\-equilibrium distribution function \refp{e2.15}
or \refp{e2.13} can be rewritten in a somewhat different form. This form is 
more convenient for the comparison with $\vrho_{\rm q}\xnt$ \refp{e2.8}, 
obtained in a usual way \cite{21,51}, when $f_1(x;t)$ is the only 
parameter of the abbreviated description. First of all, let us note that one
can include parameter $\Phi(t)$ from \refp{e2.13} into parameter
$a(x;t)$ as a term which does not depend on $x$. Parameter $a(x;t)$ in
$\vrho_{\rm q}\xnt$ can be excluded with the help of the self-consistency
condition $\la\hat{n}(x)\ra^t_{\rm q}=\la\hat{n}(x)\ra^t=f_1(x;t)$. 
Reduction of $\vrho_{\rm q}\xnt$ results in
%
%
\be
\vrho_{\rm q}\xnt=\exp\lc-U_N\lp\bmv{r}^N;t\rp\rc\prod_{j=1}^N
        \frac{f_1(x_j;t)}{u(\bmv{r}_j;t)},\label{e2.25}
\ee
where functions $u(\bmv{r}_j;t)$ are obtained from the relations
%
%
\bea
u(\bmv{r}_j;t)&=&\int\frac{\d\bmv{r}^{N-1}}{(N-1)!}
        \exp\lc-U_N\lp\bmv{r},\bmv{r}^{N-1};t\rp\rc
        \prod_{j=2}^N\frac{n(\bmv{r}_j;t)}{u(\bmv{r}_j;t)},
        \label{e2.26}\\
U_N\lp\bmv{r}^N;t\rp&=&U_N\lp\bmv{r}_1,\ldots,\bmv{r}_N;t\rp=
        \frac 12 \sum_{j\ne k=1}^N\mPhi(|\bmv{r}_j-\bmv{r}_k|)
        \beta(\bmv{r}_k;t)\nonumber,
\eea
$n(\bmv{r};t)=\la\hat{n}(\bmv{r})\ra^t=\int\d\bmv{p}\;f_1(x;t)$ is the
nonequilibrium particles concentration. In expression \refp{e2.25}, 
$U_N(\bmv{r};t)$ and $u(\bmv{r}_j;t)$ respectively, depend explicitly and 
implicitly on $n(\bmv{r};t)$ and $\beta(\bmv{r};t)$ (or 
$\la\hat{\cal E}'(\bmv{r})\ra^t$). To obtain the ordinary Bogolubov scheme 
\cite{21,51}, it is necessary to put $U_N(\bmv{r};t)=0$ in \refp{e2.25} and 
\refp{e2.26}. Then, one can define $u=\e$, and \refp{e2.25} transforms into 
the quasiequilibrium distribution \refp{e2.8}, as it should be. In a 
general case, $u(\bmv{r};t)$ is a functional of the nonequilibrium density 
of particles number $n(\bmv{r};t)$ and $\beta(\bmv{r};t)$, which is an 
analogue of the  inverse local temperature. Nevertheless, one should handle 
this analogy with care, as far as definition \refp{e2.25} can describe 
states which are far from local equilibrium. In particular, $f_1(x;t)$ can 
differ considerably from the local Maxwellian distribution.

The entropy expression \refp{e2.19} can be transformed according to the
structure of the quasiequilibrium distribution function \refp{e2.25}
%
%
\be
S(t)=\int\d\bmv{r}\;\beta(\bmv{r};t)\la\hat{\cal E}_{\rm int}(\bmv{r})\ra^t-
        \int\d x\;f_1(x;t)\ln\frac{f_1(x;t)}{u(\bmv{r};t)}.\label{e2.27}
\ee
Here the potential and kinetic parts are separated. In the case of low 
density gases, the influence of the potential energy can be neglected and 
$u(\bmv{r};t)=\e$. Then, expression \refp{e2.27} tends to the usual 
Boltzmann entropy.

Thus, determining the quasiequilibrium distribution function 
$\vrho_{\rm q}\xnt$ \refp{e2.25} and entropy $S(t)$ \refp{e2.27}
of the system, when the nonequilibrium one-particle distribution function, 
as well as the average values of densities for the number of particles, 
momentum and energy are parameters of an abbreviated description of the 
nonequilibrium state which are locally conserved, the Liouville equation 
with the source \refp{e2.7} can be presented in the form \cite{1,5}:
%
%
\be
\lp\ddt+\im L_N\rp\vrho\xnt=-\veps\lp\vrho\xnt-
        \exp\lc-U_N\lp\bmv{r}^N;t\rp\rc\prod_{j=1}^N
        \frac{f_1(x_j;t)}{u(\bmv{r}_j;t)}\rp.
        \label{e2.28}
\ee
Further, on the basis of this equation, one obtains the BBGKY hierarchy for 
nonequilibrium distribution functions of classical particles with modified 
bo\-un\-da\-ry conditions which take into account the nonequilibriumnes of 
the one-particle distribution function, as well as the local conservation 
laws. To obtain the first equation of the hierarchy, let us integrate
with respect to the variables $x^{N-1}=\big\{x_2,\ldots,x_N\big\}$ the both 
parts of \refp{e2.28}. Taking into account \refp{e2.10}, \refp{e2.11} and 
\refp{e2.17}, one obtains:
%
%
\be
\lp\ddt+\im L(1)\rp f_1(x_1;t)+\int\d x_2\;\im L(1,2)f_2(x_1,x_2;t)=0.
        \label{e2.29}
\ee
Integrating now \refp{e2.28} over the variables 
$x^{N-2}=\big\{x_3,\ldots,x_N\big\}$, after simple transformations one can 
obtain an equation for the two-particle distribution function 
$f_2(x_1,x_2;t)$ which differs from the corresponding equation in the 
Bogolubov hierarchy \cite{1,3} by a source in the right-hand side:
%
%
\bea
&&\lp\ddt+\im L_2\rp f_2(x_1,x_2;t)+\int\d x_3
        \Big(\im L(1,3)+\im L(2,3)\Big)f_3(x_1,x_2,x_3;t)={}\nonumber\\
&&-\veps\Big(f_2(x_1,x_2;t)-\g_2(\bmv{r}_1,\bmv{r}_2;t)
        f_1(x_1;t)f_1(x_2;t)\Big).\label{e2.30}
\eea
In this equation, $L_2=L(1)+L(2)+L(1,2)$ is the Liouville two-particle 
operator and $\g_2(\bmv{r}_1,\bmv{r}_2;t)$ denotes the binary coordinate 
distribution function for the quasiequilibrium state \refp{e2.25};
$f_2(x_1,x_2;t)$ is the nonequilibrium two-particle distribution function:
\[
f_2(x_1,x_2;t)=\la\hat{n}_2(x_1,x_2;t)\ra^t=
        \int\d\mGamma_{N-2}\;\vrho(x_1,x_2,x^{N-2};t).
\]
Similarly, integrating equation \refp{e2.28} over 
$x^{N-s}=\big\{x_{s+1},\ldots,x_N\big\}$ phase variables, one derives next 
equations of the chain with the corresponding sources which define boundary 
conditions for reduced distribution functions. For the $s$-particle 
nonequilibrium distribution function
\[
f_s(x^s;t)=\la\hat{n}_s(x^s)\ra^t=\int\d\mGamma_{N-s}\;
        \vrho(x_1,\ldots,x_s,x^{N-s};t)
\]
one has the following equation:
%
%
\bea
&&\lp\ddt+\im L_s\rp f_s(x^s;t)+\int\d x_{s+1}
        \sum_{j=1}^s\im L(j,s+1)f_{s+1}(x^{s+1};t)={}\nonumber\\
&&-\veps\Big(f_s(x^s;t)-\g_s(\bmv{r}_s;t)
        \prod_{j=1}^sf_1(x_j;t)\Big),\label{e2.31}
\eea
where
%
%
\be
\begin{array}{rclclcl}
\ds L_s&=&\ds\sum_{j=1}^sL(j)+\frac 12\sum_{j\ne k=1}^sL(j,k),&\quad&
        \ds \g_s(\bmv{r}^s;t)&=&\ds f_s(\bmv{r}^s;t)\Big/
        \prod_{j=1}^sn(\bmv{r}_j;t),\\
        \\
\ds\ds\hat{n}_s(\bmv{r}^s)&=&
        \ds\mathop{\sum_{j_1=1}^s\ldots\sum_{j_s=1}^s}
        \limits_{j_1\ne j_2\ne\ldots j_s}\sum_{k=1}^s
        \delta(\bmv{r}_{k}-\bmv{r}'_{j_k}),&\quad&
        f_s(\bmv{r}^s;t)&=&\ds\la\hat{n}_s(\bmv{r}^s)\ra^t_{\rm q}.
\end{array}
\label{e2.32}
\ee
As usual, we shall assume that the principle of weakening spatial 
correlations is valid for the quasiequilibrium state. Then, in the 
thermodynamic limit, the coordinate distribution functions 
$f_s(\bmv{r}^s;t)$, $\g_s(\bmv{r}^s;t)$ satisfy the boundary relations:
%
%
\bea
\lim_{(\min|\bmv{r}_j-\bmv{r}_k|)\to\infty}
        f_s(\bmv{r}_1,\ldots,\bmv{r}_s;t)&=&
        \prod_{j=1}^sn(\bmv{r}_j;t),\label{e2.33}\\
\lim_{(\min|\bmv{r}_j-\bmv{r}_k|)\to\infty}
        \g_s(\bmv{r}_1,\ldots,\bmv{r}_s;t)&=&1.\label{e2.34}
\eea
Therefore, taking into account ``slow'' hydrodynamical variables (the 
density of the interaction energy in the present case) leads to a 
modification of boundary conditions for the chain of equations 
\refp{e2.29}--\refp{e2.31} for nonequilibrium distribution functions. In 
order to reproduce the usual Bogolubov boundary conditions for the 
weakening of correlations \cite{21}, it is necessary to replace all 
$\g_s(\bmv{r}^s;t)$ by their limiting values \refp{e2.34}. Such a replacing 
is valid in the case of small density, however, new boundary conditions can 
be more useful for dense gasses, since they automatically take into 
account spatial correlations connected with the interaction of a pair of 
particles with the rest of the particles of the system. It is obvious that 
the influence of such an interaction increases as the density rises.

We note that the chain of equations \refp{e2.29}--\refp{e2.31} requires us 
to add equations for coordinate quasiequilibrium distribution functions 
which are functionals on the nonequilibrium density of the particles number 
$n(\bmv{r};t)$ and the inverse local temperature $\beta(\bmv{r};t)$. In 
particular, in \cite{53} it is shown that the binary quasiequilibrium 
distribution function $\g_2(\bmv{r}_1,\bmv{r}_2;t)$ is connected with the 
pair quasiequilibrium correlation function 
$h_2(\bmv{r}_1,\bmv{r}_2;t)=\g_2(\bmv{r}_1,\bmv{r}_2;t)-1$ which satisfies 
the Ornstein-Zernike equation:
%
%
\be
h_2(\bmv{r}_1,\bmv{r}_2;t)=c_2(\bmv{r}_1,\bmv{r}_2;t)+
        \int\d\bmv{r}_3\;c_2(\bmv{r}_1,\bmv{r}_3;t)
        h_2(\bmv{r}_2,\bmv{r}_3;t)n(\bmv{r}_3;t),\label{e2.35}
\ee
where $c_2(\bmv{r}_1,\bmv{r}_2;t)$ is a direct quasiequilibrium correlation 
function.

Thus, the chain of equations \refp{e2.29}--\refp{e2.31} is distinguished
from the usual Bogolubov hierarchy by the existence of sources in the 
right-hand sides, beginning from the second equation, and it takes into 
account both one-particle and collective hydrodynamical effects. It is 
important to investigate solutions to this chain of equations in the 
simplest approximations which lead to model kinetic equations for dense 
gases and simple liquids. We shall consider the most investigated binary 
collision approximation.

%% file: s_02_3.tex
\subsection{Binary collision approximation}

In this section, we consider as an example of the use of hierarchy 
\refp{e2.31}, the simplest approximation of binary collisions which in the 
case of the Bogolubov ordinary conditions of correlations weakening leads 
to the Boltzmann equation for $f_1(x_1;t)$. In equation \refp{e2.30} for 
$f_2(x_1,x_2;t)$ we omit the term with the three-particle distribution
function, i.e. we take into account the influence of the ``medium'' on the
evolution of the distinguished pair of particles only through the
correlation corrections in the boundary condition. Then, we arrive at the
equation:
%
%
\be
\lp\ddt+\im L_2+\veps\rp f_2(x_1,x_2;t)=
\veps \g_2(\bmv{r}_1,\bmv{r}_2;t)f_1(x_1;t)f_1(x_2;t).\label{e2.36}
\ee
The formal solution of \refp{e2.36} has the form:
%
%
\be
f_2(x_1,x_2;t)=\veps\int\limits^{0}_{-\infty}\d\tau\;
	\e^{(\veps+\im L_2)\tau}\g_2(\bmv{r}_1,\bmv{r}_2;t+\tau)
        f_1(x_1;t+\tau)f_1(x_2;t+\tau).\label{e2.37}
\ee
Following the Abel theorem \cite{9,9a,50,54,55}, this solution can be written 
in the form:
%
%
\be
f_2(x_1,x_2;t)=\lim_{\tau\to-\infty}\e^{\im L_2\tau}
	\g_2(\bmv{r}_1,\bmv{r}_2;t+\tau)f_1(x_1;t+\tau)f_1(x_2;t+\tau).
	\label{e2.38}
\ee
Substituting expression \refp{e2.38} into equation \refp{e2.29}, one obtains
a kinetic equation in the binary collision approximation
%
%
\be
\lp\ddt+\im L(1)\rp f_1(x_1;t)=I_{\rm coll}(x_1;t),
        \label{e2.39}
\ee
where
%
%
\be
I_{\rm coll}(x_1;t)=\int\d x_2\;\im L(1,2)\lim_{\tau\to-\infty}
	\e^{\im L_2\tau}\g_2(\bmv{r}_1,\bmv{r}_2;t+\tau)
        f_1(x_1;t+\tau)f_1(x_2;t+\tau)\label{e2.40}
\ee
is a collision integral. It is necessary to point out that equation
\refp{e2.35} for the pair quasiequilibrium correlation function must be 
added to kinetic equation \refp{e2.29}. Equation \refp{e2.35} takes into 
account an essential part of the many-particles correlations.

%% file: s_02_31.tex
\subsubsection{Kinetic equation of the revised Enskog theory}

In papers \cite{1,3} it was shown how the collision integral \refp{e2.40} 
in the case of a system of hard spheres transforms into the collision 
integral of the RET kinetic equation \cite{33}. For subsequent 
manipulations, it is convenient to represent the particle interaction 
potential in the form:
%
%
\be
\mPhi^{\rm hs}(|\bmv{r}_{jk}|)=\lim_{a\to+\infty}\mPhi^a(|\bmv{r}_{jk}|),
\quad\mPhi^a(|\bmv{r}_{jk}|)=\left\{
				\begin{array}{cc}
				a,\quad|\bmv{r}_{jk}|<\sigma,\\
				0,\quad|\bmv{r}_{jk}|\geqslant\sigma,\\
				\end{array}\right.\label{e3.1}
\ee
where $\sigma$ is a hard sphere diameter. Since the potential 
$\mPhi^{\rm hs}(|\bmv{r}_{jk}|)$ is strongly singular (in particular, the 
operator $\im L(1,2)$ is not well defined), we shall operate with the 
potential $\mPhi^a(|\bmv{r}_{jk}|)$, when deriving the kinetic equation, 
and set $a\to+\infty$ in the final expressions.

The limit $\tau\to-\infty$ in the collision integral \refp{e2.40} is
mathematically formal. At the physical level of description, this limit
assumes that $|\tau|\gg\tau_0$, where $\tau_0>0$ is some characteristic
time scale. Depending on the choice of $\tau_0$, we obtain different stages
of the evolution of the system (kinetic, hydrodynamic). At the kinetic 
stage, $\tau_0$ is a characteristic interaction time for which the singular
potential \refp{e3.1} is an arbitrarily small quantity $(\tau_0\to+0)$.
Therefore, the limit $\tau-\tau_0\to-\infty$ keeps its form even in the
case $\tau\to-0$, provided $\tau_0$ is of the higher order of smallness 
relatively to $\tau$ :
%
%
\be
\lim_{\tau\to -0}\left\{\lim_{\tau_0\to +0}
	\frac{\tau}{\tau_0}\right\}=-\infty.\label{e3.2}
\ee
In this case the collision integral \refp{e2.40} \cite{1,3,53} for the hard 
spheres interparticle potential of interaction transforms into the 
following final expression:
%
%
\begin{eqnarray}
&&I_{\rm coll}^{\rm hs}(x_1;t)=\sigma^2\int\d\hat{r}_{12}\;\d\bmv{v}_2\;
	\theta(\hat{\bmv{r}}_{12}\cdot\gb)
	(\hat{\bmv{r}}_{12}\cdot\gb)\times{}\label{e3.42}\\
&&\quad
\left\{\g_2^{\rm hs}(\bmv{r}_1,\bmv{r}_1+\sigma^+\hat{\bmv{r}}_{12};t) 
	f_1(\bmv{r}_1,\bmv{v}_1';t)
	f_1(\bmv{r}_1+\sigma^+\hat{\bmv{r}}_{12},\bmv{v}_2';t)-{}
	\right.\nonumber\\
&&\quad\phantom{\{}\!\left.
	\g_2^{\rm hs}(\bmv{r}_1,\bmv{r}_1-\sigma^+\hat{\bmv{r}}_{12};t)
	f_1({\bf r}_1,\bmv{v}_1;t)
	f_1(\bmv{r}_1-\sigma^+\hat{\bmv{r}}_{12},\bmv{v}_2;t)\rc,\nonumber
\end{eqnarray}
where
%
%
\be
\begin{array}{lll}
\ds\bmv{v}_1'&=&
	\ds\bmv{v}_1+\hat{\bmv{r}}_{12}(\hat{\bmv{r}}_{12}\cdot\gb),\\
\ds\bmv{v}_2'&=&
	\ds\bmv{v}_2-\hat{\bmv{r}}_{12}(\hat{\bmv{r}}_{12}\cdot\gb).\\
\end{array}\label{e3.43}
\ee
The collision integral in the form \refp{e3.42} is identical to the 
collision integral of the RET theory first introduced by van Beijeren and 
Ernst \cite{33} on the basis of a diagram method. As R\'esibois showed 
\cite{34,35}, the kinetic equation of the RET theory satisfies an
$H$-theorem.

We represent the collision integral \refp{e3.42} in a more compact form by 
using the two-particle quasi-Liouvillian (evolution pseudo-operator) 
$\hat{T}(1,2)$ \cite{25}. Then, the kinetic equation takes the form:
%
%
\be
\lp\ddt+\bmv{v}_1\dd{\bmv{r}_1}\rp f_1(x_1;t)=
	\int\d x_2\;\hat{T}(1,2)\g_2^{\rm hs}(\bmv{r}_1,\bmv{r}_2;t)
	f_1(x_1;t)f_1(x_2;t),\label{e3.44}
\ee
%
%
%
%
%
\bea
\hat{T}(1,2)&=&\sigma^2\int\d\hat{\bmv{r}}_{12}\;
	\theta(\hat{\bmv{r}}_{12}\cdot\gb)
	(\hat{\bmv{r}}_{12}\cdot\gb)\times{}\label{e3.45}\\
&&\left\{\delta(\bmv{r}_{12}+\sigma^+\hat{\bmv{r}}_{12})
	\hat{B}(\hat{\bmv{r}}_{12})-
	\delta(\bmv{r}_{12}-\sigma^+\hat{\bmv{r}}_{12})\right\},\nonumber
\eea
%
%
%
%
%
\be
\hat{B}(\hat{\bmv{r}}_{12})\Psi(\bmv{v}_1,\bmv{v}_2)=
	\Psi(\bmv{v}_1',\bmv{v}_2'),\label{e3.46}
\ee
where $\Psi$ is an arbitrary function of the velocities.

Thus, on the basis of the considered approach we have derived the kinetic
equation of the RET theory without additional phenomenological assumptions. 
We have shown that within the framework of the employed method this equation
corresponds to the simplest pair-collision approximation without allowance
for retardation in time.

%% file: s_02_32.tex
\subsubsection{Kinetic equation for a multistep potential of interaction}

\begin{figure}[hbt]
\unitlength=1pt
\begin{centering}
\begin{picture}(244,232)
\put(0,0){\epsffile[177 305 421 537]{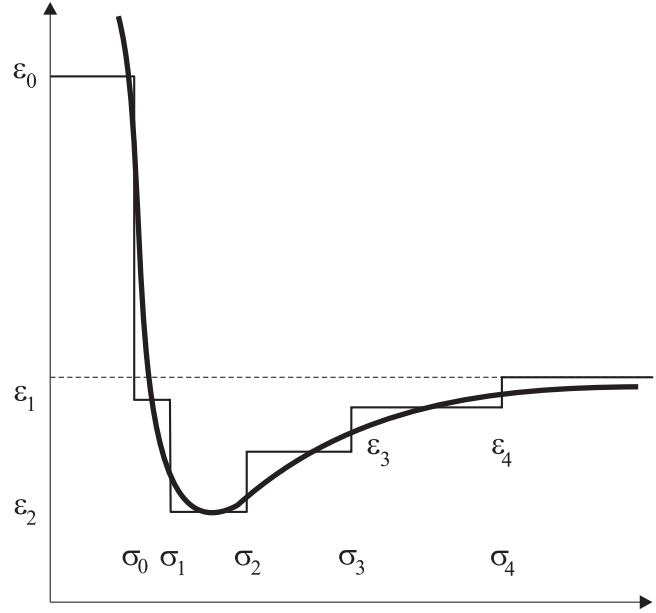}}
\end{picture}\\
\end{centering}
\caption{The multistep potential at $n^*=1$ and $m^*=3$.}
\label{f1}
\end{figure}

The collision integral \refp{e2.40} is still rather complicated to be 
written in an explicit form for an arbitrary potential of interaction. As 
it is now known \cite{29,31}, only in two particular cases, namely, for 
hard spheres and square-well potentials, this integral is  reduced to 
an analytical form. In order to extend these previous results, we consider 
the interparticle potential in the form of a multistep function
%
%
\be
\mPhi_{jk}\equiv\mPhi_{jk}^{\rm ms}=\lim_{\veps_0\to\infty}
	\mPhi_{jk}^{\veps_0}(r_{jk}),\label{e3.47}
\ee
where
%
%
\be
\mPhi_{jk}^{\veps_0}(r_{jk})=\left\{
\begin{array}{ll}
\veps_0,&\quad r_{jk}<\sigma_0;\\
\veps_l,&\quad \sigma_{l-1}<r_{jk}<\sigma_l,\quad l=1,\ldots,N^*;\\ 
0,&\quad\sigma_{N^*}<r_{jk};\\
\end{array}
\right.\label{e3.48}
\ee
and $N^*$ is the total number of walls except for a hard sphere wall. 
Potential of a hard sphere contains strong singularity (operator $L(j,k)$ is
hard to define). That is why we shall deal with the potential 
$\mPhi_{jk}^{\veps_0}(r_{jk})$ and only in final expressions we shall put
$\veps_0\to\infty$. Let us for convenience separate the systems of
attractive and repulsive walls. Let $n^*$ be the number of repulsive walls at
the distance $\sigma_{{\rm r}i}\equiv\sigma_i$ and of the
height $\vt\veps^{\rm r}_i=\veps^{\rm r}_i-\veps^{\rm r}_{i+1}>0$, where 
$\veps^{\rm r}_i\equiv\veps_i$, $i=1,\ldots,n^*$; and $m^*$ -- 
the number of attractive walls with the corresponding parameters 
$\sigma_{{\rm a}j}\equiv\sigma_{j+n^*}$, 
$\vt\veps^{\rm a}_j=\veps^{\rm a}_{j+1}-\veps^{\rm a}_j>0$, where 
$\veps^{\rm a}_j\equiv\veps_{j+n^*}$, $j=1,\ldots,m^*$; 
$\veps^{\rm r}_{n^*+1}=\veps^{\rm a}_1$, $\veps^{\rm a}_{m^*+1}=0$, 
$N^*=n^*+m^*$. Thus, the parameters $\sigma_0$ (hard sphere diameter), 
$n^*$, $\sigma_{{\rm r}i}$, $\vt\veps^{\rm r}_i$, $m^*$, 
$\sigma_{{\rm a}j}$ and $\vt\veps^{\rm a}_j$ completely determine 
the geometry of the multistep potential (see figure \ref{f1} where a 
specific case $n^*=1$, $m^*=3$ is shown and the multistep function 
approximates some real smooth  potential).

The limit $\tau\to-\infty$ in the collision integral \refp{e2.40} is a  
formal one in the mathematical sense. On the physical level of description 
this means that $|\tau|\gg\tau_0$, where $\tau_0>0$ is some characteristic 
interval of time. For different values of $\tau_0$ we can investigate
different stages of the evolution of the system. In the Boltzmann theory 
\cite{56} $|\tau|\gg\tau_{\rm c}$, where $\tau_{\rm c}$ is the time of 
binary interactions (collision time). On the other hand, $|\tau|$ is far 
less than the characteristic scale of time $\tau_{\rm m}$ for hydrodynamic 
variables. The above situation is possible because of the fact that for 
dilute gases the kinetic ($\tau\sim\tau_{\rm c}$) and hydrodynamic 
($\tau\sim\tau_{\rm m}$) stages of the evolution are far from one another 
in time, i.e. $\tau_{\rm c}\ll\tau_{\rm f}\ll\tau_{\rm m}$, where 
$\tau_{\rm f}$ is the characteristic time of free motion. The pattern is 
qualitatively different in dense gases and liquids with a realistic smooth 
potential of interaction. The kinetic and hydrodynamic stages appear to be 
closely connected. Besides, such quantities as the length of free motion 
and the time of interaction are not defined in a usual way, because all 
particles make influence on the dynamics of interaction for some chosen 
pair of particles.

However, for special types of potentials such a classification of times 
remains valid even for high densities. For a multi-step potential the 
region $\Omega$ of binary interactions consists of the following subregions 
$\ls\sigma_k-\vt r_0,\sigma_k+\vt r_0\rs,k=0,1,\ldots,N^*$, where 
$\vt r_0\to+0$ due to the singular nature of the potential under 
consideration. This potential has the finite range 
$\;\max\lc\sigma_k\rc=\sigma_{N^*}>0\;$ of action. We can introduce
the following set of specific time intervals: $\tau_0=\vt r_0/\g_0\to+0$ is 
the time of pair interactions on the walls, 
$\vt\tau=\min\lc\sigma_k-\sigma_{k-1}\rc/\g_0>0$ 
is the time of motion between the two nearest neighbouring walls and 
$t_{\rm whole}=\sigma_{N^*}/\g_0>0$ is the time of motion of the whole 
systems of walls for some pair of particles, where $\g_0$ is an average 
relative velocity of two particles. As the potential contains the 
horizontal parts, where the force of the interparticle interaction is equal 
to zero, it is also possible to introduce time $\tau_{\rm f}$ as an average 
time of free motion of particles in the system. Changing the geometry of 
the potential and increasing the density, we can make this time arbitrarily 
small in order to match the relation
%
%
\be
\tau_0\ll\tau_{\rm f}\ll\vt\tau<t_{\rm whole}<\tau_{\rm m}.\label{e3.49}
\ee
The kinetic stage of the evolution corresponds to small times of order 
$\tau\sim\tau_{\rm f}$. Therefore, as far as the relation \refp{e3.49} is 
satisfied, the formal limit $\tau\to-\infty$ should be considered as 
$\tau/\tau_0\to-\infty$ or merely as $|\tau|\gg\tau_0$. For smooth 
non-singular potentials the region of interaction has some non-zero size 
and $\tau_0$ is finite. For singular potentials the case $\tau_0\to+0$ is
possible and the limit $\tau/\tau_0\to-\infty$ remains valid even for 
$\tau\to-0$ if only $\tau_0$ is a value of the higher infinitesimal order
%
%
\be
\lim_{\tau\to-0}\lc\lim_{\tau_0\to+0}\frac{\tau}{\tau_0}\rc\to-\infty.
	\label{e3.50}
\ee
Thus, the formal limit $\tau\to-0$ can be applied to the collision integral 
\refp{e2.40} in the sense of \refp{e3.49} and \refp{e3.50}. As it is now 
well established, the limit $\tau \to -0$ leads to the kinetic equations of 
the RET \cite{3,33,36} and RDRS \cite{39} theories.

As it was shown in \cite{2,8}, the collision integral \refp{e2.40} for 
potential \refp{e3.47} taking into account \refp{e3.49}, \refp{e3.50}, can 
be presented in a compact form:
%
%
\be
I_{\rm coll}^{\rm ms}(x_1;t)=
	\int\d x_2\;\hat{T}_{12}\g_2(\bmv{r}_1,\bmv{r}_2;t)
	f_1(x_1;t)f_1(x_2;t)\label{e3.73}
\ee
in terms of the two-particle pseudo-Liouville operator
%
%
\be
\hat T_{12}=\hat T_{\rm hs}+\sum_{i=1}^{n^*}\sum_{p=b,c,d}
	\hat T_{{\rm r}i}^p+\sum_{j=1}^{m^*}\sum_{p=b,c,d}
	\hat T_{{\rm a}j}^p,\label{e3.74}
\ee
which consists of the pseudo-Liouville operators for a hard sphere wall, 
for $i$th repulsive wall and for $j$th attractive wall, correspondingly:
%
%
\bea
\hat T_{\rm hs}&=&\sigma_0^2\,\int\d\hat{\bmv{r}}_{12}
	(\hat{\bmv{r}}_{12}\cdot\gb) 
	\theta(\hat{\bmv{r}}_{12}\cdot\gb)\left\{
	\delta(\bmv{r}_{12}+\sigma_0^+\hat{\bmv{r}}_{12})
	\hat B^a(\hat{\bmv{r}}_{12})-
	\delta(\bmv{r}_{12}-\sigma_0^+\hat{\bmv{r}}_{12})\rc,
	\nonumber\\\label{e3.75}\\
\hat T_{{\rm r}i}^p&=&\sigma_{{\rm r}i}^2\int\d\hat{\bmv{r}}_{12}
	(\hat{\bmv{r}}_{12}\cdot\gb)
	\theta_{{\rm r}i}^p(\hat{\bmv{r}}_{12}\cdot\gb)\lc
	\delta(\bmv{r}_{12}+\sigma_{{\rm r}i}^{1,p}\hat{\bmv{r}}_{12})
	\hat B_{{\rm r}i}^p(\hat{\bmv{r}}_{12})-
	\delta(\bmv{r}_{12}-\sigma_{{\rm r}i}^{2,p}\hat{\bmv{r}}_{12})\rc,
	\nonumber\\\label{e3.76}\\
\hat T_{{\rm a}j}^p&=&\sigma_{{\rm a}j}^2\int\!\d\hat{\bmv{r}}_{12}
	(\hat{\bmv{r}}_{12}\cdot\gb)
	\theta_{{\rm a}j}^p(\hat{\bmv{r}}_{12}\cdot\gb)\lc
	\delta(\bmv{r}_{12}+\sigma_{{\rm a}j}^{1,p}\hat{\bmv{r}}_{12})
	\hat B_{{\rm a}j}^p(\hat{\bmv{r}}_{12})-
	\delta(\bmv{r}_{12}-\sigma_{{\rm a}j}^{2,p}\hat{\bmv{r}}_{12})\rc.
	\nonumber\\\label{e3.77}
\eea
Here,
%
%
\be
\begin{array}{ll}
\theta_{{\rm r}i}^b(\hat{\bmv{r}}_{12}\cdot\gb)=
\theta(-\hat{\bmv{r}}_{12}\cdot\gb),&\qquad
\theta_{{\rm a}j}^b(\hat{\bmv{r}}_{12}\cdot\gb)=
\theta(\hat{\bmv{r}}_{12}\cdot\gb),\\ [0.5ex]
\theta_{{\rm r}i}^c(\hat{\bmv{r}}_{12}\cdot\gb)=
\theta(\hat{\bmv{r}}_{12}\cdot\gb-\alpha^{\rm r}_i),&\qquad
\theta_{{\rm a}j}^c(\hat{\bmv{r}}_{12}\cdot\gb)=
\theta(-\hat{\bmv{r}}_{12}\cdot\gb-\alpha^{\rm a}_j),\\ [0.5ex]
\theta_{{\rm r}i}^d(\hat{\bmv{r}}_{12}\cdot\gb)=
\theta(\hat{\bmv{r}}_{12}\cdot\gb)
\theta(\alpha^{\rm r}_i-\hat{\bmv{r}}_{12}\cdot\gb),&\qquad
\theta_{{\rm a}j}^d(\hat{\bmv{r}}_{12}\cdot\gb)=
\theta(-\hat{\bmv{r}}_{12}\cdot\gb)
\theta(\alpha^{\rm r}_i+\hat{\bmv{r}}_{12}\cdot\gb)\\
\end{array}\label{e3.78}
\ee
are the corresponding unit step functions,
%
%
\bea
\sigma_{{\rm r}i}^{1,b}=\sigma_{{\rm r}i}^+,\qquad
\sigma_{{\rm r}i}^{2,b}=\sigma_{{\rm r}i}^-,\qquad
\sigma_{{\rm a}j}^{1,b}=\sigma_{{\rm a}j}^-,\qquad
\sigma_{{\rm a}j}^{2,b}=\sigma_{{\rm a}j}^+,\nonumber\\ [0.5ex]
\sigma_{{\rm r}i}^{1,c}=\sigma_{{\rm r}i}^-,\qquad
\sigma_{{\rm r}i}^{2,c}=\sigma_{{\rm r}i}^+,\qquad
\sigma_{{\rm a}j}^{1,c}=\sigma_{{\rm a}j}^+,\qquad
\sigma_{{\rm a}j}^{2,c}=\sigma_{{\rm a}j}^-,\label{e3.79}\\ [0.5ex]
\sigma_{{\rm r}i}^{1,d}=\sigma_{{\rm r}i}^+,\qquad
\sigma_{{\rm r}i}^{2,d}=\sigma_{{\rm r}i}^+,\qquad
\sigma_{{\rm a}j}^{1,d}=\sigma_{{\rm a}j}^-,\qquad
\sigma_{{\rm a}j}^{2,d}=\sigma_{{\rm a}j}^-,\nonumber
\eea
and $\hat B^p$ are operators of the velocity displacement caused by an 
interaction of the $p$-type at each of the walls, namely,
%
%
\be
\begin{array}{lllllll}
\hat B^a(\hat{\bmv{r}}_{12})\Psi(\bmv{v}_1,\bmv{v}_2)&=&
\hat B_{{\rm r}i}^d(\hat{\bmv{r}}_{12})\Psi(\bmv{v}_1,\bmv{v}_2)&=&
\hat B_{{\rm a}j}^d(\hat{\bmv{r}}_{12})\Psi(\bmv{v}_1,\bmv{v}_2)&=&
\Psi(\bmv{v}'_1,\bmv{v}'_2),\\ [0.5ex]
\hat B_{{\rm r}i}^b(\hat{\bmv{r}}_{12})\Psi(\bmv{v}_1,\bmv{v}_2)&=&
\Psi(\bmv{v}''_{{\rm r}1},\bmv{v}''_{{\rm r}2}),&&
\hat B_{{\rm a}j}^b(\hat{\bmv{r}}_{12})\Psi(\bmv{v}_1,\bmv{v}_2)&=&
\Psi(\bmv{v}''_{{\rm a}1},\bmv{v}''_{{\rm a}2}),\\ [0.5ex]
\hat B_{{\rm r}i}^c(\hat{\bmv{r}}_{12})\Psi(\bmv{v}_1,\bmv{v}_2)&=&
\Psi(\bmv{v}'''_{{\rm r}1},\bmv{v}'''_{{\rm r}2}),&&
\hat B_{{\rm a}j}^c(\hat{\bmv{r}}_{12})\Psi(\bmv{v}_1,\bmv{v}_2)&=&
\Psi(\bmv{v}'''_{{\rm a}1},\bmv{v}'''_{{\rm a}2}).\\
\end{array}\label{e3.80}
\ee
Physically, the values $\sigma_{{\rm r}i}^{1,p}$ ($\sigma_{{\rm 
a}j}^{1,p}$) \refp{e3.79} correspond to the distances between the particles 
just after the interaction of the $p$-type on the $i$th repulsive ($j$th 
attractive) wall of the potential, whereas the values $\sigma_{{\rm 
r}i}^{2,p}$ ($\sigma_{{\rm a}j}^{2,p}$) correspond to the distances just 
before the collision. 

The $H$-theorem for the kinetic equation with a multistep potential of 
interaction \refp{e3.73} was proved in \cite{8}. The normal solution by 
using the boundary conditions method \cite{38,57} was found in \cite{2,44}. 
On the basis of the solution found, numerical calculations of transport 
coefficients for liquid Argon along the liquid-vapour curve were performed.

%% file: s_02_33.tex
\subsubsection{Enskog-Landau kinetic equations for systems of charged hard 
\protect\newline spheres}

We consider kinetic equation \refp{e2.39} in the binary collision 
approximation with collision integral \refp{e2.40}, when the interaction 
potentials $\mPhi(|\bmv{r}_{jk}|)$ of classical particles at short 
distances can be modelled by the hard-sphere potential 
$\mPhi^{\rm hs}(|\bmv{r}_{jk}|)$ \refp{e3.1}, and at large distances by a 
certain long-range smooth ``tail'' $\mPhi^{\rm l}(|\bmv{r}_{jk}|)$, i.e.
%
%
\begin{equation}
\mPhi(|\bmv{r}_{jk}|)=\mPhi^{\rm hs}(|\bmv{r}_{jk}|)+
	\mPhi^{\rm l}(|\bmv{r}_{jk}|),\label{e4.1}
\end{equation}
where
%
%
\begin{equation}
\mPhi^{\rm l}(|\bmv{r}_{jk}|)=\lc
\begin{array}{ll}
0,&|\bmv{r}_{jk}|<\sigma,\\
\mPhi^{\rm l}(|\bmv{r}_{jk}|),&|\bmv{r}_{jk}|\geqslant\sigma.\\
\end{array}
\right.\label{e4.2}
\end{equation}
We note that breaking the particle interaction potential \refp{e4.1} into 
short- and long-range parts is not unique and, therefore, there arises the 
problem of optimal separation.

These problems  have often been discussed in equilibrium statistical 
mechanics \cite{58}. The similar ideas about the separation of the 
interaction potential at the derivation of kinetic equations were used by 
Rudyak \cite{59,59a}.

With allowance for \refp{e2.3} and \refp{e3.1}, the collision integral 
\refp{e2.40} for the interparticle potential \refp{e4.1}, \refp{e4.2} takes 
the form:
%
%
\begin{eqnarray}
I_{\rm coll}(x_1;t)&=&I_1(x_1;t)+I_2(x_1;t),\label{e4.3}\\
I_1(x_1;t)&=&-\lim_{a\to+\infty}
	\intl^{\sigma^+}_0\d r_{12}\;r_{12}^2\int\d\hat r_{12}
	\int\d\bmv{v}_2\;\im L^a(1,2)\lim_{\tau\to-0}\e^{\im L_2^a\tau}
	\times{}\nonumber\\
&&\g_2(\bmv{r}_1,\bmv{r}_2;t+\tau)f_1(x_1;t+\tau)f_1(x_2;t+\tau),
	\label{e4.4}\\
I_2(x_2;t)&=&-\intl_{\sigma^+}^\infty\d r_{12}\;r_{12}^2\int\d\hat r_{12}
	\int\d\bmv{v}_2\;\im L^{\rm l}(1,2)\lim_{\tau\to-\infty}
	\e^{\im\lp L_2^{(0)}+L^{\rm l}(1,2)\rp\tau}\times{}\nonumber\\
&&\g_2(\bmv{r}_1,\bmv{r}_2;t+\tau)f_1(x_1;t+\tau)f_1(x_2;t+\tau),
	\label{e4.5}
\end{eqnarray}
where
%
%
\begin{equation}
L_2^{(0)}=L(1)+L(2),\quad
L^{\rm l}(1,2)=\im\frac{\hat{\bmv{r}}_{12}}{m^*}
	\frac{\partial}{\partial r_{12}}
	\mPhi^{\rm l}(|\bmv{r}_{12}|)\lp\frac{\partial}{\partial\bmv{v}_1}- 
	\frac{\partial}{\partial\bmv{v}_2}\rp\label{e4.6}
\end{equation}
and we have used the idea of time separation of the interactions: instant 
collisions of hard spheres $(\tau\to-0)$ and an extended process of 
interaction with the long-range potential $(\tau\to-\infty)$. In accordance 
with the results obtained in subsection 3.1, the first term in the 
right-hand side of \refp{e4.3} is identical to the collision integral of 
the hard-sphere system (the right-hand side of \refp{e3.44}):
%
%
\begin{equation}
I_1(x_1;t)=\int\d x_2\;\hat 
	T(1,2)\g_2(\bmv{r}_1,\bmv{r}_2;t)f_1(x_1;t)f_1(x_2;t)\label{e4.7}
\end{equation}
with the only difference that here $\g_2(\bmv{r}_1,\bmv{r}_2;t)$ is a 
quasiequilibrium binary distribution function of the particles in the 
system with a total interparticle interaction potential \refp{e4.1}.

In the case when the long-range part of the interaction potential is absent 
$(\mPhi^{\rm l}(r)=0)$, the first part $I_1(x_1;t)$ of the collision 
integral is identical to the collision integral of the RET theory; the 
second part $I_2(x_1;t)$ is then identically equal to zero. In the second 
limiting case, when the density of the system is low (rarefied gases: 
$n\to0$, $\g_2(\bmv{r}_1,\bmv{r}_2;t)\to1$ and the hard-sphere part of the 
potential vanishes $(\sigma\to+0)$, the second part $I_2(x_1;t)$ of the 
collision integral is identical to the collision integral of the Boltzmann 
equation \cite{29}, and at the same time $I_1(x_1;t)$$\to$0.

In a special case, when the long-range interaction is weak, we make an 
expansion for $\exp\{(\im L_2^{(0)}+\im L^{\rm l}(1,2))\tau\}$, restricting 
ourselves to the term linear in $\im L^{\rm l}(1,2)$. Then, the second part 
$I_2(x_1;t)$ \refp{e4.5} reduces to the form:
%
%
\begin{eqnarray}
I_2(x_1;t)&=&I_2^{(0)}(x_1;t)+I_2^{(1)}(x_1;t),\label{e4.8}\\
I_2^{(0)}(x_1;t)&=&-\intl_{\sigma^+}^\infty\d r_{12}\;
	r_{12}^2\int\d\hat r_{12}
	\int\d\bmv{v}_2\;\im L^{\rm l}(1,2)\lim_{\tau\to-\infty}
	\e^{\im L_2^{(0)}\tau}\times{}\nonumber\\
&&\g_2(\bmv{r}_1,\bmv{r}_2;t+\tau)f_1(x_1;t+\tau)f_1(x_2;t+\tau),
	\label{e4.9}\\
I_2^{(1)}(x_1;t)&=&\intl_{\sigma^+}^\infty\d r_{12}\;
	r_{12}^2\int\d\hat r_{12}
	\int\d\bmv{v}_2\;\im L^{\rm l}(1,2)\lim_{\tau\to-\infty}
	\e^{\im L_2^{(0)}\tau}\times{}\label{e4.10}\\
&&\int\limits_0^\tau\d\tau'\;\e^{-\im L_2^{(0)}\tau'}\im L^{\rm l}(1,2)
	\e^{\im L_2^{(0)}\tau'}\g_2(\bmv{r}_1,\bmv{r}_2;t+\tau)
	f_1(x_1;t+\tau)f_1(x_2;t+\tau).\nonumber
\end{eqnarray}
The first term $I_2^{(0)}(x_1;t)$ is a generalization of the Vlasov mean 
field in KMFT \cite{36}, and the second $I_2^{(1)}(x_1;t)$ is a generalized 
Landau collision integral with allowance for retardation in time in the 
approximation of the second order on interaction. Indeed, if we set 
formally $\g_2(\bmv{r};t)\equiv1$ (rarefied gases) and $\sigma\to0$, and 
completely ignore the spatial inhomogeneity of $f_1(x_1;t)$ and the time 
retardation, then the second term $I_2^{(1)}(x_1;t)$ \refp{e4.10} is 
transformed into the ordinary Landau collision integral \cite{60}. In the 
case when $\mPhi^{\rm l}(|\bmv{r}_{12}|)$ is the Coulomb interaction 
potential, the complete collision integral \refp{e4.3}, \refp{e4.7}, 
\refp{e4.8} can be called the Enskog-Landau collision integral for a 
system of charged hard spheres which, in contrast to the ordinary Landau 
collision integral, does not diverge at short distances. However, at large 
distances we will observe a divergence, as usual. To eliminate this, it is 
necessary to take into account the effects of screening \cite{23,61,62}. To 
avoid this problem {\sl sequentially} we have to consider the kinetic 
equation with taking into account the dynamical screening effects 
\cite{61}. But this way is impossible with the Enskog-Landau kinetic 
equation. The only thing we can do for further calculation is to change the 
upper integral limit to some finite value which could have a meaning of 
the statical screening value in our system (see the following subsections). 
To {\sl solve} this problem we must consider dynamical screening effects. 

Using the boundary conditions method, a normal solution to the generalized 
Enskog-Landau kinetic equation was found in \cite{63,64}. This solution 
coincides with that obtained in \cite{3,7} for a stationary case. On the 
basis of normal solutions, numerical calculations of such transport 
coefficients as viscosity and thermal conductivity were performed for 
once-ionized Argon \cite{7} and for mixtures of ionized inert gases 
\cite{45}.

%% file: s_02_41.tex
\subsection{Modified group expansions for the construction of solutions to 
the BBGKY hierarchy}

\subsubsection{Modified group expansions}

Recently, a kinetic equation of the revised Enskog theory for a dense 
system of hard spheres \refp{e3.44} and an Enskog-Landau kinetic equation 
for a dense system of charged hard spheres with collision integrals 
\refp{e4.7}, \refp{e4.8} have been obtained from the BBGKY hierarchy in the 
binary collisions approximation \cite{3,7}. It should be noted that 
this approximation does not correspond to the usual two-particle 
approximation inherent in the Boltzmann theory, because an essential part 
of the many-particle correlations is implicitly taken into account by the 
pair quasiequilibrium distribution function $\g_2(\bmv{r}_1,\bmv{r}_2;t)$.

To analyze solutions to the BBGKY hierarchy \refp{e2.31} \cite{1,7} in 
higher approximations on interparticle correlations, it is more convenient 
to use the concept of group expansions \cite{65,66,67}. This was applied to 
the BBGKY hierarchy in previous investigations \cite{25,27,51,65,67} using 
the boundary conditions which correspond to the weakening correlations 
principle by Bogolubov \cite{21}. The same conception was envolved in 
papers by Zubarev and Novikov \cite{51}, where a diagram method for 
obtaining solutions to the BBGKY hierarchy was developed.

To analyze the BBGKY hierarchy \refp{e2.31}, we turn to the papers by 
Zubarev and Novikov \cite{51} and earlier ones by Green \cite{65} and Cohen 
\cite{66,67}, and pass from the nonequilibrium distribution functions 
$f_s(x^s;t)$ to the irreducible distribution ones $G_s(x^s;t)$, which can 
be introduced by the equalities presented in \cite{51,65}. In our case, 
with some modifications we obtain:
%
%
\bea
&&f_1(x_1;t)=G_1(x_1;t),\label{e5.1}\\
&&f_2(x_1,x_2;t)=G_2(x_1,x_2;t)+
	\g_2(\bmv{r}_1,\bmv{r}_2;t)G_1(x_1;t)G_1(x_2;t),\nonumber\\
&&f_3(x_1,x_2,x_3;t)=G_3(x_1,x_2,x_3;t)+
	\sum\nolimits_{P}G_2(x_1,x_2;t)G_1(x_3;t)+{}\nonumber\\
&&\g_3(\bmv{r}_1,\bmv{r}_2,\bmv{r}_3;t)G_1(x_1;t)G_1(x_2;t)G_1(x_3;t),
	\nonumber\\
&&\vdots\;\vdots\nonumber
\eea
Here, the position-dependent quasiequilibrium distribution functions
$\g_2(\bmv{r}_1,\bmv{r}_2;t)$, $\g_3(\bmv{r}_1,\bmv{r}_2,\bmv{r}_3;t)$,
$\g_s(\bmv{r}^s;t)$ are defined in \cite{10,11}. The modification of group 
expansions (\ref{e5.1}) consists in the fact that a considerable part of space 
time correlations is accumulated in the quasiequilibrium functions 
$\g_s(\bmv{r}^s;t)$. If all $\g_s(\bmv{r}^s;t)=1$ for $s=2,3,\ldots$, these 
group expansions coincide with those of papers \cite{51,65,66,67}. As far 
as each line in (\ref{e5.1}) brings in new functions 
$G_s(\bmv{r}^s;t)$, $s=1,2,3,\ldots$, the corresponding equations can be 
solved with respect to irreducible distribution functions and we can write 
the following:
%
%
\bea
&&G_1(x_1;t)=f_1(x_1;t),\label{e5.2}\\
&&G_2(x_1,x_2;t)=f_2(x_1,x_2;t)-
	\g_2(\bmv{r}_1,\bmv{r}_2;t)f_1(x_1;t)f_1(x_2;t),\nonumber\\
&&G_3(x_1,x_2,x_3;t)=f_3(x_1,x_2,x_3;t)-
	\sum\nolimits_{P}f_2(x_1,x_2;t)f_1(x_3;t)-{}\nonumber\\
&&h_3(\bmv{r}_1,\bmv{r}_2,\bmv{r}_3;t)f_1(x_1;t)f_1(x_2;t)f_1(x_3;t),
	\nonumber\\
&&\vdots\;\vdots .\nonumber
\eea
In (\ref{e5.1}) and (\ref{e5.2}), the symbol $\sum_P$ denotes the sum of
all different permutations of coordinates for three and more particles
%
%
\bea
h_3(\bmv{r}_1,\bmv{r}_2,\bmv{r}_3;t)&=&
	\g_3(\bmv{r}_1,\bmv{r}_2,\bmv{r}_3;t)-\g_2(\bmv{r}_1,\bmv{r}_2;t)-
	\g_2(\bmv{r}_1,\bmv{r}_3;t)-\g_2(\bmv{r}_2,\bmv{r}_3;t)\nonumber\\
&\equiv&h_3^{\prime}(\bmv{r}_1,\bmv{r}_2,\bmv{r}_3;t)-2,\label{e5.3}
\eea
where $h_3^{\prime}(\bmv{r}_1,\bmv{r}_2,\bmv{r}_3;t)$ is a three-particle 
quasiequilibrium correlation function. Now let us write the BBGKY hierarchy 
\refp{e2.31} \cite{1,3,7} for the irreducible distribution functions 
$G_s(x^s;t)$, namely, the first two equations, 
%
%
\bea
&&\lp\ddt+\im L(1)\rp G_1(x_1;t)+{}\label{e5.4}\\
&&\int\d x_2\;\im L(1,2)\g_2(\bmv{r}_1,\bmv{r}_2;t)G_1(x_1;t)G_1(x_2;t)+
	\int\d x_2\;\im L(1,2)G_2(x_1,x_2;t)=0.\nonumber
\eea
Differentiating the relation for $G_2(x_1,x_2;t)$ in (\ref{e5.2}) with
respect to time and using the second equation from the BBGKY hierarchy for
the function $f_2(x_1,x_2;t)$, we can get for the pair irreducible
distribution function $G_2(x_1,x_2;t)$ an equation, which reads:
%
%
\bea
&&\lp\ddt+\im L_2+\veps\rp G_2(x_1,x_2;t)=
	-\lp\ddt+\im L_2\rp\g_2(\bmv{r}_1,\bmv{r}_2;t)G_1(x_1;t)G_1(x_2;t)-{}
	\nonumber\\
&&\int\d x_3\;\Big\{\im L(1,3)+\im L(2,3)\Big\}\;\Big\{G_3(x_1,x_2,x_3;t)+
	\sum_PG_2(x_1,x_2;t)G_1(x_3;t)+{}\nonumber\\
&&\g_3(\bmv{r}_1,\bmv{r}_2,\bmv{r}_3;t)G_1(x_1;t)G_1(x_2;t)G_1(x_3;t)\Big\}.
	\label{e5.5}
\eea
In a similar way, we can obtain other equations for the three-particle
irreducible function $G_3(x_1,x_2,x_3;t)$ and the higher $G_s(x^s;t)$ ones.
One remembers now that the appearance\hfill of\hfill the\hfill 
quasiequilibrium\hfill distribution\hfill
functions\hfill $\g_2(\bmv{r}_1,\bmv{r}_2;t)$,\linebreak
$\g_3(\bmv{r}_1,\bmv{r}_2,\bmv{r}_3;t)$, $\g_s(\bmv{r}^s;t)$ in the 
hierarchy is closely connected with the fact that the boundary conditions 
for the solutions of the Liouville equation take into consideration both the
nonequilibrium character of the one-particle distribution function and 
the local conservation laws, which corresponds to a consistent description of
the kinetics and hydrodynamics of the system \cite{1,3}. Since in the 
present paper we analyze two first equations, (\ref{e5.4}) and (\ref{e5.5}) 
only, we will not write the others. It is important to note that, if we 
put formally $\g_s(\bmv{r}^s;t)\equiv 1$ for all $s=2,3,\ldots$ in 
(\ref{e5.4}) and (\ref{e5.5}), we come to the first two equations of the 
BBGKY hierarchy for the irreducible distribution functions $G_1(x_1;t)$ and
$G_2(x_1,x_2;t)$, which were obtained in paper \cite{51} by D.N.Zubarev
and M.Yu.Novikov. The first term in the right-hand side of (\ref{e5.5}) is a
peculiarity of (\ref{e5.4}) and (\ref{e5.5}) equation system. This is a 
term with a time derivative of the pair quasiequilibrium distribution
function $\g_2(\bmv{r}_1,\bmv{r}_2;t)$. As it was shown in \cite{1,3}, the 
binary quasiequilibrium distribution function is a functional of the local
values of temperature $\beta(\bmv{r};t)$ and mean particle density 
$n(\bmv{r};t)$. Thus, time derivatives of
$\g_2\big(\bmv{r}_1,\bmv{r}_2|\beta(t),n(t)\big)$ will conform to 
$\beta(\bmv{r};t)$ and $n(\bmv{r};t)$. These quantities, in their turn,
according to the self-consistency conditions \cite{1,7}, will be expressed
via the average energy value $\la\hat{\cal E}'(\bmv{r})\ra^t$ in an
accompanying reference frame and via $\la\hat{n}(\bmv{r})\ra^t$, which
constitute a basis of the hydrodynamical description of a nonequilibrium 
state of the system. Solving equation \refp{e5.5} for the irreducible 
quasiequilibrium two-particle distribution function $G_2(x_1,x_2;t)$ in the 
generalized polarization approximation and taking into account the first 
equation of the chain \refp{e5.4}, lead to \cite{10,11}:
%
%
\bea
&&\lp\ddt+\im L(1)\rp G_1(x_1;t)+\int\d x_2\;\im L(1,2)
	\g_2(\bmv{r}_1,\bmv{r}_2;t)G_1(x_1;t)G_1(x_2;t)={}\nonumber\\
&&\int\d x_2\intl_{-\infty}^{t}\d t'\;\e^{\veps(t'-t)}
	\im L(1,2)U(t,t')\times{}\label{e5.14}\\
&&\lp\dd{t'}+\im L_2+\El(x_1,x_2;t')\rp\g_2(\bmv{r}_1,\bmv{r}_2;t')
	G_1(x_1;t')G_1(x_2;t'),\nonumber
\eea
where
%
\bea
U(t,t')&=&\exp_+\lc-\int_{t'}^{t}\d t''\;\ls\im L_2+\El(x_1,x_2;t'')\rs\rc,
	\nonumber\\
\El(x_1,x_2;t)&=&\El(x_1;t)+\El(x_2;t).\label{e5.13}
\eea
The operator $\El(x_1;t)$ can be obtained by a variation of the Vlasov 
collision integral near the nonequilibrium distribution $G_1(x_1;t)$, 
namely,
%
%
\bea
\delta\lp\int\d x_3\;\im L(1,3)G_1(x_3;t)G_1(x_1;t)\rp&=&\label{e5.10}\\
\int\d x_3\;\im L(1,3)G_1(x_3;t)\delta G_1(x_1;t)\,&=&
	\El(x_1;t)\delta G_1(x_1;t).\nonumber
\eea
This is a kinetic equation for a nonequilibrium one-particle distribution
function with the non-Markovian collision integral in the generalized 
polarization approximation. It should be noted that the presence of the 
Vlasov operator $\El(x_1,x_2;t)$ in the collision integral \refp{e5.14} 
indicates taking into consideration collective effects. Analysis of the 
collision integral \refp{e5.14} in a general case is a rather complicated 
problem. But it is obvious that the collision integral in \refp{e5.14} or 
an expression for $G_2(x_1,x_2;t)$ with \refp{e5.5} may be much simplified 
for every physical model of a particle system or for each nonequilibrium 
state of the collision integral in \refp{e5.14}. To show this, we shall 
consider two particular cases: a hard spheres model and Coulomb plasma.

%% file: s_02_42.tex
\subsubsection{Hard spheres model in the polarization approximation}

In this subsection we shall investigate kinetic processes for a hard 
spheres model in the approximations which are higher than a binary 
collisions one. We take into account the character of the model parameters
and the results of the previous section of this article and papers 
\cite{25,48}. This investigation is convenient to carry out on the basis of 
the equation chain \refp{e5.4}, \refp{e5.5} at the formal substitution of a 
potential part of the Liouville operator $\im L(1,2)$ by the Enskog 
collision operator $\hat{T}(1,2)$ \cite{25,68}. In this case, equations 
\refp{e5.4} and \refp{e5.5} have the form:
%
%
\bea
&&\lp\ddt+\im L(1)\rp G_1(x_1;t)+{}\label{e5.15}\\
&&\int\d x_2\;\hat{T}(1,2)\g_2(\bmv{r}_1,\bmv{r}_2;t)G_1(x_1;t)G_1(x_2;t)+
	\int\d x_2\;\hat{T}(1,2)G_2(x_1,x_2;t)=0,\nonumber
\eea
%
%
%
%
%
\bea
\lefteqn{\ds\lp\ddt+\im L_2^0+\hat{T}(1,2)+
	\veps\rp G_2(x_1,x_2;t)={}}\label{e5.16}\\
&&{}-\lp\ddt+\im L_2^0+\hat{T}(1,2)\rp\g_2(\bmv{r}_1,\bmv{r}_2;t)
	G_1(x_1;t)G_1(x_2;t)\nonumber\\
&&{}-\int\d x_3\;\Big\{\hat{T}(1,3)+\hat{T}(2,3)\Big\}
	\Big\{G_3(x_1,x_2,x_3;t)+\sum_PG_2(x_1,x_2;t)G_1(x_3;t)
	\nonumber\\
&&{}+\g_3(\bmv{r}_1,\bmv{r}_2,\bmv{r}_3;t)G_1(x_1;t)
	G_1(x_2;t)G_1(x_3;t)\Big\}.\nonumber
\eea
Further, we will consider the same approximations concerning equation 
\refp{e5.16}, in which $G_3(x_1,x_2,x_3;t)$ and 
$h_3(\bmv{r}_1,\bmv{r}_2,\bmv{r}_3;t)$ are neglected. Then, if we introduce 
similarly to \refp{e5.10} the Boltzmann-Enskog collision operator 
$C(x_1;t)\!$ ($\El(x_1;t)$$\to$$C(x_1;t)\!$), equation \refp{e5.16} could 
be rewritten in the next form:
%
%
\bea
&&\delta\int\d x_3\;\hat{T}(1,3)G_1(x_1;t)G_1(x_3;t)=
	C(x_1;t)\delta G_1(x_1;t),\label{e5.17}\\
&&\lp\ddt+\im L_2^0+\hat{T}(1,2)+C(x_1,x_2;t)+
	\veps\rp G_2(x_1,x_2;t)={}
	\label{e5.18}\\
&-&\lp\ddt+\im L_2^0+\hat{T}(1,2)+C(x_1,x_2;t)\rp
	\g_2(\bmv{r}_1,\bmv{r}_2;t)G_1(x_1;t)G_1(x_2;t),\nonumber
\eea
Hence it appears that the formal solution to $G_2(x_1,x_2;t)$ reads:
%
%
\bea
&&G_2(x_1,x_2;t)=-\intl_{-\infty}^0\d t'\;\e^{\veps(t'-t)}
	U_{\rm hs}(t,t')\times{}\label{e5.19}\\
&&\lc\dd{t'}+\im L_2^0+\hat{T}(1,2)+C(x_1,x_2;t')\rc
	\g_2(\bmv{r}_1,\bmv{r}_2;t')G_1(x_1;t')G_1(x_2;t'),\nonumber
\eea
where $U_{\rm hs}(t,t')$ is an evolution operator for the system of hard
spheres:
%
%
\bea
U_{\rm hs}(t,t')&=&\exp_+\lc-\int_{t'}^{t}\d t''\;
	\ls\im L_2^0+\hat{T}(1,2)+C(x_1,x_2;t'')\rs\rc,\label{e5.20}\\
C(x_1,x_2;t)&=&C(x_1;t)+C(x_2;t).\nonumber
\eea
Now let us put \refp{e5.19} into the first equation \refp{e5.15}. Then, the
resulting equation takes the form \cite{10,11}:
%
%
\bea
&&\lp\ddt+\im L(1)\rp G_1(x_1;t)=\int\d x_2\;\hat{T}(1,2)
	\g_2(\bmv{r}_1,\bmv{r}_2;t)G_1(x_1;t)G_1(x_2;t)-{}\label{e5.21}\\
&&\int\d x_2\;\hat{T}(1,2)
	\intl_{-\infty}^0\d t'\;\e^{\veps(t'-t)}U_{\rm hs}(t,t')
	\lc\dd{t'}+\im L_2^0+\hat{T}(1,2)+C(x_1,x_2;t')\rc
	\times{}\nonumber\\
&&\g_2(\bmv{r}_1,\bmv{r}_2;t')G_1(x_1;t')G_1(x_2;t').\nonumber
\eea
This equation can be called a generalized kinetic equation for the 
nonequilibrium one-particle distribution function of hard spheres with a 
non-Markovian collision integral in the generalized polarization 
approximation. The first term in the right-hand side of this equation is the
collision integral from the revised Enskog theory \cite{1,7,33,53}. 
Neglecting time retardation effects and assuming that the operator 
$C(x_1,x_2;t)$ does not depend on time when
\[G_1(x_1;t)=f_0(\bmv{p})=n\lp\frac{m}{2\pi 
	kT}\rp^{3/2}\exp\lc-\frac{p^2}{2mkT}\rc
\]
is a local equilibrium Maxwell distribution function, the next term can
be rewritten in a simplified form:
%
%
\be
I_{\rm R}(x_1;t)=-\intl_{-\infty}^{t}\d t'\;
	\e^{\veps(t'-t)}R_0(x_1;t,t')G_1(x_1;t')
	-\intl_{-\infty}^{t}\d t'\;\e^{\veps(t'-t)}R_1(x_1;t,t')G_1(x_1;t'),
\label{e5.22}
\ee
where
%
%
\bea
R_0(x_1;t,t')&=&\int\d x_2\,\hat{T}(1,2)
	\exp\lc(t'-t)\ls\im L_2^0+\hat{T}(1,2)+C(x_1,x_2)\rs\rc
	\times{}\nonumber\\
&&\ls\im L_2^0+C(x_1,x_2)\rs\g_2(\bmv{r}_1,\bmv{r}_2;t')G_1(x_2;t'),
	\label{e5.23}\\ [1ex]
R_1(x_1;t,t')&=&\int\d x_2\,\hat{T}(1,2)
	\exp\lc(t'-t)\ls\im L_2^0+\hat{T}(1,2)+C(x_1,x_2)\rs\rc
	\times{}\nonumber\\
&&\hat{T}(1,2)\g_2(\bmv{r}_1,\bmv{r}_2;t')G_1(x_2;t'),\label{e5.24}
\eea
$R_1(x_1;t,t')$ is a generalized ring operator. The kinetic equation
\refp{e5.21} together with \refp{e5.23} and \refp{e5.24} is the 
generalization of the kinetic equation for a system of hard spheres which 
was obtained by Bogolubov in \cite{48,68}. It coincides with the case, when
the quasiequilibrium pair distribution function of hard spheres is set 
formally to be a unity.

%% file: s_02_43.tex
\subsubsection{Coulomb plasma in the polarization approximation}

Here we shall study an electron gas, which is contained in a homogeneous
positively charged equilibrating background. This background can be created,
for example, by hard motionless ions. Then, electrons interact
according to the Coulomb law:
\[
\mPhi(|\bmv{r}_{12}|)=\frac{e^2}{|\bmv{r}_1-\bmv{r}_2|}=
        \frac{e^2}{|\bmv{r}_{12}|},
\]
the Fourier transform of which exists in the form of a real function
$\mPhi(|\bmv{k}|)$:
%
%
\be
\frac{e^2}{r_{12}}=\int\frac{\d\bmv{k}}{(2\pi)^3}\mPhi(|\bmv{k}|)
        \e^{\im\bmv{k}\cdot\bmv{r}_{12}},\quad
        \mPhi(|\bmv{k}|)=\mPhi(k)=\frac{4\pi e^2}{k^2},
        \label{e5.25}
\ee
here $\bmv{k}$ is a wavevector, $e$ is an electron charge. Let us
consider equation chain \refp{e5.4}, \refp{e5.5}, when 
$G_3(x_1,x_2,x_3;t)=0$, $\g_3(\bmv{r}_1,\bmv{r}_2,\bmv{r}_3;t)=0$ in the 
homogeneous case,
when $G_1(x_1;t)=G_1(\bmv{p}_1;t)$ and pair distribution functions 
depend on $|\bmv{r}_{12}|$. Following the Bogolubov method \cite{21}, we
shall assume that the one-particle distribution function $G_1(\bmv{p}_1;t)$
is calculated in the ``zeroth'' order on the interaction constant $q$, pair
distribution functions $G_2(\bmv{r}_{12},\bmv{p}_1,\bmv{p}_2;t)$ and
$\g_2(\bmv{r}_{12};t)$ in the first order on $q$, and $G_3(x_1,x_2,x_3;t)$,
$\g_3(\bmv{r}_1,\bmv{r}_2,\bmv{r}_3;t)$ $\sim$ $q^2$,
where $q=\frac{e^2}{r_{\rm d}}\mTheta$, $r_{\rm d}=\sqrt{\mTheta/4\pi e^2n}$
is the Debye radius, $n=N/V$, $\mTheta=k_{\rm B}T$, $k_{\rm B}$ is the
Boltzmann constant, $T$ is thermodynamic temperature. Therefore, to obtain 
an equation for $G_2(\bmv{r}_{12},\bmv{p}_1,\bmv{p}_2;t)$ in the first 
approximation on the interaction constant $q$ without time retardment 
effects, it is necessary to retain all integral terms, but omit all the 
others. In this case, using the Fourier transform with respect to spatial 
coordinates for a homogeneous Coulomb electron gas, the set of equations 
\refp{e5.4}, \refp{e5.5} yields:
\bean
\ddt G_1(\bmv{p}_1;t)&=&-\dd{\bmv{p}_1}\!\int\d\bmv{k}\;\d\bmv{p}_2\;
        \im\mPhi(|\bmv{k}|)\g_2(\bmv{k};t)
        G_1(\bmv{p}_1;t)G_1(\bmv{p}_2;t)\\
&&-\dd{\bmv{p}_1}\!\int\d\bmv{k}\;\d\bmv{p}_2\;
        \im\mPhi(|\bmv{k}|)G_2(\bmv{k},\bmv{p}_1,\bmv{p}_2;t),
\eean
or
%
%
\bea
\ddt G_1(\bmv{p}_1;t)&=&\dd{\bmv{p}_1}G_1(\bmv{p}_1;t)\int\d\bmv{k}\;\bmv{k}
        \mPhi(|\bmv{k}|)\Imm\g_2(\bmv{k};t)\nonumber\\
&+&\dd{\bmv{p}_1}\int\d\bmv{k}\;\bmv{k}\mPhi(|\bmv{k}|)\Imm
        G_2(\bmv{k},\bmv{p}_1;t)\label{e5.26}
\eea
and an equation for $G_2(\bmv{k},\bmv{p}_1,\bmv{p}_2;t)$:
%
%
\bea
&&\text{\small$\ds\lp\ddt+\im\bmv{k}\frac{\bmv{p}_{12}}{m}+\veps\rp
        G_2(\bmv{k},\bmv{p}_1,\bmv{p}_2;t)={}$}\label{e5.27}\\
&&\text{\small$\ds\im\bmv{k}\mPhi(|\bmv{k}|)\lc\dd{\bmv{p}_1}G_1(\bmv{p}_1;t)
        \int\d\bmv{p}_3\;G_2(\bmv{k},\bmv{p}_2,\bmv{p}_3;t)-
        \dd{\bmv{p}_2}G_1(\bmv{p}_2;t)\int\d\bmv{p}_3\;
        G_2(\bmv{k},\bmv{p}_1,\bmv{p}_3;t)\rc\!+{}$}\nonumber\\
&&\text{\small$\ds\im\bmv{k}\mPhi(|\bmv{k}|)\lc\dd{\bmv{p}_1}G_1(\bmv{p}_1;t)
        \g_2(-\bmv{k};t)G_1(\bmv{p}_2;t)-
        \dd{\bmv{p}_2}G_1(\bmv{p}_2;t)\g_2(\bmv{k};t)G_1(\bmv{p}_1;t)\rc,$}
        \nonumber
\eea
$\veps\to+0$, and $G_2(\bmv{k},\bmv{p}_1;t)=
\int\d\bmv{p}_2\;G_2(\bmv{k},\bmv{p}_1,\bmv{p}_2;t)$; 
$\Imm\g_2(\bmv{k};t)$, $\Imm G_2(x_1,x_2;t)$ are imaginary parts of the
corresponding distribution functions. The following properties should be
noted:
\bean
G_2(-\bmv{k},\bmv{p}_1,\bmv{p}_2;t)&=&G_2^*(\bmv{k},\bmv{p}_1,\bmv{p}_2;t),\\
\g_2(-\bmv{k};t)&=&\g_2^*(\bmv{k};t),
\eean
where $*$ denotes a complex conjugation. The solution to \refp{e5.27}, 
neglecting time retardment effects, reads:
%
%
\bea
&&G_2(\bmv{k},\bmv{p}_1,\bmv{p}_2;t)=\label{e5.28}\\
&&\frac{\bmv{k}\mPhi(|\bmv{k}|)}{\ds\bmv{k}\cdot\frac{\bmv{p}_{12}}{m}-
        \im0}\lc\dd{\bmv{p}_1}G_1(\bmv{p}_1;t)
        G_2(-\bmv{k},\bmv{p}_2;t)-\dd{\bmv{p}_2}G_1(\bmv{p}_2;t)
        G_2(\bmv{k},\bmv{p}_1;t)\rc+{}\nonumber\\
&&\frac{\bmv{k}\mPhi(|\bmv{k}|)}{\ds\bmv{k}\cdot\frac{\bmv{p}_{12}}{m}-
        \im0}\lc\dd{\bmv{p}_1}G_1(\bmv{p}_1;t)\g_2(-\bmv{k};t)
        G_1(\bmv{p}_2;t)-\dd{\bmv{p}_2}G_1(\bmv{p}_2;t)\g_2(\bmv{k};t)
        G_1(\bmv{p}_1;t)\rc.\nonumber
\eea
It should also be noted that equation \refp{e5.26} contains an
imaginary part of the irreducible pair nonequilibrium distribution function,
to be integrated with respect to momentum of the second particle. Now one
integrates equation \refp{e5.28} over all the values of momentum $\bmv{p}_2$
and defines in such a way some function $G_2(\bmv{k},\bmv{p}_1;t)$:
%
%
\bea
&&\text{\small$\ds\ls 1+\int\d\bmv{p}_2\;\frac{\bmv{k}\mPhi(|\bmv{k}|)}
        {\ds\bmv{k}\cdot\frac{\bmv{p}_{12}}{m}-\im 0}
        \dd{\bmv{p}_2}G_1(\bmv{p}_2;t)\rs
        G_2(\bmv{k},\bmv{p}_1;t)={}$}\label{e5.29}\\
&&\text{\small$\ds\dd{\bmv{p}_1}G_1(\bmv{p}_1;t)\int\d\bmv{p}_2\;
        \frac{\bmv{k}\mPhi(|\bmv{k}|)}
        {\ds\bmv{k}\cdot\frac{\bmv{p}_{12}}{m}-\im 0}
        G_2(-\bmv{k},\bmv{p}_2;t)+{}$}\nonumber\\
&&\text{\small$\ds\int\d\bmv{p}_2\;\frac{\bmv{k}\mPhi(|\bmv{k}|)}
        {\ds\bmv{k}\cdot\frac{\bmv{p}_{12}}{m}-\im 0}
        \lc\dd{\bmv{p}_1}G_1(\bmv{p}_1;t)\g_2(-\bmv{k};t)
        G_1(\bmv{p}_2;t)-\dd{\bmv{p}_2}G_1(\bmv{p}_2;t)\g_2(\bmv{k};t)
        G_1(\bmv{p}_1;t)\rc.$}\nonumber
\eea
Further, we should exclude from \refp{e5.29} the term with
$G_2(-\bmv{k},\bmv{p}_2;t)$. To do this, we follow Lenard \cite{46,48} and
integrate equation \refp{e5.29} over the momentum component 
$\bmv{p}_{1\perp}$, which is perpendicular to wavevector $\bmv{k}$. The 
resulting expression then reads:
%
%
\bea
&&\text{\small$\ds\ls 1+\mPhi(|\bmv{k}|)\Chi(\bmv{k},p_1;t)\rs
        G_2(\bmv{k},p_1;t)=\dd{\bmv{p}_1}G_1(p_1;t)\int\d\bmv{p}_2\;
        \frac{\bmv{k}\mPhi(|\bmv{k}|)}{\ds k\cdot\frac{p_{12}}{m}-\im 0}
        G_2(-\bmv{k},\bmv{p}_2;t)+{}$}\nonumber\\
&&\text{\small$\ds\int\d\bmv{p}_2\;\frac{\bmv{k}\mPhi(|\bmv{k}|)}
        {\ds k\cdot\frac{p_{12}}{m}-\im 0}
        \lc\dd{\bmv{p}_1}G_1(p_1;t)\g_2(-\bmv{k};t)G_1(\bmv{p}_2;t)-
        \dd{\bmv{p}_2}G_1(\bmv{p}_2;t)\g_2(\bmv{k};t)G_1(p_1;t)\rc,$} .
        \nonumber\\\label{e5.30}
\eea
%
%
%
Here the following conventional designations have been introduced:
%
%
\be
\begin{array}{rclclcl}
\Chi(\bmv{k},p_1;t)&=&\ds\int\d\bmv{p}_2\;\frac{\bmv{k}}
        {\ds k\cdot\frac{p_{12}}{m}-\im 0}\dd{\bmv{p}_2}G_1(\bmv{p}_2;t),
        &\qquad&\ds p_1&=&\ds\frac{\bmv{p}_1\cdot\bmv{k}}{k},\\
G_1(p_1;t)&=&\ds\int\d\bmv{p}_{1\perp}\;G_1(\bmv{p}_1;t),
	&\qquad&\ds p_2&=&\ds\frac{\bmv{p}_2\cdot\bmv{k}}{k},\\
G_2(\bmv{k},p_1;t)&=&\ds\int\d\bmv{p}_{1\perp}\;G_2(\bmv{k},\bmv{p}_1;t),
	&\qquad&\ds k&=&|\bmv{k}|.\\
\end{array}\label{e5.31}
\ee
Now we multiply the both equations, \refp{e5.29} and \refp{e5.30}, by
$\dd{p_1}G_1(p_1;t)$ and by $\dd{p_1}G_1(\bmv{p}_1;t)$, respectively, and
subtract them:
%
%
\bea
&&\,\Big(1+\mPhi(|\bmv{k}|)\Chi(\bmv{k},p_1;t)\Big)
        \ls G_2(\bmv{k},\bmv{p}_1;t)\dd{p_1}G_1(p_1;t)-
        G_2(\bmv{k},p_1;t)\dd{p_1}G_1(\bmv{p}_1;t)\rs={}\nonumber\\
&&\mPhi(|\bmv{k}|)\Chi(\bmv{k},p_1;t)\g_2(\bmv{k};t)
        \ls G_1(p_1;t)\dd{p_1}G_1(\bmv{p}_1;t)-
        G_1(\bmv{p}_1;t)\dd{p_1}G_1(p_1;t)\rs.\label{e5.32}
\eea
If we extract the imaginary part of this equation, one can find the unknown
quantity $\Imm G_2(\bmv{k},\bmv{p}_1;t)$, provided
$\Imm G_2(\bmv{k},p_1;t)=0$ \cite{10,11}:
%
%
\bea
\dd{p_1}G_1(p_1;t)\Imm G_2(\bmv{k},\bmv{p}_1;t)=
        \frac{\mPhi(|\bmv{k}|)\Imm
        \ls\Chi(\bmv{k},p_1;t)\g_2(\bmv{k};t)\rs}
        {\left|1+\mPhi(|\bmv{k}|)\Chi(\bmv{k},p_1;t)\right|^2}
        \nonumber\\
{}\times\ls G_1(p_1;t)\dd{p_1}G_1(\bmv{p}_1;t)-G_1(\bmv{p}_1;t)
        \dd{p_1}G_1(p_1;t)\rs.\label{e5.33}
\eea
Since $\Imm \Chi(\bmv{k},p_1;t)$=-$\pi\dd{p_1}G_1(p_1;t)$
\cite{62}, putting an expression for $\Imm G_2(\bmv{k},\bmv{p}_1;t)$ into 
equation \refp{e5.26} gives the generalized Bogolubov-Lenard-Balescu 
kinetic equation for an electron gas in an equilibrating background:
%
%
\bea
\ddt G_1(\bmv{p}_1;t)&=&\dd{\bmv{p}_1}G_1(\bmv{p}_1;t)
        \int\d\bmv{k}\;\bmv{k}\mPhi(|\bmv{k}|)\Imm\g_2(\bmv{k};t)+{}
        \label{e5.34}\\
&&\dd{\bmv{p}_1}\int\d\bmv{p}_2\;Q(\bmv{p}_1,\bmv{p}_2;t)
        \ls\dd{\bmv{p}_1}-\dd{\bmv{p}_2}\rs G_1(\bmv{p}_1;t)G_1(\bmv{p}_2;t),
        \nonumber
\eea
where $Q(\bmv{p}_1,\bmv{p}_2;t)$ is a second rank tensor
%
%
\be
Q(\bmv{p}_1,\bmv{p}_2;t)=-\pi\int\d\bmv{k}\;
        \frac{\left|\mPhi(|\bmv{k}|)\right|^2\bmv{k}\cdot\bmv{k}}
        {\left|1+\mPhi(|\bmv{k}|)\Chi(\bmv{k},p_1;t)\right|^2}
        \Imm\g_2(\bmv{k};t)\delta\big(\bmv{k}\cdot(\bmv{p}_1-\bmv{p}_2)\big),
        \label{5.35}
\ee
which coincides with $Q(\bmv{p}_1,\bmv{p}_2)$ \cite{62} at
$\Imm\g_2(\bmv{k};t)=1$. In this case, the kinetic equation \refp{e5.34}
transforms into the well-known Lenard-Balescu equation \cite{46,47,48,62}.
Evidently, the generalized Bogolubov-Lenard-Balescu kinetic equation
\refp{e5.34} claims the description of a dense electron gas, since in both 
the generalized mean field and the generalized Bogolubov-Lenard-Balescu
collision integrals, many-particle correlations are treated by the imaginary
part of $\g_2(\bmv{k};t)$. Nevertheless, the problem of divergence in the
collision integral of equation \refp{e5.34} at small distances 
($k\to\infty$) still remains. There are papers where the divergence of 
collision integrals is avoided with the help of a special choice of the 
differential cross section (quantum systems \cite{69}), or via a 
combination of simpler collision integrals (classical systems \cite{23}). 
These generalizations for collision integrals are attractive by their 
simplicity and helpful for ideal plasma. But, contrary to the obtained 
by us Bogolubov-Lenard-Balescu kinetic equation, they do not work for 
nonideal plasma. In accordance with the proposed structure of the collision 
integral
\[
I_{\rm total}=I_{\rm Boltzmann}-I_{\rm Landau}+I_{\rm Lenard-Balescu}
\]
the influence of particles interaction on plasma energy will be defined by
the correlation function $\g_2(r)$. Its asymptotic is
\[
\lim_{r\to\infty}\ls
        \underbrace{\exp\lc-\frac{e_ae_b}{rk_{\rm B}T}\rc-1}_{\rm Boltzmann}+
        \underbrace{\frac{e_ae_b}{rk_{\rm B}T}}_{\rm Landau}-
        \underbrace{\frac{e_ae_b}{rk_{\rm B}T}
        \e^{-r/r_{\rm D}}}_{\rm Lenard-Balescu}
        \rs=\frac{1}{r^2}\qquad(!),
\]
where $r_{_{\rm D}}$ denotes the Debye radius. In other combinations one 
arrives at false expressions for thermodynamic functions \cite{23}. 
Dynamical screening, which appears in the generalized 
Bogolubov-Lenard-Balescu collision integral obtained by us, is free of these
discrepancies. Generally speaking, the problem of divergency could be solved 
within the framework of a charged hard spheres model, combining the results 
of this section and the preciding one. But this step is an intricate and 
complicated problem and needs a separate consideration.

Evidently, an investigation of the obtained kinetic equation is important 
in view of its solutions and studying transport coefficients and time 
correlation functions for model systems.

In view of the dense systems study, where the consideration of spatial 
interparticle correlations is important, the BBGKY hierarchy \refp{e5.4}, 
\refp{e5.5} with the modified boundary conditions and group expansions has 
quite a good perspective. The kinetic equation \refp{e5.21}--\refp{e5.24} 
is a generalization of the Bogolubov one \cite{48,68} for a system of hard 
spheres. M.Ernst and J.Dorfman \cite{70} investigated collective modes 
in an inhomogeneous gas and showed that the solution of a dispersion 
equation for hydrodynamic modes leads to a nonanalytic frequency dependence 
on a wavevector. This is connected with the fact that the ring operator 
for inhomogeneous systems at small wavenumbers has a term proportional to 
$\sqrt{k}$. Similar investigations of collective modes and time correlation 
functions in the hydrodynamic region were carried out by Bogolubov 
\cite{68}. Nevertheless, it is necessary to carry out analogous 
investigations of hydrodynamic collective modes and time correlation 
functions on the basis of kinetic equation \refp{e5.21}, taking into 
account \refp{e5.22}--\refp{e5.24}, where some part of space correlations 
is considered in the pair quasiequilibrium function 
$\g_2(\bmv{r}_1,\bmv{r}_2;t)$. Obviously, these results may appear to be 
good for very dense gases, which could be described by a hard spheres 
model. An important factor is that in the kinetic equation  
\refp{e5.21}--\refp{e5.24}, as well as in the generalized 
Bogolubov-Lenard-Balescu one, collective effects are taken into account 
both via the Vlasov mean field and the binary quasiequilibrium correlation 
function which is a functional of nonequilibrium values of temperature and 
a chemical potential.

Transferring the obtained results to quantum systems is not obvious. Such a
procedure is rather complicated and needs additional investigations. 
Nevertheless, some steps in this way have been already done by Morozov and 
R\"opke \cite{9,9a,15}.

%% file: s_03_1.tex
\subsection{Overview}

One of the main problems of the nonequilibrium statistical mechanics of
liquids is an investigation of collective excitations, time correlation
functions and transport coefficients because, using these quantities, we
can compare the corresponding theoretical results with the data on light and 
neutron scattering as well as with the results obtained by the method of 
molecular dynamics (MD) \cite{71,72,73,74,75,76,77}. At present, one can 
distinguish three regions in these investigations.

The first one is the usual hydrodynamics in a linear approximation when the
time and spatial evolution of small excitations in liquids can be described
in terms of hydrodynamic modes (in terms of eigenvalues and eigenvectors of
the linear hydrodynamics equations) \cite{25,30,75,78,79,80}. Such 
hydrodynamic modes are the heat mode $z_{\rm H}(\bmv{k})=-D_{\rm T}k^2$, two
sound modes $z_{\pm}(\bmv{k})=\pm\im ck-\Gamma k^2$ and two viscosity modes
$z_{\nu_{1,2}}=-\nu k^2$ (see, for example, \cite{79}). In our notations
$\bmv{k}$ denotes a wavenumber which describes spatial dispersion of
excitations, $D_{\rm T}=\lambda/nc_{p}$ is a coefficient of thermal
diffusion, $\nu=\eta/mn$ is the kinematical viscosity,
$\Gamma=\frac23\nu+\frac13\kappa/mn+\frac12(c_p/c_V-1)D_{\rm T}$ is the
constant of sound decrement. $\lambda$, $\kappa$ and $\eta$ are the thermal
conductivity, bulk and shear viscosity coefficients defined by the
Green-Kubo formulas, $n$ is the average density, $m$ is the mass of a
separate particle, $c_p$ and $c_V$ are specific heats at constant pressure
and volume, $c$ is the sound velocity. In this region, where
$|\bmv{k}|^{-1}$ is much larger than atomic sizes of liquid $\sigma$,
and $\omega\tau\ll1$, where $\omega$ is frequency and $\tau$ is the
characteristic correlation time, the dynamical structure factor
$S(\bmv{k};\omega)$ is well described by the Landau-Plachek formula
\cite{25,79,82} in terms of Brillouin lines which corresponds to heat and
sound modes. With increasing wavevector values $\bmv{k}$, the precision of
the hydrodynamic description decreases, since correlations at short times
and small spatial distances which are inherent in neutron scattering in 
liquids are not described by linear hydrodynamic equations.

Extensions of the usual hydrodynamics were performed on the
basis of modern methods of nonequilibrium statistical mechanics by Zwanzig
and Mori \cite{83,84,85} with the use of the method of projection 
operators, by Sergeev and Tishchenko \cite{86,87} using the NSO method 
\cite{49,50} and by Tserkovnikov with the use of the method of Green 
functions \cite{88,89,90}. Transport equations were obtained for the 
average values of the densities of mass, momentum and energy. These 
equations generalize the hydrodynamic ones. In these equations, 
thermodynamic quantities $c_p$, $c_V$ depend on wavevector $\bmv{k}$, 
whereas the transport coefficients $\lambda$, $\kappa$ and $\eta$ depend on 
$\bmv{k}$, as well as on frequency $\omega$. Moreover, new generalized 
transport coefficients appear in the generalized hydrodynamic equations. 
These coefficients describe dynamical correlations between the thermal and 
viscous motions which vanish in the limit $k\to0$ and $\omega\to0$. This is 
the second region, namely, the region of the generalized hydrodynamics
\cite{25,75,81,85,58,87,88,89,90,91,92}. Papers by G\"otze, L\"ucke, Bosse
and others \cite{93,94,95,96,97,98,99} are of  special interest. In these
papers, an investigation of spectra for fluctuations of densities for the 
number of particles, their longitudinal and transverse fluxes for small as
well as for intermediate values of $\bmv{k}$ and $\omega$ is performed. For
Argon \cite{71,74} and Rubidium \cite{72,73,74}, a good coincidence with 
the experimental data was obtained. It is necessary to point out papers by
Yulmetyev and Shurygin \cite{100,101,101a,101b,102} where the dynamical 
structure factor $S(\bmv{k};\omega)$ for Rubidium \cite{101,101a,101b} and 
Argon \cite{100,102} was investigated by the projection operator method 
taking into account non-Markovian effects. The results obtained in these 
papers are valid in the low-frequency approximation.

The third region of investigations on collective modes and time correlation
functions in dense gases and liquids is connected with the kinetic theory
\cite{75,103,104,105} on the basis of the method of projection operators by
Mori and its generalizations \cite{75,77,85}. In this approach, the
nonequilibrium one-particle distribution function in phase space of
coordinates and momenta is a variable for an abbreviated description of a
nonequilibrium state of the system. At the same time, collective effects
arise explicitly in memory functions. Approximate calculations of memory
functions (expansions on density, weak interaction) in the hydrodynamic
limit were performed in papers by Mazenko \cite{105,106,107,108,108a,109},
Forster and Martin \cite{104}, Forster \cite{110} and others
\cite{98,111,112}. John and Forster unified paper \cite{113} and
proposed a formalism similar to that of the generalized hydrodynamics
\cite{103,104,105,109,110}. This formalism is distinguished by the fact
that the nonequilibrium one-particle distribution function in phase space
of coordinates and momenta, together with the density of total energy are 
included into a set of variables of an abbreviated description. The
results of this theory for the dynamical structure factor 
$S(\bmv{k};\omega)$ of Argon agree well with the experimental data and MD
calculations in the regions of small and intermediate values of $\bmv{k}$
and $\omega$. The same formalism was used later in papers by Sjodin and
Sjolander \cite{114} in which the dynamical structure factor 
$S(\bmv{k};\omega)$ for Rubidium was investigated. The authors of this
paper touched a number of interesting questions about the influence of 
one-particle motion and temperature fluctuations on $S(\bmv{k};\omega)$ in 
the regions of intermediate and great values of wavenumber $\bmv{k}$, as 
well as about the existence of short-length collective modes \cite{73}. 
Such questions and a number of others arise in discussing the results of 
neutron scattering with $\ell\sim\sigma$, ($|\bmv{k}|=2\pi/\ell$), where 
the kinetic theory of liquids is developed insufficiently. From the 
experimental results of neutron scattering in Rubidium  \cite{73} 
it follows that there are sharp side peaks for $S(\bmv{k};\omega)$ at 
points $\omega=\pm ck$, where $c$ is an adiabatic velocity of sound. Such 
side peaks could be identified as Brillouin ones. However, the 
Landau-Plachek formula is not valid in this region, where $\ell\sim\sigma$.

During the last period, the question of the existence of short-length
collective modes, beginning from the papers by de Schepper and Cohen
\cite{115,116}, is intensively discussed
\cite{117,118,119,120,121,122,123,124,125,126,127,128,129,130}. The
proposed in \cite{115,116} approach is based on the kinetic equation of the
revised Enskog theory (RET) \cite{1,3,33}, which was derived by us
sequentially in subsection 2.3.1. The linearized kinetic equation of RET
for hard spheres, which was used in \cite{115,116}, was obtained in papers
by Mazenko and others as well \cite{108,108a,112}. Although the kinetic 
theory considered by de Schepper and Cohen for short-length collective 
modes is related to a model system of hard spheres, it has explained
many questions which are significant for understanding collective modes
in the region of large $\bmv{k}$. First of all, it proposed a concept of 
extending hydrodynamic modes into the region of large wavevectors 
\cite{116,117}. Solving the problem on eigenvalues for the generalized 
Enskog operator (taking into account a mean field) for hard spheres has 
shown \cite{117,127} the existence of five generalized (extended) 
hydrodynamic and kinetic modes. The generalized hydrodynamic modes are a 
generalization of hydrodynamic modes to the region of large $\bmv{k}$ and 
tend to zero at $k\to0$. The kinetic modes at $k\to0$ take positive nonzero 
values. The dynamical structure factor $S(\bmv{k};\omega)$ is presented in 
this approach in the form of a sum of spectral terms which correspond to 
the Lorentzian form of lines. Secondly, to confirm the theoretical 
predictions of papers 
\cite{115,116,117,118,119,120,121,122,123,124,125,126,127,128,129,130}, 
experiments on neutron scattering in liquid Argon \cite{118}, Neon 
\cite{120}, as well as MD calculations for a system of hard spheres 
\cite{121} and for a liquid with the Lennard-Jones-like interparticle 
potential of interaction \cite{123} were made. In \cite{118} short-length 
sound modes and their decrement in the region $\ell\sim\sigma$ for a liquid 
Argon were investigated. The obtained results \cite{126} for a 
nonanalytical dependence of the sound dispersion on $\bmv{k}$ were compared 
with the experimental data and the results of the theory of coupled modes 
\cite{131}. Eigenvalues for the heat mode $z_{\rm H}(\bmv{k})$, obtained on
the basis of the generalized Enskog equation, were investigated in 
\cite{125}. The calculations performed agree quantitatively with the MD data
for hard spheres \cite{120} and with the data on neutron scattering for 
Argon \cite{118,119,132} and Crypton \cite{133}.

Interesting results of the investigation of the dynamical structure factor
$S(\bmv{k};\omega)$ and time correlation functions of flow particle and 
enthalpy densities, as well as collective modes for a liquid with the 
Lennard-Jones potential of interaction were obtained in \cite{134}. There 
a system of equations for time correlation functions of densities for the 
number of particles, momentum and enthalpy, the generalized stress tensor 
and the flow of energy obtained by the Mori projection operator method was 
used. The basis of this system are equations of a generalized hydrodynamic 
description obtained for the first time in \cite{91} for simple liquid and 
ionic systems with the use of the NSO method \cite{135}. Solving the system 
of equations for time correlation functions in the Markovian approximation 
on eigenvalues in the hydrodynamic limit $k\to0$, $\omega\to0$ showed 
\cite{134} that apart from the eigenvalues which correspond to pure 
hydrodynamic modes, in particular, to heat $z_{\rm H}(\bmv{k})$ and two 
sound modes $z_\pm(\bmv{k})$, there exist two eigenvalues that correspond 
to kinetic modes. They differ from zero in the limit $k\to0$. The analogous 
results were obtained in our paper \cite{136}. Further, such an approach 
was developed for a Lennard-Jones fluid in a number of works
\cite{137,138,139,140,141,142,143,144,145,146,147,148,149}.
The concept of generalized collective modes was developed \cite{137,138}
on the basis of equations of extended hydrodynamics and the relevant
equilibrium static quantities which normally are obtained within the
method of molecular dynamics (MD). This approach gives the possibility
to calculate generalized-mode spectra of a Lennard-Jones fluid using nine- 
and four-mode descriptions for longitudinal \cite{137,142} and transverse 
\cite{137,141} fluctuations, respectively. The investigation of generalized 
transport coefficients dependent on wavevector and frequency was carried 
out for the first time in \cite{143}. Further, this approach was developed 
for mixtures \cite{149,150,151} and polar liquids, in particular, for 
Stockmayer models and TIP4P water 
\cite{152,153,154,155,156,157,158,159,160}.

Problems of the description of kinetic and hydrodynamic fluctuations,
taking into account correlations in a simple fluid, were discussed in
papers by Peletminskii, Sokolovskii and Slusarenko \cite{161,162}. They used
a functional hypothesis and the method of an abbreviated description
\cite{163}.

On the basis of general equations of fluctuational hydrodynamics obtained
in \cite{161}, we carried out investigations of liquid hydrodynamics near 
the equilibrium without taking into account far correlations and spatial
dispersion of transport coefficients \cite{164,164a}.

It is necessary to point out a series of papers by Balabanyan 
\cite{165,166,166a,166b,166c} devoted to the investigation of classical 
time correlation and Poisson Green functions and spectra of collective 
modes on the basis of the Boltzmann kinetic equation by the method of 
moments for a system of hard spheres and Maxwellian molecules.

The problem of the construction of interpolation formulas for 
density-density and flow-flow correlation functions which are valid for a
wide range of frequencies and wavevectors was considered in papers by 
Tserkovnikov \cite{89,90,167,168}. Here a method of Green functions was
applied to the molecular hydrodynamics of quantum Bose-systems. These 
questions were presented in detail in \cite{167,168} for a weakly nonideal 
Bose-gas. The obtained interpolation formulas for the Green functions of 
transverse components of flow density \cite{167}, as well as of 
fluctuations of the number of particles and energy \cite{168}, appear to be 
valid in the hydrodynamic region and in the region of great frequencies and 
small wavelengths. These results are obviously valid in the classical case 
as well.

From the presented survey, one can draw the following conclusions: firstly,
for a more detailed study of collective modes, generalized transport
coefficients and time correlation functions in liquids and dense gases, it
is necessary to have a theory in which kinetic and hydrodynamic processes
are considered simultaneously; secondly, at the present time, a consistent 
theory for the description of light and neutron scattering in the whole 
range of wavevector $\bmv{k}$ and frequency $\omega$ has not been 
formulated yet.

In our papers \cite{1,3,5,7} we proposed an approach in which the kinetics 
and hydrodynamics of transport processes in dense gases and liquids are 
considered as coupled. As a result, we obtained a system of coupled 
generalized equations for the nonequilibrium one-particle distribution 
function and the mean density of the total energy. It can be applied to the 
description of nonequilibrium states of the system of particles which are 
as close as they are far from the equilibrium state. In the next subsections of 
this section we shall present this approach in comparison with other 
theories and investigate time correlation functions, generalized transport 
coefficients and collective modes in comparison with the molecular 
hydrodynamics \cite{75} and its extended versions 
\cite{134,136,137,138,139,140,141,142,143,144,145,146,147}.

%% file: s_03_2.tex
\subsection{The generalized kinetic equation for the nonequilibrium 
one-particle distribution function with taking into account transport equations 
for the mean density of energy}

According to the formulation of a modified boundary condition for the 
Liouville equation \refp{e2.4} with choosing parameters of an abbreviated
description $\la \hat n_{1}(x) \ra^{t}=f_{1}(x;t)$ and 
$\la\hat{\cal E}(\bmv{r})\ra^{t}$ (the definition of the quasiequilibrium 
distribution function is presented by equation \refp{e2.12}), the 
dependence of the the quasiequilibrium distribution function on time is 
defined by variation in time of the average values 
$\la\hat{n}_{1}(x)\ra^{t}$ and $\la\hat{\cal E}(\bmv{r})\ra^{t}$:
%
%
\be
\vrho\xnt=\vrho\lp\ldots,\{\la\hat{n}_{1}(x)\ra^{t},
	\la\hat{\cal E}(\bmv{r})\ra^{t}\}\ldots\rp.\label{e6.1}
\ee
Time evolution of the parameters of an abbreviated description 
$f_1(x;t)$, $\la\hat{\cal E}(\bmv{r})\ra^{t}$ is described by transport 
equations. For obtaining these equations it is more convenient to use the 
method of projection operators, which was widely used by Robertson 
\cite{169,170} and modified by Kawasaki and Gunton \cite{171}. We shall 
reformulate the NSO method in order to take into account projection operators.

Let us write the Liouville equation with source \refp{e2.4} in the form:
%
%
\be
\lp\ddt+\im L_{N}+\veps\rp\vt\vrho\xnt=-\lp\ddt+\im L_{N}\rp
	\vrho_{\rm q}\xnt,\label{e6.2}
\ee
introducing $\vt\vrho\xnt=\vrho\xnt-\vrho_{\rm q}\xnt$. We shall use the 
quasiequilibrium distribution function $\vrho_{\rm q}\xnt$ defined as 
follows:
%
%
\be
\vrho_{\rm q}\xnt=\exp\lc-\Phi(t)-
	\int\d\bmv{r}\;\beta(\bmv{r};t)\hat{\cal E}(\bmv{r})-
	\int\d x\;b(x;t)\hat{n}_1(x)\rc,\label{e6.3}
\ee
which is found from \refp{e2.12}, taking into account \refp{e2.13}, by 
redefinition of the parameter conjugated to $\la\hat{n}_{1}(x)\ra^{t}$:
%
%
\be
b(x;t)=a(x;t)-\beta(\bmv{r};t)\frac{p^{2}}{2m}.\label{e6.4}
\ee
Parameters $\beta(\bmv{r};t)$ and $b(x;t)$ in \refp{e6.3} are defined 
from the corresponding self-consistency conditions:
%
%
\be
\begin{array}{rcl}
\la\hat{\cal E}(\bmv{r})\ra^t&=&\la\hat{\cal E}(\bmv{r})\ra^t_{\rm q},\\
\la\hat{n}_{1}(x)\ra^t&=&\la\hat{n}_{1}(x)\ra^t_{\rm q}.\\
\end{array}\label{e6.5}
\ee

Taking into account the structure of the quasiequilibrium distribution
function \refp{e6.3}, the time derivative $\ddt$ of this function in the
right-hand side of equation \refp{e6.2} can be presented as
%
%
\be
\ddt\vrho_{\rm q}\xnt=-\SP_{\rm q}(t)\im L_{N}\vrho\xnt,\label{e6.6}
\ee
where $\SP_{\rm q}(t)$ is the Kawasaki-Gunton projection operator, defined
as follows:
%
%
\bea
&&\SP_{\rm q}(t)\vrho'={}\label{e6.7}\\
&&\lc\vrho_{\rm q}\xnt-\int\d\bmv{r}\;
	\frac{\partial\vrho_{\rm q}\xnt}
	{\partial\la\hat{\cal E}(\bmv{r})\ra^t}\la\hat{\cal E}(\bmv{r})\ra^t-
	\int\d x\;\frac{\partial\vrho_{\rm q}\xnt}
	{\partial\la\hat{n}_1(x)\ra^t}\la\hat{n}_1(x)\ra^t\rc
	\int\d\mGamma_N\;\vrho'+{}\nonumber\\
&&\int\d\bmv{r}\;\frac{\partial\vrho_{\rm q}\xnt}
	{\partial\la\hat{\cal E}(\bmv{r})\ra^t}
	\int\d\mGamma_N\;\hat{\cal E}(\bmv{r})\vrho'+
	\int\d x\;\frac{\partial\vrho_{\rm q}\xnt}
	{\partial\la\hat{n}_1(x)\ra^t}\int\d\mGamma_N\;\hat{n}_1(x)\vrho'.
	\nonumber
\eea
This operator acts only on distribution functions and has the properties:
\bean
\SP_{\rm q}(t)\vrho(x^{N};t')&=&\vrho_{\rm q}(x^{N};t),\\
\SP_{\rm q}(t)\vrho_{\rm q}(x^{N};t')&=&\vrho_{\rm q}(x^{N};t),\\
\SP_{\rm q}(t)\SP_{\rm q}(t')&=&\SP_{\rm q}(t).
\eean
Taking into account \refp{e6.6}, the Liouville equation \refp{e2.4} 
transforms into the form
%
%
\be
\lp\ddt-\Big(1-\SP_{\rm q}(t)\Big)\im L_N+\veps\rp\vt\vrho\xnt=
	-\Big(1-\SP_{\rm q}(t)\Big)\im L_N\vrho_{\rm q}\xnt,
	\label{e6.8}
\ee
the formal solution to which is
%
%
\be
\vt\vrho\xnt=-\intl_{-\infty}^t\d t'\;\e^{\veps(t'-t)}T(t,t') 
	\Big(1-\SP_{\rm q}(t')\Big)\im L_N\vrho_{\rm q}\xntp,\label{e6.9}
\ee
and one finds an expression for the nonequilibrium distribution function
%
%
\be
\vrho\xnt=\vrho_{\rm q}\xnt-\intl_{-\infty}^{t}\d t'\;
	\e^{\veps(t'-t)}T(t,t')\Big(1-\SP_{\rm q}(t')\Big)
	\im L_N\vrho_{\rm q}\xntp,\label{e6.10}
\ee
where
%
%
\be
T(t,t')=\exp_{+}\lc-\int_{t'}^t\d t''\;\Big(1-\SP_{\rm q}(t'')\Big)
	\im L_N\rc\label{e6.11}
\ee
is a generalized evolution operator with taking account the Kawasaki-Gunton 
projection operator, $\exp_+$ is an ordering exponent. Solution 
\refp{e6.10} is exact. It corresponds to the idea of an abbreviated 
description of a nonequilibrium state of the system \refp{e6.1}. Acting by 
the operators $\big(1-\SP_{\rm q}(t')\big)$ and $\im L_N$ on $\vrho_{\rm 
q}(x^{N};t')$ in the right-hand side of expression \refp{e6.10}, we write 
an expression for the nonequilibrium distribution function in the explicit 
form:
%
%
\bea
\vrho\xnt=\vrho_{\rm q}\xnt&+&
	\int\d\bmv{r}\intl_{-\infty}^t\d t'\;\e^{\veps(t'-t)}
	T(t,t')I_{\cal E}(\bmv{r},t')\beta(\bmv{r};t')\vrho_{\rm q}\xntp
	\qquad\quad\nonumber\\
&-&\int\d x\;\intl_{-\infty}^t\d t'\;\e^{\veps(t'-t)}
	T(t,t')I_{n}(x,t')b(x;t')\vrho_{\rm q}\xntp,
	\label{e6.12}
\eea
where
%
%
\bea
I_{\cal E}(\bmv{r},t')&=&\Big(1-\SP(t')\Big)\dot{\hat{\cal E}}(\bmv{r}),
	\label{e6.13}\\
I_{n}(x,t')&=&\Big(1-\SP(t')\Big)\dot{\hat n}_1(x)\label{e6.14}
\eea
are generalized flows, 
$\dot{\hat{\cal E}}(\bmv{r})=\im L_{N}\hat{\cal E}(\bmv{r})$, 
$\dot{\hat n}_1(x)=\im L_N\hat{n}_1(x)$. $\SP(t')$ is a generalized Mori 
projection operator which acts on dynamical variables $\hat{A}(\bmv{r})$ 
and has the following structure:
%
%
\bea
&\SP(t)\hat A(\bmv{r}')=\la\hat A(\bmv{r}')\ra_{\rm q}^t+{}\label{e6.15}\\
&\ds\int\d\bmv{r}\;\frac{\partial\la\hat A(\bmv{r}')\ra_{\rm q}^t}
	{\partial\la\hat{\cal E}(\bmv{r})\ra^t}
	\lp\hat{\cal E}(\bmv{r})-\la\hat{\cal E}(\bmv{r})\ra^t\rp+
	\int\d x\;\frac{\partial\la\hat A(\bmv{r}')\ra_{\rm q}^t}
	{\partial\la\hat n_1(x)\ra^t}\lp\hat n_1(x)-\la\hat n_1(x)\ra^t\rp.
	\nonumber
\eea
The projection operator $\SP(t)$ has the following properties:
%
%
\be
\begin{array}{rclcrcl}
\SP(t)\hat{\cal E}(\bmv{r})&=&\hat{\cal E}(\bmv{r}),
	&\qquad&\SP(t)(1-\SP(t))&=&0,\\
\SP(t)\hat n_1(x)&=&\hat n_1(x),
	&\qquad&\SP(t)\SP(t')&=&\SP(t).\\
\end{array}\label{e6.16}
\ee
With the help of the solution to the Liouville equation \refp{e6.12}, one 
obtains a system of coupled equations for the nonequilibrium one-particle 
distribution function $f_1(x;t)$ and the average density of energy 
$\la\hat{\cal E}(\bmv{r})\ra^t$. To do it explicitly, it is necessary to 
calculate
\bean
\ddt\la\hat n_1(x)\ra^t&=&\ddt f_1(x;t)=\la\dot{\hat n}_1(x)\ra^{t},\\
\ddt\la\hat{\cal E}(\bmv{r})\ra^t&=&\la\dot{\hat{\cal E}}(\bmv{r})\ra^t.
\eean
To this end let us use the equalities
\bean
\La\Big(1-\SP(t)\Big)\dot{\hat n}_1(x)\Ra^t&=&
	\la\dot{\hat n}_1(x)\ra^t-
	\la\dot{\hat n}_1(x)\ra_{\rm q}^t,\\
\La\Big(1-\SP(t)\Big)\dot{\hat{\cal E}}(\bmv{r})\Ra^t&=&
	\la\dot{\hat{\cal E}}(\bmv{r})\ra^t-
	\la\dot{\hat{\cal E}}(\bmv{r})\ra_{\rm q}^t
\eean
and perform averaging in the right-hand side with the nonequilibrium
distribution function \refp{e6.12}. Then we obtain, calculating 
$\la\dot{\hat n}_1(x)\ra_{\rm q}^t$, 
$\la\dot{\hat{\cal E}}(\bmv{r})\ra_{\rm q}^t$ \cite{5}:
%
%
\bea
\ddt f_1(x;t)+\frac{\bmv{p}}{m}\dd{\bmv{r}}f_1(x;t)&=&
	\int\d x'\;\dd{\bmv{r}}\mPhi(|\bmv{r}-\bmv{r}'|)
	\dd{\bmv{p}}\g_2(\bmv{r},\bmv{r}';t)f_1(x;t)f_1(x';t)+{}\nonumber\\
&&\int\d x'\intl_{-\infty}^t\d t'\;\e^{\veps(t'-t)}
	\vphi_{nn}(x,x';t,t')b(x';t')+{}\nonumber\\
&&\int\d\bmv{r}'\intl_{-\infty}^t\d t'\;\e^{\veps(t'-t)}
	\vphi_{n{\cal E}}(x,\bmv{r}';t,t')\beta(\bmv{r}';t'),\label{e6.17}
\eea
%
%
%
%
%
\bea
-\ddt\la\hat{\cal E}(\bmv{r})\ra^{t}&=&\frac{1}{2}\int\d\bmv{r'}\;
	\d\bmv{p}\;\d\bmv{p'}\;\frac{\bmv{p}}{m}\dd{\bmv{r}}
	\mPhi(|\bmv{r}-\bmv{r}'|)\g_2(\bmv{r},\bmv{r}';t)f_1(x;t)f_1(x';t)
	\label{e6.18}\\
&&+\int\d\bmv{r}'\;\d\bmv{p}\;\d\bmv{p'}\;\frac{\bmv{p}}{m}
	\dd{\bmv{r}}\ls\frac{p^{2}}{2m}+\frac12\mPhi(|\bmv{r}-\bmv{r}'|)\!\rs
	\!\g_2(\bmv{r},\bmv{r}';t)f_1(x;t)f_1(x';t)\nonumber\\
&&-\int\d x'\intl_{-\infty}^t\d t'\;\e^{\veps(t'-t)}
	\vphi_{{\cal E}n}(\bmv{r},x';t,t')b(x';t')\nonumber\\
&&-\int\d\bmv{r}'\intl_{-\infty}^t\d t'\;\e^{\veps (t'-t)}
	\vphi_{{\cal E}{\cal E}}(\bmv{r},\bmv{r}';t,t')\beta(\bmv{r}';t'),
	\nonumber
\eea
where $\g_2(\bmv{r},\bmv{r}';t)$ is a binary quasiequilibrium 
distribution function (2.32), and
%
%
\bea
\vphi_{nn}(x,x';t,t')&=&\int\d\mGamma_N\;I_n(x;t)T(t,t')I_n(x';t')
	\vrho_{\rm q}(x^N;t'),\label{e6.19}\\
\vphi_{n{\cal E}}(x,\bmv{r}';t,t')&=&\int\d\mGamma_N\;I_n(x;t)T(t,t')
	I_{\cal E}(\bmv{r}';t')\vrho_{\rm q}(x^N;t'),\label{e6.20}\\
\vphi_{{\cal E}n}(\bmv{r},x';t,t')&=&\int\d\mGamma_N\;I_{\cal E}(\bmv{r};t)
	T(t,t')I_n(x';t')\vrho_{\rm q}(x^N;t'),\label{e6.21}\\
\vphi_{{\cal E}{\cal E}}(\bmv{r},\bmv{r}';t,t')&=&\int\d\mGamma_N\;
	I_{{\cal E}}(\bmv{r};t)T(t,t')I_{{\cal E}}(\bmv{r}';t')
	\vrho_{\rm q}(x^N;t')\label{e6.22}
\eea
are generalized transport kernels which describe the kinetic and 
hydrodynamic dissipative processes in a system. We obtained generalized 
transport equations for a nonequilibrium one-particle distribution function 
and the average energy density, which describe both strong and weak 
nonequilibrium processes. From these equations one can derive interesting 
limiting cases. First of all, if two last terms in the right-hand side of 
equation \refp{e6.17} are not taken into account (these terms describe 
non-Markovian dissipative processes), then one obtains a generalized 
kinetic equation in the mean field approximation \cite{5,36}:
%
%
\be
\ddt f_1(x;t)+\frac{\bmv{p}}{m}\dd{\bmv{r}}f_1(x;t)=
	\int\d x'\;\dd{\bmv{r}}\mPhi(|\bmv{r}-\bmv{r}'|)\dd{\bmv{p}}
	\g_2(\bmv{r},\bmv{r}';t)f_1(x;t)f_1(x';t).\label{e6.23}
\ee
From equation \refp{e6.23} at $\g_2(\bmv{r},\bmv{r}';t)=1$ we have a usual 
Vlasov kinetic equation for a nonequilibrium one-particle distribution 
function. The kinetic equation \refp{e6.23} for $f_1(x;t)$ must be 
complemented by an equation for the average energy density \refp{e6.18}, 
without taking into account two last terms which describe non-Markovian 
dissipative processes. It is important to remember here that the binary 
quasiequilibrium distribution function $\g_2(\bmv{r},\bmv{r}';t)$ is a 
functional of the local values of inversed temperature $\beta(\bmv{r};t)$ 
and average density $n(\bmv{r};t)$: 
$\g_2(\bmv{r},\bmv{r}';t)=\g_2(\bmv{r},\bmv{r}'|\beta(t),n(t))$. If spatial 
correlations connected with an interaction between a separated group of 
particles and the media are neglected (that is valid in the case of small 
densities), i.e. formally putting for $\vrho_{\rm q}\xnt$ in \refp{e2.22} the 
density of interaction energy $\hat{\cal E}^{\rm int}(\bmv{r})$ to be equal 
to zero ($U_N(\bmv{r}^N;t)=0$), then \refp{e2.22} transforms into a 
quasiequilibrium distribution \refp{e2.5} that corresponds to the usual 
boundary conditions of the weakening of correlations by Bogolubov \cite{21} 
for the solution of the Liouville equation. In this case, from the system 
of equations \refp{e6.17}, \refp{e6.18}, we have a generalized kinetic 
equation for the nonequilibrium one-particle distribution function which 
was obtained earlier by Zubarev and Novikov \cite{51}
%
%
\bea
\ddt f_1(x;t)+\frac{\bmv{p}}{m}\dd{\bmv{r}}f_1(x;t)&=&
	\int\d x'\;\dd{\bmv{r}}\mPhi(|\bmv{r}-\bmv{r}'|)
	\dd{\bmv{p}}f_1(x;t)f_1(x';t)+{}\nonumber\\
&&\int\d x'\intl_{-\infty}^t\d t'\;\e^{\veps(t'-t)}
	\vphi_{nn}(x,x';t,t')a(x';t').\label{e6.24}
\eea
However, in this equation, contrary to the kinetic equation for $f_1(x;t)$ 
of paper \cite{51}, we used a projection procedure for the exclusion of 
time derivatives for parameters $a(x;t)$ which are determined from the 
self-consistency condition \refp{e6.5} and in our case, according to 
\refp{e2.12}, \refp{e2.22}, \refp{e2.23} and $u=\e$, they are equal to
$\e^{-a(x;t)}=f_1(x;t)/\e$. In the kinetic equation \refp{e6.24}, the 
transport kernel has the following structure:
%
%
\be
\vphi_{nn}(x,x';t,t')=\int\d\mGamma_N\;\Big(1-\SP'(t)\Big)\dot{\hat n}_1(x)
	T(t,t')\Big(1-\SP'(t')\Big)\dot{\hat n}_1(x')
	\prod^N_{j=1}\frac{f_1(x_j;t)}{\e},\label{e6.25}
\ee
where the projection operator $\SP'(t)$ is defined as
%
%
\be
\SP'(t)\hat A(x')=\la\hat A(x')\ra_{\rm q}^t+
	\int\d x\;\frac{\partial\la\hat A(x')\ra_{\rm q}^t}
	{\partial\la\hat n_1(x)\ra^t}\lp\hat n_1(x)-\la\hat n_1(x)\ra^t\rp,
	\label{e6.26}
\ee
and the average values $\la\hat{A}(x')\ra^{t}_{\rm q}$ are calculated with 
the help of the quasiequilibrium distribution function \refp{e2.5}. The 
kernel $\vphi_{nn}(x,x';t,t')$ describes kinetic processes in the system.

A coupled system of equations for the nonequilibrium one-particle 
distribution function and the density of the total energy \refp{e6.17}, 
\refp{e6.18} is strongly nonlinear. It takes into account complicated 
kinetic and hydrodynamic processes and such a system can be used for 
describing both strongly and weakly nonequilibrium states.

In the next subsection we apply equations \refp{e6.17}, \refp{e6.18} to the 
investigation of nonequilibrium states of the system, which are close to 
equilibrium. In this case the equations are simplified significantly.

%% file: s_03_3.tex
\newcommand{\sump}{{\mathop{\sum_{\bmv{k}}}\nolimits'}}

\subsection{Kinetics and hydrodynamics of nonequilibrium state near 
equilibrium}

Let us assume that the average energy density 
$\la\hat{\cal E}(\bmv{r})\ra^t$, the nonequilibrium one-particle 
distribution function $f_1(x;t)$ and parameters $\beta(\bmv{r};t)$, 
$b(x;t)$ deviate slightly from their equilibrium values. Then, being 
restricted
to the linear approximation \cite{5}, the quasiequilibrium distribution 
function $\vrho_{\rm q}\xnt$ \refp{e6.3} can be expanded over deviations of 
the parameters $\beta(\bmv{r};t)$, $b(x;t)$ from their equilibrium values:
%
%
\be
\vrho_{\rm q}\xnt=\vrho_0(x^N)
	\ls 1-\int\d\bmv{r}\;\delta\beta(\bmv{r};t)\hat{\cal E}(\bmv{r})-
	\int\d x\;\delta b(x;t)\hat n_1(x)\rs,\label{e6.27}
\ee
where $\vrho_0(x^N)=Z^{-1}\e^{-\beta(H-\mu N)}$ is an equilibrium 
distribution of particles, $Z=\int\d\mGamma_N\;\e^{-\beta(H-\mu N)}$ is a 
grand partition function, $\beta=1/k_{\rm B}T$ is an equilibrium value for 
inverse temperature, $k_{\rm B}$ is the Boltzmann constant, 
$\delta\beta(\bmv{r};t)=\beta(\bmv{r};t)-\beta$, 
$\delta b(x;t)=b(x;t)+\beta\mu$, $\mu$ is an equilibrium value of the 
chemical potential. In formula \refp{e6.27} it is convenient to transform 
the dynamical variables $\hat{\cal E}(\bmv{r})$, 
$\hat n_1(x)=\hat n_1(\bmv{r},\bmv{p})$ and parameters 
$\delta\beta(\bmv{r};t)$, $\delta b(x;t)$ into Fourier-components. Then we 
obtain:
%
%
\be
\vrho_{\rm q}\xnt=\vrho_0(x^N)
	\ls 1-\sump\delta\beta_{-\bmv{k}}(t)\hat{\cal E}_{\bmv{k}}-
	\sump\int\d\bmv{p}\;\delta b_{-\bmv{k}}(\bmv{p};t)
	\hat n_{\bmv{k}}(\bmv {p})\rs,\label{e6.28}
\ee
where $\sum_{\bmv{k}}^{\prime}=\sum_{\bmv{k}(\bmv{k}\neq 0)}$ and
\[
\begin{array}{rclcrcl}
\hat{\cal E}_{\bmv{k}}&=&\int\d\bmv{r}\;\e^{-\im\bmv{k}\cdot\bmv{r}}
	\hat{\cal E}(\bmv{r}),&\qquad&
\delta\beta_{-\bmv{k}}(t)&=&\int\d\bmv{r}\;\e^{\im\bmv{k}\cdot\bmv{r}}
	\delta\beta(\bmv{r};t),\\
\hat n_{\bmv{k}}(\bmv{p})&=&\int\d\bmv{r}\;\e^{-\im\bmv{k}\cdot\bmv{r}}
	\hat n_1(\bmv{r},\bmv{p}),&\qquad&
\delta b_{-\bmv{k}}(\bmv{p};t)&=&\int\d\bmv{r}\;\e^{\im\bmv{k}\cdot\bmv{r}}
	\delta b(x;t).\\
\end{array}
\]
Using the self-consistency conditions \refp{e6.5}, let us consequently 
exclude parameters $\delta\beta_{-\bmv{k}}(t)$, $\delta 
b_{-\bmv{k}}(\bmv{p};t)$ in equation \refp{e6.28}. From 
$\la\hat n_{\bmv{k}}(\bmv{p})\ra^t=
\la\hat n_{\bmv{k}}(\bmv{p})\ra_{\rm q}^t$ one finds:
%
%
\be
\int\d\bmv{p}'\;\Phi_{\bmv{k}}(\bmv{p},\bmv{p}')
	\delta b_{-\bmv{k}}(\bmv{p}';t)=-
	\la\hat n_{\bmv{k}}(\bmv{p})\ra^t-
	\la\hat n_{\bmv{k}}(\bmv{p})\hat{\cal E}_{-\bmv{k}}\ra_0
	\delta\beta_{\bmv{k}}(t),\label{e6.29}
\ee
where $\la\ldots\ra_0$ denotes averaging over the equilibrium 
distribution, namely $\la\ldots\ra_{0}=$ 
$\int\d\mGamma_N\;\ldots\vrho_0(x^{N})$, 
$\la\hat n_{\bmv{k}}(\bmv{p})\ra_0=0$, $(\bmv{k}\neq 0);$
%
%
\be
\Phi_{\bmv{k}}(\bmv{p},\bmv{p}')=
	\la\hat n_{\bmv{k}}(\bmv{p})\hat n_{-\bmv{k}}(\bmv{p}')\ra_0=
	n\delta(\bmv{p}-\bmv{p}')f_0(p')+n^2f_0(p)f_0(p')h_2(\bmv{k})
	\label{e6.30}
\ee
is an equilibrium correlation function, $n=N/V$, 
$f_0(p)=(\beta/2\pi m)^{3/2}\e^{-\beta p^{2}/2m}$ is a Maxwell 
distribution, $h_2(\bmv{k})$ is a Fourier-image of the function 
$\g_2(\bmv{R})-1$:
%
%
\be
h_2(\bmv{k})=\int\d\bmv{R}\;\e^{-\im\bmv{k}\cdot\bmv{R}}
	\Big(\g_2(\bmv{R})-1\Big),\label{e6.31}
\ee
$\g_2(\bmv{R})=\g_2(|\bmv{r}-\bmv{r}'|)$ is a pair equilibrium distribution 
function. Now let us define the function 
$\Phi^{-1}_{\bmv{k}}(\bmv{p},\bmv{p}')$ inversed to the function 
$\Phi_{\bmv{k}}(\bmv{p},\bmv{p}')$ by means of the relation
%
%
\be
\int\d\bmv{p}'\;\Phi^{-1}_{\bmv{k}}(\bmv{p}'',\bmv{p}')
	\Phi_{\bmv{k}}(\bmv{p}',\bmv{p})=\delta(\bmv{p}''-\bmv{p}).
	\label{e6.32}
\ee
Taking into account \refp{e6.30}, from \refp{e6.32} we find 
$\Phi^{-1}_{\bmv{k}}(\bmv{p}'',\bmv{p}')$ in an explicit form
%
%
\be
\Phi^{-1}_{\bmv{k}}(\bmv{p}'',\bmv{p}')=\frac{\delta(\bmv{p}''-\bmv{p}')}
	{nf_0(p'')}-c_2(\bmv{k}),\label{e6.33}
\ee
where $c_2(\bmv{k})$ denotes a direct correlation function which is 
connected with the correlation function $h_2(\bmv{k})$ as: 
$h_2(\bmv{k})=c_2(\bmv{k})[1-nc_2(\bmv{k})]^{-1}$. Let us multiply the 
left-hand side of \refp{e6.29} by the function 
$\Phi^{-1}_{\bmv{k}}(\bmv{p}'',\bmv{p}')$ and integrate it with respect to 
$\bmv{p}'$. Taking into account \refp{e6.32}, we find:
\[
\delta b_{-\bmv{k}}(\bmv{p};t)=-\int\d\bmv{p}'\;
	\Phi^{-1}_{\bmv{k}}(\bmv{p},\bmv{p}')
	\la\hat n_{\bmv{k}}(\bmv{p}')\ra^{t}-
	\delta\beta_{\bmv{k}}(t)\int\d\bmv{p}'\;
	\Phi^{-1}_{\bmv{k}}(\bmv{p},\bmv{p}')
	\la\hat n_{\bmv{k}}(\bmv{p}')\hat{\cal E}_{-\bmv{k}}\ra_0.
\]
Substituting the obtained value of the parameter 
$\delta b_{-\bmv{k}}(\bmv{p};t)$ in \refp{e6.28} we find after some 
transformations:
%
%
\be
\vrho_{\rm q}\xnt=\vrho_0(x^N)\!
	\ls 1-\sump\delta\beta_{-\bmv{k}}(t)\hat h^{\rm int}_{\bmv{k}}+
	\sump\!\int\d\bmv{p}\;\d\bmv{p}'\;\la\hat n_{\bmv{k}}(\bmv{p}')\ra^t
	\Phi^{-1}_{\bmv{k}}(\bmv{p}',\bmv{p})\hat n_{\bmv{k}}(\bmv{p})\rs\!,
	\label{e6.34}
\ee
where
%
%
\be
\hat h^{\rm int}_{\bmv{k}}=\hat{\cal E}_{\bmv{k}}-
	\int\d\bmv{p}\;\d\bmv{p}'\;
	\la\hat{\cal E}_{\bmv{k}}\hat n_{-\bmv{k}}(\bmv{p}')\ra_0
	\Phi^{-1}_{\bmv{k}}(\bmv{p}',\bmv{p})\hat n_{\bmv{k}}(\bmv{p})=
	\hat{\cal E}^{\rm int}_{\bmv{k}}-
	\la\hat{\cal E}^{\rm int}_{\bmv{k}}\hat n_{-\bmv{k}}\ra_0
	S_2^{-1}(\bmv{k})\hat n_{\bmv{k}},\label{e6.35}
\ee
%
%
%
%
%
\be
\hat{\cal E}^{\rm int}_{\bmv{k}}=\frac12\sum_{l\neq j=1}^N
	\mPhi(|\bmv{r}_{lj}|)\e^{-\im\bmv{k}\cdot\bmv{r}_{l}},\qquad
	\hat n_{\bmv{k}}=\sum_{l=1}^N\e^{-\im\bmv{k}\cdot\bmv{r}_{l}}
	\label{e6.36}
\ee
are the Fourier-components of densities for the interaction energy and the 
number of particles, respectively. Further, it is more convenient, instead 
of the dynamical variable of energy $\hat{\cal E}_{\bmv{k}}$, to use the 
variable $\hat h^{\rm int}_{\bmv{k}}$ \refp{e6.35} which is orthogonal to 
$\hat n_{\bmv{k}}(\bmv{p})$ by means of the equality:
%
%
\be
\la\hat h^{\rm int}_{\bmv{k}}\hat n_{-\bmv{k}}\ra_0=0.\label{e6.37}
\ee
From the structure of the dynamical variable $\hat h^{\rm int}_{\bmv{k}}$ 
\refp{e6.35} it can be seen that it corresponds to a potential part of 
the Fourier-component of the generalized enthalpy $\hat h_{\bmv{k}}$, which 
is introduced in molecular hydrodynamics \cite{5,75,77}:
%
%
\be
\hat h_{\bmv{k}}=\hat{\cal E}_{\bmv{k}}-
	\la\hat{\cal E}_{\bmv{k}}\hat n_{-\bmv{k}}\ra_0
	\hat n_{\bmv{k}}=\hat h^{\rm kin}_{\bmv{k}}+
	\hat h^{\rm int}_{\bmv{k}},\label{e6.38}
\ee
where
%
%
\be
\hat h^{\rm kin}_{\bmv{k}}=\hat{\cal E}^{\rm kin}_{\bmv{k}}-
	\la\hat{\cal E}^{\rm kin}_{\bmv{k}}\hat n_{-\bmv{k}}\ra_0
	S_2^{-1}(\bmv{k})\hat n_{\bmv{k}}\label{e6.39}
\ee
is a kinetic part of the generalized enthalpy,
\[
\hat{\cal E}^{\rm kin}_{\bmv{k}}=\sum_{l=1}^N\frac{p_l^2}{2m}
	\e^{-\bmv{k}\cdot\bmv{r}_l}
\]
is the Fourier-component of the kinetic energy density.
$S_2=\la\hat n_{\bmv{k}}\hat n_{-\bmv{k}}\ra_0$ is a static structure 
factor of the system. Taking into account the orthogonality of the dynamical
variables $\hat h^{\rm int}_{\bmv{k}}$ and $\hat n_{\bmv{k}}$ \refp{e6.37}, 
from the self-consistency condition
$\la\hat h^{\rm int}_{\bmv{k}}\ra^t=
\la\hat h^{\rm int}_{\bmv{k}}\ra_{\rm q}^t$ that is equivalent to 
$\la\hat{\cal E}_{\bmv{k}}\ra^t=\la\hat{\cal E}_{\bmv{k}}\ra_{\rm q}^t$ 
one defines the parameter $\delta\beta_{\bmv{k}}(t)$:
%
%
\be
\delta\beta_{\bmv{k}}(t)=-\la\hat h^{\rm int}_{\bmv{k}}\ra^t
	\Phi^{-1}_{hh}(\bmv{k}),\label{e6.40}
\ee
where 
$\Phi_{hh}(\bmv{k})=
\la\hat h^{\rm int}_{\bmv{k}}\hat h^{\rm int}_{-\bmv{k}}\ra_0$. Finally, 
let us substitute \refp{e6.40} into \refp{e6.34}. As a result, we obtain 
a quasiequilibrium distribution function in the linear approximation:
%
%
\be
\text{\small$\ds\vrho_{\rm q}\xnt=\vrho_0(x^N)
\ds\ls 1+\sump\la\hat h^{\rm int}_{\bmv{k}}\ra^t
	\Phi^{-1}_{hh}(\bmv{k})\hat h^{\rm int}_{\bmv{k}}+
	\sump\!\int\d\bmv{p}\;\d\bmv{p}'\;\la\hat n_{\bmv{k}}(\bmv{p}')\ra^t
	\Phi^{-1}_{\bmv{k}}(\bmv{p}',\bmv{p})
	\hat n_{\bmv{k}}(\bmv{p})\rs\!.$}\label{e6.41}
\ee
Taking into account \refp{e6.41}, the nonequilibrium distribution function
$\vrho(x^N;t)$ \refp{e6.10} or \refp{e6.12} has the following form in this 
approximation \cite{5}:
%
%
\bea
&&\vrho\xnt=\vrho_0(x^N)\times{}\label{e6.42}\\
&&\bigg[ 1+\sump\la\hat h^{\rm int}_{\bmv{k}}\ra^t
	\Phi^{-1}_{hh}(\bmv{k})\hat h^{\rm int}_{\bmv{k}}+{}\nonumber\\
&&\phantom{\bigg[ 1+{}}\sump\int\d\bmv{p}\;\d\bmv{p}'\;
	\la\hat n_{\bmv{k}}(\bmv{p}')\ra^t
	\Phi^{-1}_{\bmv{k}}(\bmv{p}',\bmv{p})
	\hat n_{\bmv{k}}(\bmv{p})-{}\nonumber\\
&&\phantom{\bigg[ 1+{}}\sump\intl_{-\infty}^t\d t'\;\e^{\veps(t'-t)}
	\la\hat h^{\rm int}_{\bmv{k}}\ra^{t'}
	\Phi^{-1}_{hh}(\bmv{k})T_0(t,t')I_h^{\rm int}(-\bmv{k})-{}\nonumber\\
&&\phantom{\bigg[ 1+{}}\sump\int\d\bmv{p}\;\d\bmv{p}'
	\intl_{-\infty}^t\d t'\;\e^{\veps(t'-t)}
	\la\hat n_{\bmv{k}}(\bmv{p}')\ra^{t'}
	\Phi^{-1}_{\bmv{k}}(\bmv{p}',\bmv{p})
	T_0(t,t')I_n(-\bmv{k},\bmv{p})\bigg],\nonumber
\eea
where
%
%
\bea
I_n(\bmv{k},\bmv{p})&=&(1-\SP_0)\dot{\hat n}_{\bmv{k}}(\bmv{p}),
	\label{e6.43}\\
I_h^{\rm int}(\bmv{k})&=&(1-\SP_0)\dot{\hat h}{}^{\rm int}_{\bmv{k}}
	\label{e6.44}
\eea
are generalized fluxes in the linear approximation,
$\dot{\hat n}_{\bmv{k}}(\bmv{p})=\im L_N\hat n_{\bmv{k}}(\bmv{p})$, 
$\dot{\hat{h}}{}^{\rm int}_{\bmv{k}}=\im L_N\hat h^{\rm int}_{\bmv{k}}$,
$T_0(t,t')=\e^{(t-t')(1-\SP_0)\im L_N}$ is a time evolution operator with 
the projection operator $\SP_0$ which is a linear approximation of the 
generalized Mori projection operator $\SP(t)$ \refp{e6.15}, with taking 
into account orthogonalization of variables 
$\{\hat{\cal E}_{\bmv{k}},\hat{n}_{\bmv{k}}(\bmv{p})\}\to
\{\hat{h}_{\bmv{k}}^{\rm int},\hat{n}_{\bmv{k}}(\bmv{p})\}$. 
According to the structure \refp{e6.41}, $\SP_0$ acts on the dynamical 
variables $\hat{A}_{\bmv{k}}$
%
%
\be
\SP_0\hat{A}_{\bmv{k}'}=\sum_{\bmv{k}}
	\la\hat{A}_{\bmv{k}'}\hat h^{\rm int}_{-\bmv{k}}\ra_0
	\Phi^{-1}_{hh}(\bmv{k})\hat h^{\rm int}_{\bmv{k}}+
	\sum_{\bmv{k}}\int\d\bmv{p}\;\d\bmv{p}'\;
	\la\hat{A}_{\bmv{k}'}\hat n_{-\bmv{k}}(\bmv{p}')\ra_0
	\Phi^{-1}_{\bmv{k}}(\bmv{p}',\bmv{p})
	\hat n_{\bmv{k}}(\bmv{p}).\label{e6.45}
\ee

For kinetics and hydrodynamics of nonequilibrium processes which are
close to an equilibrium state, the generalized transport equations 
\refp{e6.17}, \refp{e6.18} in the linear approximation for the 
nonequilibrium distribution function $\vrho\xnt$ \refp{e6.42} transform 
into a transport equation for $f_{\bmv{k}}(\bmv{p};t)=\la\hat 
n_{\bmv{k}}(\bmv{p})\ra^t$,
$h^{\rm int}_{\bmv{k}}(t)=\la\hat h^{\rm int}_{\bmv{k}}\ra^t$
%
%
\[
\ddt f_{\bmv{k}}(\bmv{p};t)+
	\frac{\im\bmv{k}\cdot\bmv{p}}{m}f_{\bmv{k}}(\bmv{p};t)=-
	\frac{\im\bmv{k}\cdot\bmv{p}}{m}nf_0(p)c_2(\bmv{k})
	\int\d\bmv{p}'\;f_{\bmv{k}}(\bmv{p}';t)+
	\im\Omega_{nh}(\bmv{k},\bmv{p})h^{\rm int}_{\bmv{k}}(t)-{}
\]
\be
\int\d\bmv{p}'\intl_{-\infty}^t\d t'\;\e^{\veps(t'-t)}
	\vphi_{nn}(\bmv{k},\bmv{p},\bmv{p}';t,t')f_{\bmv{k}}(\bmv{p}';t')-
	\intl_{-\infty}^t\d t'\;\e^{\veps(t'-t)}
	\vphi_{nh}(\bmv{k},\bmv{p};t,t')h^{\rm int}_{\bmv{k}}(t'),
	\label{e6.46}
\ee
%
%
%
%
%
\bea
&&\ddt h^{\rm int}_{\bmv{k}}(t)=\int\d\bmv{p}\;
	\im\Omega_{hn}(\bmv{k},\bmv{p})f_{\bmv{k}}(\bmv{p};t)-
	\label{e6.47}\\
&&\int\d\bmv{p}\intl_{-\infty}^t\d t'\;\e^{\veps(t'-t)}
	\vphi_{hn}(\bmv{k},\bmv{p};t,t')f_{\bmv{k}}(\bmv{p};t')-
	\intl_{-\infty}^t\d t'\;\e^{\veps(t'-t)}
	\vphi_{hh}(\bmv{k};t,t')h^{\rm int}_{\bmv{k}}(t'),\nonumber
\eea
where $\im\Omega_{nh}(\bmv{k},\bmv{p})$, $\im\Omega_{hn}(\bmv{k},\bmv{p})$ 
are  normalized static correlation functions:
%
%
\bea
\im\Omega_{nh}(\bmv{k},\bmv{p}\phantom{'})&=&
	\la\dot{\hat n}_{\bmv{k}}(\bmv{p})
	\hat h^{\rm int}_{-\bmv{k}}\ra_0\Phi^{-1}_{hh}(\bmv{k}),
	\label{e6.48}\\
\im\Omega_{hn}(\bmv{k},\bmv{p})&=&\int\d\bmv{p}'\;
	\la\dot{\hat h}^{\rm int}_{\bmv{k}}\hat n_{-\bmv{k}}(\bmv{p}')\ra_0
	\Phi^{-1}_{\bmv{k}}(\bmv{p}',\bmv{p})\label{e6.49}
\eea
and
%
%
\bea
\vphi_{nn}(\bmv{k},\bmv{p},\bmv{p}';t,t')&=&\int\d\bmv{p}''\;
	\la I_n(\bmv{k},\bmv{p})T_0(t,t')I_n(-\bmv{k},\bmv{p}'')\ra_0
	\Phi^{-1}_{\bmv{k}}(\bmv{p}'',\bmv{p}'),\label{e6.50}\\
\vphi_{hn}(\bmv{k},\bmv{p};t,t')&=&\int\d\bmv{p}'\;
	\la I_h^{\rm int}(\bmv{k})T_0(t,t')I_n(-\bmv{k},\bmv{p}')\ra_0
	\Phi^{-1}_{\bmv{k}}(\bmv{p}',\bmv{p}),\label{e5.51}\\
\vphi_{nh}(\bmv{k},\bmv{p};t,t')&=&
	\la I_n(\bmv{k},\bmv{p})T_0(t,t')I_h^{\rm int}(-\bmv{k})\ra_0
	\Phi^{-1}_{hh}(\bmv{k}),\label{e6.52}\\
\vphi_{hh}(\bmv{k};t,t')&=&
	\la I_h^{\rm int}(\bmv{k})T_0(t,t')I_h^{\rm int}(-\bmv{k})\ra_0
	\Phi^{-1}_{hh}(\bmv{k})\label{e6.53}
\eea
are generalized transport kernels (memory functions) which describe 
kinetic and hydrodynamic processes. The system of transport equations 
\refp{e6.46}, \refp{e6.47} is closed. Eliminating 
$h^{\rm int}_{\bmv{k}}(t)$, it is possible to obtain a closed kinetic 
equation for a Fourier component of the nonequilibrium one-particle 
distribution function. For this purpose one uses a Laplace transform with 
respect to time, assuming that at $t>0$ the quantities 
$f_{\bmv{k}}(\bmv{p};t=0)$, $h^{\rm int}_{\bmv{k}}(t=0)$ are known
%
%
\be
A(z)=\im\int^{\infty}_0\d t\;\e^{\im zt}A(t),
	\qquad z=\omega+\im\veps,\qquad \veps\to+0.\label{e6.54}
\ee
Then, equations (6.46) and (6.47) are presented in the form:
%
%
\bea
&&zf_{\bmv{k}}(\bmv{p};z)+
	\frac{\im\bmv{k}\cdot\bmv{p}}{m}f_{\bmv{k}}(\bmv{p};z)=-
	\frac{\im\bmv{k}\cdot\bmv{p}}{m}nf_0(p)c_2(\bmv{k})
	\int\d\bmv{p}'\;f_{\bmv{k}}(\bmv{p}';z)+{}\label{e6.55}\\
&&\Sigma_{nh}(\bmv{k},\bmv{p};z)h^{\rm int}_{\bmv{k}}(z)-
	\int\d\bmv{p}'\;\vphi_{nn}(\bmv{k},\bmv{p},\bmv{p}';z)
	f_{\bmv{k}}(\bmv{p}';z)+f_{\bmv{k}}(\bmv{p};t=0),\nonumber
\eea
%
%
%
%
%
\be
zh^{\rm int}_{\bmv{k}}(z)=\int\d\bmv{p}'\;\Sigma_{hn}(\bmv{k},\bmv{p}';z)
	f_{\bmv{k}}(\bmv{p}';z)-
	\vphi_{hh}(\bmv{k};z)h^{\rm int}_{\bmv{k}}(z)+
	h^{\rm int}_{\bmv{k}}(t=0),\label{e6.56}
\ee
where
%
%
\bea
\Sigma_{nh}(\bmv{k},\bmv{p};z)&=&\im\Omega_{nh}(\bmv{k},\bmv{p})-
	\vphi_{nh}(\bmv{k},\bmv{p};z),
	\label{e6.57}\\
\Sigma_{hn}(\bmv{k},\bmv{p};z)&=&\im\Omega_{hn}(\bmv{k},\bmv{p})-
	\vphi_{hn}(\bmv{k},\bmv{p};z).\label{e6.58}
\eea
Let us solve equation \refp{e6.56} with respect to 
$h^{\rm int}_{\bmv{k}}(z)$ and substitute the result into \refp{e6.55}. 
Then one obtains a closed kinetic equation for $f_{\bmv{k}}(\bmv{p};z)$ (at 
$h^{\rm int}_{\bmv{k}}(t=0)=0$):
%
%
\bea
&&zf_{\bmv{k}}(\bmv{p};z)+
	\frac{\im\bmv{k}\cdot\bmv{p}}{m}f_{\bmv{k}}(\bmv{p};z)=-
	\frac{\im\bmv{k}\cdot\bmv{p}}{m}nf_0(p)c_2(\bmv{k})
	\int\d\bmv{p}'\;f_{\bmv{k}}(\bmv{p}';z)-{}\nonumber\\
&&\int\d\bmv{p}'\;D_{nn}(\bmv{k},\bmv{p},\bmv{p}';z)
	f_{\bmv{k}}(\bmv{p}';z)+f_{\bmv{k}}(\bmv{p};t=0),\label{e6.59}
\eea
where
%
%
\be
D_{nn}(\bmv{k},\bmv{p},\bmv{p}';z)=\vphi_{nn}(\bmv{k},\bmv{p},\bmv{p}';z)-
	\Sigma_{nh}(\bmv{k},\bmv{p};z)\frac{1}{z+\vphi_{hh}(\bmv{k};z)}
	\Sigma_{hn}(\bmv{k},\bmv{p}';z)\label{e6.60}
\ee
is a generalized transport kernel of kinetic processes, which is 
renormalized taking into account the processes of transport of the 
potential energy of interaction between the particles. If we put formally 
in $D_{nn}(\bmv{k},\bmv{p},\bmv{p}';z)$ that 
$\hat h^{\rm int}_{-\bmv{k}}=0$ (such equality is valid when the 
contribution of the average potential energy of interaction is much smaller 
than that of the average kinetic energy), then from \refp{e6.59} one 
obtains a kinetic equation for $f_{\bmv{k}}(\bmv{p};z)$,
%
%
\bea
&&zf_{\bmv{k}}(\bmv{p};z)+
	\frac{\im\bmv{k}\cdot\bmv{p}}{m}f_{\bmv{k}}(\bmv{p};z)=-
	\frac{\im\bmv{k}\cdot\bmv{p}}{m}nf_0(p)c_2(\bmv{k})
	\int\d\bmv{p}'\;f_{\bmv{k}}(\bmv{p}';z)-{}\nonumber\\
&&\int\d\bmv{p}'\;\vphi'_{nn}(\bmv{k},\bmv{p},\bmv{p}';z)
	f_{\bmv{k}}(\bmv{p}';z)+f_{\bmv{k}}(\bmv{p};t=0).\label{e6.61}
\eea
Equation \refp{e6.61} was derived for the first time by using the Mori 
projection operators method in \cite{103,104,105}, when the microscopic 
phase density $\hat n_{\bmv{k}}(\bmv {p})$ was a parameter of an 
abbreviated description. In this case the memory function 
$\vphi'_{nn}(\bmv{k},\bmv{p},\bmv{p}';z)$ has the following structure:
%
%
\be
\vphi'_{nn}(\bmv{k},\bmv{p},\bmv{p}';t,t')=\int\d\bmv{p}''\;
	\la I^0_n(\bmv{k},\bmv{p})T'_0(t,t') I^0_n(-\bmv{k},\bmv{p}'')\ra_0
	\Phi^{-1}_{\bmv{k}}(\bmv{p}'',\bmv{p}'),\label{e6.62}
\ee
where
%
%
\be
I^0_n(\bmv{k},\bmv{p})=(1-\SP'_0)\dot{\hat n}_{\bmv{k}}(\bmv{p})
	\label{e6.63}
\ee
is a generalized flow, $\SP'_0$ is the Mori projection operator
introduced in \cite{103,104,105}:
%
%
%
%
%
\be
\SP'_0\hat{A}_{\bmv{k}'}=\sum_{\bmv{k}}\int\d\bmv{p}\;\d\bmv{p}'\;
	\la \hat{A}_{\bmv{k}'}\hat n_{-\bmv{k}}(\bmv{p}')\ra_0
	\Phi^{-1}_{\bmv{k}}(\bmv{p}',\bmv{p})
	\hat n_{\bmv{k}}(\bmv{p}),\label{e6.64}
\ee
$T'_{0}(t,t')=\e^{(t-t')(1-\SP'_{0})\im L_{N}}$ is the corresponding time 
evolution operator. The kinetic equation \refp{e6.59} for the Fourier 
component of the nonequilibrium one-particle distribution function 
stimulates investigations of the dynamical structure factor 
$S(\bmv{k};\omega)$, time correlation functions for transverse and 
longitudinal flows of the particles, diffusion coefficients, viscosities 
for dense gases and liquids 
\cite{75,77,103,104,105,106,107,108,108a,109,110,111,112,113,114}. In 
papers by Mazenko, there was performed a derivation of the linearized 
Boltzmann-Enskog equation by expanding memory functions 
$\vphi'_{nn}(\bmv{k},\bmv{p},\bmv{p}';t,t')$ on density. In the case of 
weak interactions a Fokker-Planck equation was obtained. Mazenko 
\cite{109} and Forster \cite{110,131} calculated the generalized viscosity 
coefficient with the help of kinetic equation \refp{e6.59}. Self-diffusion 
of a one-component plasma \cite{172}, as well as spectra of mass and charge 
fluctuations in ionic solutions were investigated on the basis of this 
equation. In actual calculations, certain approximations for the memory 
functions $\vphi'_{nn}(\bmv{k},\bmv{p},\bmv{p}';t,t')$ were used. However, 
the main drawback of the kinetic equation \refp{e6.59} lies in its 
non-self-consistency with the conservation law of the total energy for 
dense gases and liquids when the contribution of the interaction energy between 
the particles to thermodynamic quantities and transport coefficients 
becomes important. This fact was also pointed out earlier in the paper by 
John and Forster \cite{113}, where they performed an investigation of 
the dynamical structure factor $S(\bmv{k};\omega)$ in the intermediate 
region of wavevector $\bmv{k}$ and frequency values $\omega$ for a simple 
liquid on the basis of the set of dynamical variables of an abbreviated 
description $\hat n_{\bmv{k}}(\bmv{p})$ and $\hat{\cal E}_{\bmv{k}}$.

In the next subsection, on the basis of a system of transport equations
for Fourier components of the nonequilibrium one-particle distribution
function and the potential part of enthalpy \refp{e6.46}, \refp{e6.47}, we 
shall obtain equations for time correlation functions. We  shall also
investigate the spectrum of collective excitations and the structure of 
generalized transport coefficients.

%% file: s_03_4.tex
\subsection{Time correlation functions, collective modes and generalized
transport coefficients}

With the help of combined equations \refp{e6.46}, \refp{e6.47} one 
obtains a system for time correlation functions:
%
%
\bea
\Phi_{nn}(\bmv{k},\bmv{p},\bmv{p}';t)&=&\int\d\bmv{p}''\;
	\la\hat n_{\bmv{k}}(\bmv{p};t)\hat n_{-\bmv{k}}(\bmv{p}'';0)\ra_0
	\Phi^{-1}_{\bmv{k}}(\bmv{p}'',\bmv {p}'),\label{e6.65}\\
\Phi_{hn}(\bmv{k},\bmv{p};t)&=&\int\d\bmv{p}'\;
	\la\hat h^{\rm int}_{\bmv{k}}(t)\hat n_{-\bmv{k}}(\bmv{p}';0)\ra_0
	\Phi^{-1}_{\bmv{k}}(\bmv{p}',\bmv {p}),\label{e6.66}\\
\Phi_{nh}(\bmv{k},\bmv{p};t)&=&
	\la\hat n_{\bmv{k}}(\bmv{p};t)\hat h^{\rm int}_{-\bmv{k}}(0)\ra_0
	\Phi^{-1}_{hh}(\bmv{k}),\label{e6.67}\\
\Phi_{hh}(\bmv{k};t)&=&
	\la\hat h^{\rm int}_{\bmv{k}}(t)\hat h^{\rm int}_{-\bmv{k}}(0)\ra_0
	\Phi^{-1}_{hh}(\bmv{k}),\label{e6.68}
\eea
where 
$\hat n_{\bmv{k}}(\bmv{p};t)=\e^{-\im L_Nt}\hat n_{\bmv{k}}(\bmv{p};0)$, 
$\hat h^{\rm int}_{\bmv{k}}(t)=\e^{-\im L_Nt}\hat h^{\rm int}_{\bmv{k}}(0)$.

One uses the Fourier transform with respect to time
\[
\la a\ra_{\omega}=\int_{-\infty}^{\infty}\d t\;\e^{\im\omega t}\la a\ra^t.\]
Then we write the system of equations \refp{e6.46}, \refp{e6.47} in the 
form:
%
%
\bea
\!\!\!\!\!\!\!\!\!\!\!\!-\im\omega\la\hat n_{\bmv{k}}(\bmv{p})\ra_\omega=&&
	\!\!\!\!\!\!\int\d\bmv{p}'\;
	\Sigma_{nn}(\bmv{k},\bmv{p},\bmv{p}';\omega+\im\veps)
	\la\hat n_{\bmv{k}}(\bmv{p}')\ra_\omega+
	\Sigma_{nh}(\bmv{k},\bmv{p};\omega+\im\veps) 
	\la\hat h^{\rm int}_{\bmv{k}}\ra_\omega,\label{e6.69}\\
\!\!\!\!\!\!\!\!\!\!\!\!-\im\omega\la\hat h^{\rm int}_{\bmv{k}}\ra_\omega=&&
	\!\!\!\!\!\!\int\d\bmv{p}'\;
	\Sigma_{hn}(\bmv{k},\bmv{p}';\omega+\im\veps)
	\la\hat n_{\bmv{k}}(\bmv{p}')\ra_\omega-
	\vphi_{hh}(\bmv{k};\omega+\im\veps)
	\la\hat h^{\rm int}_{\bmv{k}}\ra_\omega,\label{e6.70}
\eea
where
%
%
\bea
\Sigma_{nn}(\bmv{k},\bmv{p},\bmv{p}';\omega+\im\veps)&=&
	\im\Omega_{nn}(\bmv{k},\bmv{p},\bmv{p}')-
	\vphi_{nn}(\bmv{k},\bmv{p},\bmv{p}';\omega+\im\veps),\label{e6.71}\\
\Sigma_{nh}(\bmv{k},\bmv{p};\omega+\im\veps)&=&
	\im\Omega_{nh}(\bmv{k},\bmv{p})-
	\vphi_{nh}(\bmv{k},\bmv{p};\omega+\im\veps),
	\label{e6.72}\\
\Sigma_{hn}(\bmv{k},\bmv{p};\omega+\im\veps)&=&
	\im\Omega_{hn}(\bmv{k},\bmv{p})-
	\vphi_{hn}(\bmv{k},\bmv{p};\omega+\im\veps).\label{e6.73}
\eea
It is more convenient to present the system of equations \refp{e6.69}, 
\refp{e6.70} in a matrix form:
%
%
\be
-\im\omega\la\tilde{a}_{\bmv{k}}\ra_{\omega}=
	\tilde{\Sigma}(\bmv{k};\omega+\im\veps)
	\la\tilde{a}_{\bmv{k}}\ra_{\omega},\label{e6.74}
\ee
where $\tilde{a}_{\bmv{k}}=\mbox{col}(\hat{n}_{\bmv{k}}(\bmv{p}),
\hat{h}^{\rm int}_{\bmv{k}})$ is a vector-column and
%
%
\bea
\tilde{\Sigma}(\bmv{k};\omega+\im\veps)&=&
\left[
\begin{array}{lr}
\int\d\bmv{p}'\;\Sigma_{nn}(\bmv{k},\bmv{p},\bmv{p}';\omega+\im\veps)
	&\quad\Sigma_{nh}(\bmv{k},\bmv{p};\omega+\im \veps)\\ [1ex]
\int\d\bmv{p}'\;\Sigma_{hn}(\bmv{k},\bmv{p}';\omega+\im\veps)
	&-\vphi_{hh}(\bmv{k};\omega+\im\veps)\\
\end{array}
\right],\label{e6.75}\\ [1ex]
\tilde{\Sigma}(\bmv{k};\omega+\im\veps)&=&
	\int_0^{\infty}\d t\;\e^{\im(\omega+\im\veps)t}
	\tilde{\Sigma}(\bmv{k};t).\nonumber
\eea
Now, one uses the solution to the Liouville equation in approximation 
\refp{e6.41} without introducing the projection operator $\SP_{\rm q}(t)$:
\[
\vrho\xnt=\vrho_{\rm q}\xnt-\intl_{-\infty}^t\d t'\;\e^{\veps(t'-t)}
	\e^{\im L_N(t'-t)}\lp\ddt'+\im L_N\rp\vrho_{\rm q}\xntp.
\]
Then, from the self-consistency conditions
$\la\tilde{a}_{\bmv{k}}\ra^t=\la\tilde{a}_{\bmv{k}}\ra^t_{\rm q}$ one 
obtains a system of equations which connects the average values
$\la\hat n_{\bmv{k}}(\bmv{p})\ra_\omega$ and  
$\la\hat h^{\rm int}_{\bmv{k}}\ra_\omega$ with spectral functions of time 
correlation functions:
\bean
\im\omega&&
\left[
\begin{array}{ll}
\int\d\bmv{p}'\;\Phi_{nn}(\bmv{k},\bmv{p},\bmv{p}';\omega+\im\veps)&
	\phantom{-{}}\Phi_{nh}^{\rm int}(\bmv{k},\bmv{p};\omega+\im\veps)
	\\ [1ex]
\int\d\bmv{p}'\;\Phi_{hn}^{\rm int}(\bmv{k},\bmv{p}';\omega+\im\veps)&
	-\Phi_{hh}^{\rm int}(\bmv{k};\omega+\im\veps)\\
\end{array}
\right]
\times
\left[
\begin{array}{l}
\la\hat n_{\bmv{k}}(\bmv{p})\ra_\omega\\ [1ex]
\la\hat h^{\rm int}_{\bmv{k}}\ra_\omega\\ 
\end{array}
\right]={}\\ [2ex]
&&\left[
\begin{array}{ll}
\int\d\bmv{p}'\;\Phi_{nn}(\bmv{k},\bmv{p},\bmv{p}')&\phantom{-{}}0\\ [1ex]
0&-\Phi_{hh}(\bmv{k})\\
\end{array}
\right]-{}\\ [2ex]
\im(\omega+\im\veps)&&
\left[
\begin{array}{ll}
\int\d\bmv{p}'\;\Phi_{nn}(\bmv{k},\bmv{p},\bmv{p}';\omega+\im\veps)&
	\phantom{-{}}\Phi_{nh}^{\rm int}(\bmv{k},\bmv{p};\omega+\im\veps)
	\\ [1ex]
\int\d\bmv{p}'\;\Phi_{hn}^{\rm int}(\bmv{k},\bmv{p}';\omega+\im\veps)&
	-\Phi_{hh}^{\rm int}(\bmv{k};\omega+\im\veps)\\
\end{array}
\right]
\times
\left[
\begin{array}{l}
\la\hat n_{\bmv{k}}(\bmv{p})\ra_\omega\\ [1ex]
\la\hat h^{int}_{\bmv{k}}\ra_\omega\\ 
\end{array}
\right].
\eean
Such a system can be presented in the next compact form:
%
%
\be
\im\omega\tilde{\Phi}(\bmv{k};\omega+\im\veps)
	\la\tilde{a}_{\bmv{k}}\ra_{\omega}=
	\ls\tilde{\Phi}(\bmv{k})-
	\im(\omega+\im\veps)\tilde{\Phi}(\bmv{k};\omega+\im\veps)\rs
	\la\tilde{a}_{\bmv{k}}\ra_{\omega}.\label{e6.76}
\ee
Let us multiply equation \refp{e6.74} by the matrix 
$\tilde{\Phi}(\bmv{k};\omega+\im\veps)$ and compare the result with 
equation \refp{e6.76}. So we find
%
%
\bea
z\tilde{\Phi}(\bmv{k};z)&=&\tilde{\Sigma}(\bmv{k};z)\tilde{\Phi}(\bmv{k};z)-
	\tilde{\Phi}(\bmv{k}),\qquad z=\omega+\im\veps,\label{e6.77}\\
\tilde{\Phi}(\bmv{k};z)&=&\im\int_0^{\infty}\d t\;\e^{\im zt}
	\tilde{\Phi}(\bmv{k};t),\qquad\qquad\! \veps \to 0\nonumber
\eea
or in an explicit form:
%
%
\bea
\text{\small$\ds z\Phi_{nn}(\bmv{k},\bmv{p},\bmv{p}';z)$}
	&\text{\small$\ds=$}&\text{\small$\ds\int\d\bmv{p}''\;
	\Sigma_{nn}(\bmv{k},\bmv{p},\bmv{p}'';z)
	\Phi_{nn}(\bmv{k},\bmv{p}'',\bmv{p}';z)+{}$}\label{e6.78}\\
&&\text{\small$\ds\Sigma_{nh}(\bmv{k},\bmv{p};z)
	\Phi_{hn}^{\rm int}(\bmv{k},\bmv{p}';z)-
	\Phi_{nn}(\bmv{k},\bmv{p},\bmv{p}'),$}\nonumber\\
\text{\small$\ds z\Phi_{nh}^{\rm int}(\bmv{k},\bmv{p};z)$}
	&\text{\small$\ds=$}&\text{\small$\ds\int\d\bmv{p}''\;
	\Sigma_{nn}(\bmv{k},\bmv{p},\bmv{p}'';z)
	\Phi_{nh}^{\rm int}(\bmv{k},\bmv{p}'';z)+
	\Sigma_{nh}(\bmv{k},\bmv{p};z)\Phi_{hh}^{\rm int}(\bmv{k};z),$}
	\label{e6.79}\\
\text{\small$\ds z\Phi_{hn}^{\rm int}(\bmv{k},\bmv{p}';z)$}
	&\text{\small$\ds=$}&\text{\small$\ds\int\d\bmv{p}''\;
	\Sigma_{hn}(\bmv{k},\bmv{p}'';z)
	\Phi_{nn}^{\rm int}(\bmv{k},\bmv{p}'',\bmv{p}';z)-
	\vphi_{hh}(\bmv{k};z)\Phi_{hn}^{\rm int}(\bmv{k},\bmv{p}';z),$}
	\label{e6.80}\\
\text{\small$\ds z\Phi_{hh}^{\rm int}(\bmv{k};z)$}
	&\text{\small$\ds=$}&\text{\small$\ds\int\d\bmv{p}''\;
	\Sigma_{hn}(\bmv{k},\bmv{p}'';z)
	\Phi_{nh}^{\rm int}(\bmv{k},\bmv{p}'';z)-
	\vphi_{hh}(\bmv{k};z)\Phi_{hh}^{\rm int}(\bmv{k};z)-
	\Phi_{hh}(\bmv{k}),$}\nonumber\\\label{e6.81}
\eea
where the condition 
$\Phi_{hn}(\bmv{k},\bmv{p}')=\Phi_{nh}(\bmv{k},\bmv{p})=0$ is taken into 
account.

Therefore, we obtained a system of equations for normalized
correlation functions \refp{e6.65}--\refp{e6.68} for the description of 
nonequilibrium states of the system when kinetic and hydrodynamic processes 
are considered simultaneously. A similar system of equations for time
correlation functions in the method of projection operators was 
obtained in the paper by John and Forster \cite{113}. In this paper the 
parameters of an abbreviated description were orthogonal dynamical 
variables $\hat n_{\bmv{k}}(\bmv{p})$ and $\hat{\cal E}_{\bmv{k}}$. 
Solutions of such an equation system in the hydrodynamic limit were 
found using the projection procedure for the first five moments of the 
nonequilibrium one-particle distribution function. In view of this, the 
dynamical structure factor $S(\bmv{k};\omega)$ for  intermediate values 
of $\bmv{k}$ and $\omega$ was investigated using a parametric approximation 
of memory functions for liquid Argon. However, in this paper the question 
concerning collective modes and generalized transport coefficients, when 
kinetics and hydrodynamics are connected between themselves, is not 
discussed.

In order to solve the system of equations \refp{e6.78}--\refp{e6.81} we 
also apply the projection procedure \cite{104}. Let us introduce the 
dimensionless momentum $\bmv{\xi}=\frac{\bmv{k}}{mv_0}$, 
$v^{2}_0=(m\beta)^{-1}$. Then the system of equations \refp{e6.77} can be 
rewritten in the matrix form:
%
%
\be
z\tilde{\Phi}(\bmv{k};\bmv{\xi},\bmv{\xi}';z)-
	\tilde{\Sigma}(\bmv{k};\bmv{\xi},\bmv{\xi}'';z)
	\tilde{\Phi}(\bmv{k};\bmv{\xi}'',\bmv{\xi}';z)=-
	\tilde{\Phi}(\bmv{k};\bmv{\xi},\bmv{\xi}'),\label{e6.82}
\ee
where it is clear that the integration must be performed with respect to 
the repeating indices $\bmv{\xi}''$. Further, let us introduce the scalar 
product of two functions, $\phi(\bmv{\xi})$ and $\psi(\bmv{\xi})$, as
%
%
\be
\la\phi|\psi\ra=\int\d\bmv{\xi}\;\phi^*(\bmv{\xi})f_0(\xi)\psi(\bmv{\xi}).
	\label{e6.83}
\ee
Then, the matrix element for some ``operator" $M$ can be determined as
%
%
\be
\la\phi|M|\psi\ra=\int\d\bmv{\xi}\;\d\bmv{\xi}'\;\phi^*(\bmv{\xi})
	M(\bmv{\xi},\bmv{\xi}')f_0(\xi')\psi(\bmv{\xi}').\label{e6.84}
\ee
Let $\phi(\bmv{\xi})=\{\phi_{\mu}(\bmv{\xi})\}$ be the orthogonalized basis 
of functions with the weight $f_0(\xi)$, so that the following condition is 
satisfied:
%
%
\be
\la\phi_{\nu}|\phi_{\mu}\ra=\delta_{\nu\mu},\qquad
\sum_{\nu}|\phi_{\nu}\ra\la\phi_{\nu}|=1,\label{e6.85}
\ee
where
%
%
\be
\phi_{\mu}(\bmv{\xi})=\phi_{lmn}(\bmv{\xi})=(l!m!n!)^{-1/2}
	\bar{H}_{l}(\xi_{x})\bar{H}_{m}(\xi_{y})\bar{H}_{n}(\xi_{z}),
	\label{e6.86}
\ee
$\bar{H}_{l}(\xi)=2^{-l/2}H_{l}(\xi/2)$, ${H}_{l}(\xi)$ is a Hermite 
polynomial. Then, each function in the system of equations \refp{e6.82}, 
which depends on momentum variables $\bmv{\xi}$, $\bmv{\xi}'$, can be 
expanded over functions $\phi_{\mu}(\bmv{\xi})$ in the series:
%
%
\bea
\tilde{\Phi}(\bmv{k};\bmv{\xi},\bmv{\xi}';z)&=&
	\sum_{\nu,\mu}\phi_{\nu}^*(\bmv{\xi})
	\tilde{\Phi}_{\nu\mu}(\bmv{k};z)
	\phi_{\mu}(\bmv{\xi}')f_0(\xi'),\label{e6.87}\\
\tilde{\Sigma}(\bmv{k};\bmv{\xi},\bmv{\xi}';z)&=&
	\sum_{\nu,\mu}\phi_{\nu}^*(\bmv{\xi})
	\tilde{\Sigma}_{\nu\mu}(\bmv{k};z)
	\phi_{\mu}(\bmv{\xi}')f_0(\xi'),\label{e6.88}
\eea
where
%
%
\bea
\!\!\!\!\!\!\!\!\!\!\!\!
\tilde{\Phi}_{\nu\mu}(\bmv{k};z)=
	\la\phi_\nu|\tilde{\Phi}(\bmv{k};\bmv{\xi},\bmv{\xi}';z)|\phi_\mu\ra
	&=&\!\!\!\int\d\bmv{\xi}\;\d\bmv{\xi}'\;
	\phi_\nu^*(\bmv{\xi})f_0(\xi)
	\tilde{\Phi}(\bmv{k};\bmv{\xi},\bmv{\xi}';z)
	\phi_\mu(\bmv{\xi}'),\label{e6.89}\\
\!\!\!\!\!\!\!\!\!\!\!\!
\tilde{\Sigma}_{\nu\mu}(\bmv{k};z)=
	\la\phi_\nu|\tilde{\Sigma}(\bmv{k};\bmv{\xi},\bmv{\xi}';z)|\phi_\mu\ra
	&=&\!\!\!\int\d\bmv{\xi}\;\d\bmv{\xi}'\;
	\phi_\nu^*(\bmv{\xi})f_0(\xi)
	\tilde{\Sigma}(\bmv{k};\bmv{\xi},\bmv{\xi}';z)
	\phi_\mu(\bmv{\xi}').\label{e6.90}
\eea
Let us substitute expansions \refp{e6.87}--\refp{e6.90} into equation 
\refp{e6.82}. As a result, one obtains:
%
%
\be
z\tilde{\Phi}_{\nu\mu}(\bmv{k};z)-\sum_{\gamma}
	\tilde{\Sigma}_{\nu \gamma}(\bmv{k};z)
	\tilde{\Phi}_{\gamma\mu}(\bmv{k};z)=-
	\tilde{\Phi}_{\nu\mu}(\bmv{k}).\label{e6.91}
\ee
In actual calculations, a finite number of functions from the set 
$\phi_\nu(\bmv{\xi})$ is used. Taking into account this fact, let us 
introduce the projection operator $P$ which projects arbitrary functions 
$\psi(\bmv{\xi})$ onto a finite set of functions $\phi_{\mu}(\bmv{\xi})$:
%
%
\be
P=\sum_{\nu=1}^{n}|\phi_{\nu}\ra\la\phi_{\nu}|=1-Q,\qquad
	P\la\psi|=\sum_{\nu=1}^n\la\psi|\phi_\nu\ra\la\phi_\nu|.\label{e6.92}
\ee
Here $n$ denotes a finite number of functions. Then, from \refp{e6.91} we 
obtain a system of equations for a finite set of functions 
$\phi_\mu(\bmv{\xi})$,
%
%
\be
\sum_{\gamma=1}^n\ls z\bar{\delta}_{\nu\gamma}-
	\im\tilde{\Omega}_{\nu\gamma}(\bmv{k})+
	\tilde{D}_{\nu\gamma}(\bmv{k};z)\rs
	\tilde{\Phi}_{\gamma\mu}(\bmv{k};z)=-
	\tilde{\Phi}_{\nu\mu}(\bmv{k}),\label{e6.93}
\ee
where
%
%
\be
\tilde{D}_{\nu\mu}(\bmv{k};z)=\la\phi_\nu|\tilde{\vphi}(\bmv{k};z)+
	\tilde{\Sigma}(\bmv{k};z)Q\ls z-Q\tilde{\Sigma}(\bmv{k};z)Q\rs^{-1}
	Q\tilde{\Sigma}(\bmv{k};z)|\phi_\mu\ra\label{e6.94}
\ee
are generalized hydrodynamic transport kernels and
%
%
\be
\im\tilde{\Omega}_{\nu\mu}(\bmv{k})=
	\la\phi_\nu|\im\tilde{\Omega}(\bmv{k})|\phi_\mu\ra\label{e6.95}
\ee
is a frequency matrix. Note that matrices $\im\tilde{\Omega}(\bmv{k})$ 
and $\tilde{\vphi}(\bmv{k};z)$ are defined according to \refp{e6.48}, 
\refp{e6.49} and \refp{e6.50}--\refp{e6.53}.

Let us find solutions to the system of equations \refp{e6.93} in the
hydrodynamic region when a set of functions $\phi_{\mu}(\bmv{\xi})$ 
present five moments of a one-particle distribution function:
%
%
\be
\phi_1(\bmv{\xi})=1,\quad\phi_2(\bmv{\xi})=\xi_z,\quad
\phi_3(\bmv{\xi})=\frac{1}{\sqrt{6}}(\xi^2-3),\quad\phi_4(\bmv{\xi})=\xi_x,
\quad\phi_5(\bmv{\xi})=\xi_y.\label{e6.96}
\ee
Then, the following relations are fulfilled:
%
%
\bea
\la 1|\hat n_{\bmv{k}}(\bmv{\xi})&=&\int\d\bmv{\xi}\;
	\hat n_{\bmv{k}}(\bmv{\xi})=\hat n_{\bmv{k}},\nonumber\\
\la\xi_\gamma|\hat n_{\bmv{k}}(\bmv{\xi})&=&\int\d\bmv{\xi}\;
	\hat n_{\bmv{k}}(\bmv{\xi})\;\xi_\gamma=
	\hat p_{\bmv{k}}^\gamma,\label{e6.97}\\
\la 6^{-1/2}(\xi^2-3)|\hat n_{\bmv{k}}(\bmv{\xi})&=&\int\d\bmv{\xi}\;
	\hat n_{\bmv{k}}(\bmv{\xi})\;6^{-1/2}(\xi^2-3)=
	\hat{\cal E}_{\bmv{k}}^{\rm kin}-3\hat n_{\bmv{k}}\beta^{-1}=
	\hat h^{\rm kin}_{\bmv{k}},\nonumber
\eea
for the Fourier components of densities for the number of particles, 
momentum and the kinetic part of generalized enthalpy. Besides that, the 
microscopic conservation laws for densities of the number of particles and 
momentum can be written in the form:
%
%
\be
\begin{array}{rcl}
\la 1|\dot{\hat n}_{\bmv{k}}(\bmv{\xi})&=&
	-\im k_\gamma\hat p_{\bmv{k}}^\gamma m^{-1},\\ [1ex]
\la\xi_\alpha|\dot{\hat n}_{\bmv{k}}(\bmv{\xi})&=&
	-\im k_\gamma\widehat{\tensor{T}}_{\bmv{k}}{\!\!}^{\gamma\alpha},\\
\end{array}\label{e6.98}
\ee
where $\widehat{\tensor{T}}_{\bmv{k}}{\!\!}^{\gamma\alpha}$ is a 
Fourier component of the stress-tensor.

If we choose the direction of wavevector $\bmv{k}$ along $oz$-axis, then
$\phi_\nu(\bmv{\xi})$, $\nu=1,2,3$ will correspond to longitudinal modes, 
whereas $\phi_\nu(\bmv{\xi})$ at $\nu=4,5$  will be related to transverse 
modes.

From the system of equations \refp{e6.93}, at $\nu=4,5$, 
$\phi_4(\bmv{\xi})=\xi_x$, $\phi_5(\bmv{\xi})=\xi_y$, one obtains an 
equation for the Fourier component of the time correlation function 
connected with the transverse component of the momentum density 
$\Phi_{44}(\bmv{k};z)$. From this equation one finds:
%
%
\be
\Phi_{44}(\bmv{k};z)=\Phi_{pp}^\perp(\bmv{k};z)=-
	\frac{1}{z+D_{pp}^\perp(\bmv{k};z)},\label{e6.99}
\ee
where
%
%
\bea
\Phi_{pp}^\perp(\bmv{k};z)&=&
	\la\xi_x|\Phi_{nn}(\bmv{k},\bmv{\xi},\bmv{\xi}';z)|\xi_x'\ra,
	\label{e6.100}\\
D_{pp}^\perp(\bmv{k};z)&=&D_{pp}^{\perp(\rm kin)}(\bmv{k};z)+
	D_{pp}^{\perp(\rm int)}(\bmv{k};z),\label{e6.101}\\
D_{pp}^{\perp(\rm kin)}(\bmv{k};z)&=&
	\la\xi_x|\vphi_{nn}(\bmv{k},\bmv{\xi},\bmv{\xi}';z)|\xi_x'\ra,
	\label{e6.102}\\
\!\!\!\!\!\!\!\!\!\!\!\!
D_{pp}^{\perp(\rm int)}(\bmv{k};z)&=&
	\la\xi_x|\ls\tilde{\Sigma}(\bmv{k},\bmv{\xi},\bmv{\xi}';z)
	Q\ls z-Q\tilde{\Sigma}(\bmv{k},\bmv{\xi},\bmv{\xi}';z)Q\rs^{-1}
	Q\tilde{\Sigma}(\bmv{k},\bmv{\xi},\bmv{\xi}';z)\rs_{nn}|\xi_x'\ra,
	\nonumber\\\label{e6.103}\\
D_{pp}^\perp(\bmv{k};z)&=&\im k^2\eta(\bmv{k};z)(mn)^{-1},\label{e6.104}
\eea
where $\eta(\bmv{k};z)$ denotes a generalized coefficient of shear 
viscosity. Such a coefficient consists of two main contributions. The 
first one is $D_{pp}^{\perp(\rm kin)}(\bmv{k};z)$, 
whereas the second contribution $D_{pp}^{\perp(\rm int)}(\bmv{k};z)$
describes a relation of kinetic and hydrodynamic processes. If the last 
term is neglected, which formally corresponds to $\hat h^{\rm 
int}_{\bmv{k}}=0$, then one obtains an expression for the generalized 
coefficient of shear viscosity $\eta(\bmv{k};z)$, obtained earlier by the 
authors of \cite{109,110} when solving the equation
%
%
\bea
z\Phi_{nn}(\bmv{k},\bmv{p},\bmv{p}';z)-
	\int\d\bmv{p}''\;\im\Omega_{nn}(\bmv{k},\bmv{p},\bmv{p}'')
	\Phi_{nn}(\bmv{k},\bmv{p}'',\bmv{p}';z)&+&\label{e6.105}\\
\int\d\bmv{p}''\;\vphi_{nn}'(\bmv{k},\bmv{p},\bmv{p}'';z)
	\Phi_{nn}(\bmv{k},\bmv{p}'',\bmv{p}';z)&=&-
	\Phi_{nn}(\bmv{k},\bmv{p},\bmv{p}'),\nonumber
\eea
by using the method of projections \cite{104} in the hydrodynamic region. 
Equation \refp{e6.105} corresponds to kinetic equation \refp{e6.61} for the 
Fourier component of the nonequilibrium one-particle distribution function. 
In $\vphi_{nn}'(\bmv{k},\bmv{p},\bmv{p}';z)$ the projection is performed on 
a space of dynamics of the ``slow" variable $\hat n_{\bmv{k}}(\bmv{p})$ 
connected with the microscopic conservation laws for the particle-number 
density and momentum \refp{e6.98}. In our approach in 
$D_{pp}^\perp(\bmv{k};z)$, the projection is carried out on an extended 
space of dynamics of the ``slow" variables $\hat n_{\bmv{k}}(\bmv{p})$ and 
$\hat h^{\rm int}_{\bmv{k}}$. Moreover, $\hat h^{\rm int}_{\bmv{k}}$ is a 
hydrodynamic variable, whereas $\hat n_{\bmv{k}}(\bmv{p})$ is a kinetic one 
and only its average value with five moments \refp{e6.96} can be related to 
hydrodynamic quantities.

If we put $\nu=1,2,3$, $\phi_1(\bmv{\xi})=1$, $\phi_2(\bmv{\xi})=\xi_z$, 
$\phi_3(\bmv{\xi})-6^{-1/2}(\xi^2-3)$ in the system of equation 
\refp{e6.93}, then we obtain:
%
%
\bea
z\Phi_{na}(\bmv{k};z)-\im\Omega_{np}(\bmv{k})\Phi_{pa}(\bmv{k};z)&=&-
	\Phi_{na}(\bmv{k}),\label{e6.106}\\ [1ex]
z\Phi_{pa}(\bmv{k};z)-\im\Omega_{pn}(\bmv{k})\Phi_{na}(\bmv{k};z)+
	D_{pp}^{||}(\bmv{k};z)\Phi_{pa}(\bmv{k};z)&-&\label{e6.107}\\
	\Sigma_{ph^{\rm kin}}(\bmv{k};z)\Phi_{h^{\rm kin}a}(\bmv{k};z)-
	\Sigma_{ph^{\rm int}}(\bmv{k};z)\Phi_{h^{\rm int}a}(\bmv{k};z)
	&=&-\Phi_{pa}(\bmv{k}),\nonumber\\ [1ex]
z\Phi_{h^{\rm kin}a}(\bmv{k};z)-
	\Sigma_{h^{\rm kin}p}(\bmv{k};z)\Phi_{pa}(\bmv{k};z)
	&+&\label{e6.108}\\
D_{h^{\rm kin}h^{\rm kin}}(\bmv{k};z)\Phi_{h^{\rm kin}a}(\bmv{k};z)+
D_{h^{\rm kin}h^{\rm int}}(\bmv{k};z)\Phi_{h^{\rm int}a}(\bmv{k};z)
	&=&-\Phi_{h^{\rm kin}a}(\bmv{k}),\nonumber\\ [1ex]
z\Phi_{h^{\rm int}a}(\bmv{k};z)-
	\Sigma_{h^{\rm int}p}(\bmv{k};z)\Phi_{pa}(\bmv{k};z)
	&+&\label{e6.109}\\
D_{h^{\rm int}h^{\rm kin}}(\bmv{k};z)\Phi_{h^{\rm kin}a}(\bmv{k};z)+
D_{h^{\rm int}h^{\rm int}}(\bmv{k};z)\Phi_{h^{\rm int}a}(\bmv{k};z)
	&=&-\Phi_{h^{\rm int}a}(\bmv{k}),\nonumber
\eea
where $a=\{\hat n_{\bmv{k}},\hat{\bmv{p}}_{\bmv{k}},
\hat h^{\rm kin}_{\bmv{k}},\hat h^{\rm int}_{\bmv{k}}\}$ and
%
%
\be
\begin{array}{lclcl}
\Sigma_{ph^{\rm kin}}(\bmv{k};z)&=&\im\Omega_{ph^{\rm kin}}(\bmv{k})&-&
	D_{ph^{\rm kin}}(\bmv{k};z),\\
\Sigma_{ph^{\rm int}}(\bmv{k};z)&=&\im\Omega_{ph^{\rm int}}(\bmv{k})&-&
	D_{ph^{\rm int}}(\bmv{k};z),\\
\Sigma_{h^{\rm kin}p}(\bmv{k};z)&=&\im\Omega_{h^{\rm kin}p}(\bmv{k})&-&
	D_{h^{\rm kin}p}(\bmv{k};z),\\
\Sigma_{h^{\rm int}p}(\bmv{k};z)&=&\im\Omega_{h^{\rm int}p}(\bmv{k})&-&
	D_{h^{\rm int}p}(\bmv{k};z),\\
\end{array}\label{e6.110}
\ee
$\im\Omega_{ab}(\bmv{k})$ and $D_{ab}(\bmv{k};z)$ are determined according 
to \refp{e6.95}, \refp{e6.94}. From the system of equations 
\refp{e6.106}--\refp{e6.109} one can define the Fourier components of the 
particle number density correlation functions
%
%
\be
\Phi_{nn}(\bmv{k};z)=\Phi_{11}(\bmv{k};z)=
	\la 1|\Phi_{nn}(\bmv{k},\bmv{\xi},\bmv{\xi}';z)|1'\ra, ,\label{e6.111}
\ee
as well as of the longitudinal component of the momentum density
%
%
\be
\Phi_{pp}^{||}(\bmv{k};z)=\Phi_{22}(\bmv{k};z)=
	\la\xi_z|\Phi_{nn}(\bmv{k},\bmv{\xi},\bmv{\xi}';z)|\xi_z'\ra,
	\label{e6.112}
\ee
for the kinetic part of generalized enthalpy
%
%
\be
\Phi_{h^{\rm kin}h^{\rm kin}}(\bmv{k};z)=\Phi_{33}(\bmv{k};z)=
	\la 6^{-1/2}(\xi^2-3)|\Phi_{nn}(\bmv{k},\bmv{\xi},\bmv{\xi}';z)
	|6^{-1/2}((\xi')^2-3)\ra\label{e6.113}
\ee
as well as for the potential part of generalized enthalpy
$\Phi_{h^{\rm int}h^{\rm int}}(\bmv{k};z)$ and cross correlation 
functions, especially $\Phi_{h^{\rm int}h^{\rm kin}}(\bmv{k};z)$, 
$\Phi_{nh^{\rm kin}}(\bmv{k};z)$,
$\Phi_{nh^{\rm int}}(\bmv{k};z)$, $\Phi_{ph^{\rm kin}}(\bmv{k};z)$,
$\Phi_{ph^{\rm int}}(\bmv{k};z)$. It is important to point out that the 
system of equations \refp{e6.106}--\refp{e6.109} corresponds to the system 
of equations for Fourier components of the average values of densities for 
the number of particles $\la\hat n_{\bmv{k}}\ra_z$, longitudinal momentum 
$\la\hat{\bmv{p}}_{\bmv{k}}\ra_z$, kinetic 
$\la\hat h_{\bmv{k}}^{\rm kin}\ra_z$ and potential 
$\la\hat h_{\bmv{k}}^{\rm int}\ra_z$ parts of generalized enthalpy:
%
%
\bea
z\la\hat n_{\bmv{k}}\ra_z-
	\im\Omega_{np}(\bmv{k})\la\hat{\bmv{p}}_{\bmv{k}}\ra_z
	&=&-\la\hat n_{\bmv{k}}(t=0)\ra,\label{e6.114}\\ [1ex]
z\la\hat{\bmv{p}}_{\bmv{k}}\ra_z-
\im\Omega_{pn}(\bmv{k})\la\hat n_{\bmv{k}}\ra_z
	&+&\label{e6.115}\\
\!\!\!\!\!\!\!\!\!\!\!\!
D_{pp}^{||}(\bmv{k};z)\la\hat{\bmv{p}}_{\bmv{k}}\ra_z-
\Sigma_{ph^{\rm kin}}(\bmv{k};z)\la\hat h_{\bmv{k}}^{\rm kin}\ra_z-
\Sigma_{ph^{\rm int}}(\bmv{k};z)\la\hat h_{\bmv{k}}^{\rm int}\ra_z
	&=&-\la\hat{\bmv{p}}_{\bmv{k}}(t=0)\ra,\nonumber\\ [1ex]
z\la\hat h_{\bmv{k}}^{\rm kin}\ra_z-
	\Sigma_{h^{\rm kin}p}(\bmv{k};z)\la\hat{\bmv{p}}_{\bmv{k}}\ra_z
	&+&\label{e6.116}\\
D_{h^{\rm kin}h^{\rm kin}}(\bmv{k};z)\la\hat h_{\bmv{k}}^{\rm kin}\ra_z+
D_{h^{\rm kin}h^{\rm int}}(\bmv{k};z)\la\hat h_{\bmv{k}}^{\rm int}\ra_{z}
	&=&-\la\hat h_{\bmv{k}}^{\rm kin}(t=0)\ra,\nonumber\\ [1ex]
z\la\hat h_{\bmv{k}}^{\rm int}\ra_z-
	\Sigma_{h^{\rm int}p}(\bmv{k};z)\la\hat{\bmv{p}}_{\bmv{k}}\ra_z
	&+&\label{e6.117}\\
D_{h^{\rm int}h^{\rm kin}}(\bmv{k};z)\la\hat h_{\bmv{k}}^{\rm kin}\ra_z+
D_{h^{\rm int}h^{\rm int}}(\bmv{k};z)\la\hat h_{\bmv{k}}^{\rm int}\ra_z
	&=&-\la \hat h_{\bmv{k}}^{\rm int}(t=0)\ra.\nonumber
\eea
This system of equations is similar in construction to the equations of 
molecular hydrodynamics \cite{75}. The difference consists in the fact that 
instead of the equations for the Fourier component of the mean enthalpy 
density $\la \hat h_{\bmv{k}}\ra^{t}$ which is introduced in molecular 
hydrodynamics, there are two connected equations for the Fourier components 
of mean values of the kinetic and potential parts of enthalpy density. 
Moreover, instead of the generalized thermal conductivity which appears in 
molecular hydrodynamics, the dissipation of energy flows is described 
in equations \refp{e6.114}--\refp{e6.117} by a set of generalized 
transport kernels $D_{h^{\rm kin}h^{\rm kin}}(\bmv{k};z)$, $D_{h^{\rm 
kin}h^{\rm int}}(\bmv{k};z)$, $D_{h^{\rm int}h^{\rm kin}}(\bmv{k};z)$, 
$D_{h^{\rm int}h^{\rm int}}(\bmv{k};z)$. Obviously, transport kernels 
give more detailed information on the dissipation of energy flows in 
the system because they describe the time evolution of dynamical 
correlations between the kinetic and potential flows of enthalpy density.

Solving the system of equation \refp{e6.106}--\refp{e6.109} at $a=n$, one 
obtains an expression for the correlation function ``density-density" 
$\Phi_{nn}(\bmv{k};z)$
%
%
\be
\Phi_{nn}(\bmv{k};z)=-S_2(\bmv{k})
	\ls z-\frac{\im\Omega_{np}(\bmv{k})\im\Omega_{pn}(\bmv{k})}
	{z+\bar{D}_{pp}^{||}(\bmv{k};z)}\rs^{-1},\label{e6.118}
\ee
where
%
%
\bea
\bar{D}_{pp}^{||}(\bmv{k};z)=D_{pp}^{||}(\bmv{k};z)-
	\bar{\Sigma}_{ph^{\rm kin}}(\bmv{k};z)
	\ls z+\bar{D}_{h^{\rm kin}h^{\rm kin}}(\bmv{k};z)\rs^{-1}
	\bar{\Sigma}_{h^{\rm kin}p}(\bmv{k};z)\,&-&\nonumber\\
\Sigma_{ph^{\rm int}}(\bmv{k};z)
	\,\ls z\,+\,D_{h^{\rm int}h^{\rm int}}(\bmv{k};z)\rs^{-1}
	\Sigma_{h^{\rm int}p}(\bmv{k};z),\label{e6.119}
\eea
%
%
%
%
%
\bea
\bar{\Sigma}_{ph^{\rm kin}}(\bmv{k};z)&=&
	\Sigma_{ph^{\rm kin}}(\bmv{k};z)-{}\label{e6.120}\\
&&\Sigma_{ph^{\rm int}}(\bmv{k};z)
	\ls z+D_{h^{\rm int}h^{\rm int}}(\bmv{k};z)\rs^{-1}
	D_{h^{\rm int}h^{\rm kin}}(\bmv{k};z),\nonumber\\ [1ex]
\bar{\Sigma}_{h^{\rm kin}p}(\bmv{k};z)&=&
	\Sigma_{h^{\rm kin}p}(\bmv{k};z)-{}\label{e6.121}\\
&&D_{h^{\rm kin}h^{\rm int}}(\bmv{k};z)
	\ls z+D_{h^{\rm int}h^{\rm int}}(\bmv{k};z)\rs^{-1}
	\Sigma_{h^{\rm int}p}(\bmv{k};z),\nonumber\\ [1ex]
\bar{D}_{h^{\rm kin}h^{\rm kin}}(\bmv{k};z)&=&
	D_{h^{\rm kin}h^{\rm kin}}(\bmv{k};z)-{}\label{e6.122}\\
&&D_{h^{\rm kin}h^{\rm int}}(\bmv{k};z)
	\ls z+D_{h^{\rm int}h^{\rm int}}(\bmv{k};z)\rs^{-1}
	D_{h^{\rm int}h^{\rm kin}}(\bmv{k};z).\nonumber
\eea
In expressions \refp{e6.119}--\refp{e6.122} we can observe an interesting 
renormalization of the functions $\Sigma_{ab}$ and $D_{ab}$ via the 
generalized transport kernels for fluctuations of flows of the potential 
part of enthalpy density. It is important to point out that in the 
mode-coupling theory developed by G\"otze \cite{93,94}, an expression for 
$\Phi_{nn}(\bmv{k};z)$ has the same form as in \refp{e6.118}. However, 
$\bar{D}_{pp}^{||}(\bmv{k};z)$ is connected only with the generalized shear 
viscosity $\eta^{||}(\bmv{k};z)$, since the densities of the number of 
particles $\hat n_{\bmv{k}}$ and momentum $\hat{\bmv{p}}_{\bmv{k}}$ are 
included in the set of variables of an abbreviated description. In our case 
$\bar{D}_{pp}^{||}(\bmv{k};z)$ takes into account both thermal and viscous 
dynamical correlation processes. Excluding from \refp{e6.118} the imaginary 
part $\Phi_{nn}^{||}(\bmv{k};\omega)$ of the correlation function 
$\Phi_{nn}(\bmv{k};z)$, one obtains an expression for the dynamical 
structure factor $S(\bmv{k};\omega)$ in which  contributions of transport 
kernels corresponding to the kinetic and potential parts of the enthalpy 
density $\hat h_{\bmv{k}}$ are separated. It is evident that the main 
contribution of the transport kernel 
$D_{h^{\rm int}h^{\rm int}}(\bmv{k};z)$ to the $S(\bmv{k};\omega)$ for 
liquids was in the hydrodynamical region (the region of small values of 
wavevector $\bmv{k}$ and frequency $\omega$), whereas 
$D_{h^{\rm kin}h^{\rm kin}}(\bmv{k};z)$ will contribute to the kinetic 
region (orders of interatomic distances, small correlation times). In the 
region of intermediate values of wavevector $\bmv{k}$ and frequency 
$\omega$, it is necessary to take into account all the transport kernels 
$\Sigma_{ph^{\rm kin}}(\bmv{k};z)$, $\Sigma_{ph^{\rm int}}(\bmv{k};z)$, 
$D_{h^{\rm kin}h^{\rm kin}}(\bmv{k};z)$,
$D_{h^{\rm int}h^{\rm kin}}(\bmv{k};z)$, 
$D_{h^{\rm int}h^{\rm int}}(\bmv{k};z)$. Since it is impossible to perform 
exact calculations of the described above functions, it is necessary in
each separate region to accept models corresponding to the physical processes. 
Obviously, it is necessary to provide modelling on the level of generalized 
transport kernels $\vphi_{nn}(\bmv{k},\bmv{p},\bmv{p}';t,t')$, 
$\vphi_{nh}(\bmv{k},\bmv{p};t,t')$, $\vphi_{hn}(\bmv{k},\bmv{p}';t,t')$,
$\vphi_{hh}(\bmv{k};t,t')$ \refp{e6.50}--\refp{e6.53}. In particular, we 
can accept for $\vphi_{nn}(\bmv{k},\bmv{p},\bmv{p}';t,t')$ the 
Boltzmann-Enskog model for hard spheres or the Fokker-Planck model for the 
case of weak correlation, as in the papers by Mazenko \cite{107,109,130}. 
Such approximations can be applicable to kinetic processes at interatomic 
distances (region of large values of wavevector $\bmv{k}$) and small 
correlation times (region of large frequencies $\omega$) of interatomic 
collisions. Besides, it is necessary to take into account information about 
interactions of particles at short distances. In the transport kernels 
$\vphi_{nn}(\bmv{k};\bmv{\xi},\bmv{\xi}';t,t')$, 
$\vphi_{nh}(\bmv{k};\bmv{\xi};t,t')$, 
$\vphi_{hn}(\bmv{k};\bmv{\xi}';t,t')$$\vphi_{hh}(\bmv{k};t,t')$ it is 
important to distinguish processes connected with a type of particles 
interaction at short and long distances, analogously to the case of 
Enskog-Landau kinetic equation for a model of charged hard spheres. The 
problems of modelling for these kernels are complicated by the absence of a 
small parameter in terms of which the perturbation theory might be 
developed.

The transport kernel $\vphi'_{nn}(\bmv{k},\bmv{p},\bmv{p}';t,t')$ 
\refp{e6.62} and its moments on space $\![1,p_{l},\frac{p^{2}}{2m}]$ were
analyzed in the limit $z\to\infty$ and the hydrodynamic limit $k\to0$, 
$z\to0$ in the paper by Forster \cite{110}. The modelling problems of 
transport kernels for intermediate values of $\bmv{k}$ and $\omega$ are 
reflected in the details of the description of spectra for collective 
excitations, as well as in the dynamical structure factor. In the limit 
$k\to0$, $\omega\to0$, the cross correlations between the kinetic and 
potential flows of energy and shear flows become not so important and the 
system of equations \refp{e6.118}--\refp{e6.122} gives a spectrum of the 
collective modes of molecular hydrodynamics \cite{75,77}. For intermediate 
values of $\bmv{k}$ and $\omega$, the spectrum of collective modes can be 
found from the condition
%
%
\begin{equation}
\left[
\begin{array}{lrrr}
z&\im\Omega_{np}(\bmv{k})&0&0\\
\im\Omega_{pn}(\bmv{k})&z+D_{pp}^{||}(\bmv{k};z)&
	\Sigma_{ph^{\rm kin}}(\bmv{k};z)&\Sigma_{ph^{\rm int}}(\bmv{k};z)\\
0&\Sigma_{h^{\rm kin}p}(\bmv{k};z)&z+D_{h^{\rm kin}h^{\rm kin}}(\bmv{k};z)&
	D_{h^{\rm kin}h^{\rm int}}(\bmv{k};z)\\
0&\Sigma_{h^{\rm int}p}(\bmv{k};z)&D_{h^{\rm int}h^{\rm kin}}(\bmv{k};z)&
	z+D_{h^{\rm int}h^{\rm int}}(\bmv{k};z)\\
\end{array}
\right]=0,\label{e6.123}
\end{equation}
in which contributions of the kinetic and potential parts of generalized
enthalpy are separated. This will be reflected in the structure of a heat
mode at concrete model calculations of the wavevector- and
frequency-dependent transport kernels 
$D_{h^{\rm kin}h^{\rm kin}}(\bmv{k};z)$,
$D_{h^{\rm kin}h^{\rm int}}(\bmv{k};z)$, 
$D_{h^{\rm int}h^{\rm kin}}(\bmv{k};z)$,
$D_{h^{\rm int}h^{\rm int}}(\bmv{k};z)$ depending on $\bmv{k}$ and $\omega$.

The system of equations \refp{e6.93} for time correlation functions allows 
an extended description of collective modes in liquids, which includes
both hydrodynamic and kinetic processes. Including on the basis of 
functions $\phi_\nu(\xi)$ \refp{e6.96} some additional functions,
%
%
\begin{equation}
\psi_Q^l(\bmv{\xi})=\frac{1}{5}(\xi^2-5)\xi_l,\qquad
	\psi_{\Pi}^{lj}(\bmv{\xi})=\frac{\sqrt{2}}{2}(\xi_l\xi_j-
	\frac13\xi^2\delta_{lj}),\label{e6.124}
\ee
which corresponds to a 13-moments approximation of Grad, one obtains a system 
of equations for time correlation functions of an extended set of 
hydrodynamic variables $\{\hat n_{\bmv{k}}$, $\hat{\bmv{p}}_{\bmv{k}}$, 
$\hat h^{\rm kin}_{\bmv{k}}$, $\widehat{\tensor{\Pi}}_{\bmv{k}}$, 
$\hat{\bmv{Q}}_{\bmv{k}}$, $\hat h^{\rm int}_{\bmv{k}}\}$:
%
%
\be
z\tilde\Phi^{\rm H}(\bmv{k};z)-
	\im\tilde\Omega^{\rm H}(\bmv{k})\tilde\Phi^{\rm H}(\bmv{k};z)+
	\tilde\vphi^{\rm H}(\bmv{k};z)\tilde\Phi^{\rm H}(\bmv{k};z)=-
	\tilde\Phi^{\rm H}(\bmv{k}),\label{e6.125}
\ee
where
%
%
\be
\tilde\Phi^{\rm H}(\bmv{k};z)=
\left[
\begin{array}{llllll}
\Phi_{nn}&\Phi_{np}&\Phi_{nh^{\rm kin}}&\Phi_{n\Pi}&\Phi_{nQ}&
	\Phi_{nh^{\rm int}}\\
\Phi_{pn}&\Phi_{pp}&\Phi_{ph^{\rm kin}}&\Phi_{p\Pi}&\Phi_{pQ}&
	\Phi_{ph^{\rm int}}\\
\Phi_{h^{\rm kin}n}&\Phi_{h^{\rm kin}p}&\Phi_{h^{\rm kin}h^{\rm kin}}&
	\Phi_{h^{\rm kin}\Pi}&\Phi_{h^{\rm kin}Q}&
	\Phi_{h^{\rm kin}h^{\rm int}}\\
\Phi_{\Pi n}&\Phi_{\Pi p}&\Phi_{\Pi h^{\rm kin}}&\Phi_{\Pi\Pi}&\Phi_{\Pi Q}&
	\Phi_{\Pi h^{\rm int}}\\
\Phi_{Qn}&\Phi_{Qp}&\Phi_{Qh^{\rm kin}}&\Phi_{Q\Pi}&\Phi_{QQ}&
	\Phi_{Qh^{\rm int}}\\
\Phi_{h^{\rm int}n}&\Phi_{h^{\rm int}p}&\Phi_{h^{\rm int}h^{\rm kin}}&
	\Phi_{h^{\rm int}\Pi}&\Phi_{h^{\rm int}Q}&
	\Phi_{h^{\rm int}h^{\rm int}}\\ 
\end{array}
\right]_{(\bmv{k};z)}\label{e6.126}
\ee     
is a matrix of Laplace images of the time correlation functions, 
$\widehat{\tensor{\Pi}}_{\bmv{k}}\!\!=\!\!
\int\!\!\d\bmv{\xi}\varphi_{\Pi}(\bmv{\xi})\hat n_{\bmv{k}}(\bmv{\xi})$,
$\hat{\bmv{Q}}_{\bmv{k}}=\int\d\bmv{\xi}\;\varphi_Q(\bmv{\xi})
\hat n_{\bmv{k}}(\bmv{\xi})$,
%
%
%
%
%
\be
\im\tilde\Omega^{\rm H}(\bmv{k})=
\left[
\begin{array}{llllll}
0&\im\Omega_{np}&0&0&0&0\\
\im\Omega_{pn}&0&\im\Omega_{ph^{\rm kin}}&\im\Omega_{p\Pi}&0&
	\im\Omega_{ph^{\rm int}}\\
0&\im\Omega_{h^{\rm kin}p}&0&0&\im\Omega_{h^{\rm kin}Q}&0\\
0&\im\Omega_{\Pi p}&0&0&\im\Omega_{\Pi Q}&0\\
0&0&\im\Omega_{Qh^{\rm kin}}&\im\Omega_{Q\Pi}&0&\im\Omega_{Qh^{\rm int}}\\
0&\im\Omega_{h^{\rm int} p}&0&0&\im\Omega_{h^{\rm int} Q}&0\\
\end{array}
\right]_{(\bmv{k})}\label{e6.127}
\ee     
is an extended hydrodynamic frequency matrix,
%
%
\be
\tilde\varphi^{\rm H}(\bmv{k};z)=
\left[
\begin{array}{llllll}
0&0&0&0&0&0\\
0&D^H_{pp}&D^H_{ph^{\rm kin}}&D^H_{p\Pi}&D^H_{pQ}&D^H_{ph^{\rm int}}\\
0&D_{h^{\rm kin}p}&D_{h^{\rm kin}h^{\rm kin}}&D_{h^{\rm kin}\Pi}&
	D_{h^{\rm kin}Q}&D_{h^{\rm kin}h^{\rm int}}\\
0&D_{\Pi p}&D_{\Pi h^{\rm kin}}&D_{\Pi\Pi}&D_{\Pi Q}&D_{\Pi h^{\rm int}}\\
0&D_{Qp}&D_{Qh^{\rm kin}}&D_{Q\Pi}&D_{QQ}&D_{Qh^{\rm int}}\\
0&D_{h^{\rm int}p}&D_{h^{\rm int}h^{\rm kin}}&D_{h^{\rm int}\Pi}&
	D_{h^{\rm int}Q}&D_{h^{\rm int}h^{\rm int}}\\ 
\end{array}
\right]_{(\bmv{k};z)}\label{e6.128}
\ee     
is a matrix of generalized memory functions, elements of which are 
transport kernels (3.94) projected on the basis of functions 
$\phi_\nu(\bmv{\xi})$ \refp{e6.96}, \refp{e6.124}. For such a description, 
the spectrum of generalized collective modes of the system is determined
for intermediate $\bmv{k}$ and $\omega$ by the relation 
$\det\left|z\tilde I-\im\tilde\Omega^{\rm H}(\bmv{k})+
\tilde\varphi^{\rm H}(\bmv{k})\right|=0$ which takes into account kinetic 
and hydrodynamic processes. In the hydrodynamic limit $k\to0$, 
$\omega\to0$, when the contribution of cross dissipative correlations 
between the kinetic and hydrodynamic processes practically vanishes, the 
system of equations for the time correlation function \refp{e6.125}, after 
some transformations, can be reduced to a system of equations for time 
corelation functions of densities for the number of particles 
$\hat n_{\bmv{k}}$, momentum $\hat{\bmv{p}}_{\bmv{k}}$, total enthalpy 
$\hat h_{\bmv{k}}$, the generalized stress tensor 
$\widehat{\tensor{\pi}}_{\bmv{k}}=
(1-\SP_{\rm H})\im L_N\hat{\bmv{p}}_{\bmv{k}}$ and the generalized enthalpy 
flow $\hat{\bmv{q}}_{\bmv{k}}=(1-\SP_{\rm H})\im L_N\hat h_{\bmv{k}}$, 
where $\SP_{\rm H}$ is the Mori operator constructed on the dynamical 
variables $\{\hat n_{\bmv{k}}$, $\hat{\bmv{p}}_{\bmv{k}}$, 
$\hat h_{\bmv{k}}\}$. For such a system of equations, the spectrum of 
collective excitations is determined from \cite{134,136}:
%
%
\be
\left|
\begin{array}{lllrr}
z&\im\Omega_{np}&0&0&0\\
\im\Omega_{pn}&z&\im\Omega_{ph}&\im\Omega_{p\pi}&0\\
0&\im\Omega_{hp}&z&0&0\\
0&\im\Omega_{\pi p}&0&z+\vphi_{\pi\pi}&\im\Omega_{\pi Q}+\vphi_{\pi Q}\\
0&0&0&\im\Omega_{Q\pi}+\varphi_{Q\pi}&\varphi_{QQ}\\
\end{array}
\right|_{(\bmv{k};z)}=0.\label{e6.129}
\ee     
In the hydrodynamic limit this gives: the heat mode 
%
%
%
\be
z_{\rm H}(k)=D_{\rm T}k^2+{\cal O}(k^4),\label{e6.130}
\ee
two complex conjugated sound modes 
%
%
%
\be
z_{\pm}(k)=\pm\im\omega_{\rm S}(k)+z_{\rm S}(k),\label{e6.131}
\ee
where $\omega_{\rm S}(k)=ck+o(k^3)$ is a frequency of sound propagation, 
$z_{\rm S}(k)=\Gamma k^2+{\cal O}(k^4)$ is a frequency of sound damping 
with the damping coefficient $\Gamma$; two nonvanishing in the limit 
$k\to0$ kinetic modes
%
%
\be
\begin{array}{rcl}
z_{\pi}(k)&=&\ds\varphi_{\pi\pi}(0)+{\cal O}(k^2),\\
z_Q(k)&=&\ds\varphi_{QQ}(0)+{\cal O}(k^2).\\
\end{array}\label{e6.132}
\ee     
Here $D_{\rm T}$ denotes a thermal diffusion coefficient
\[
D_{\rm T}=\frac{v^2_{{\rm T}Q}}{\gamma\varphi_{QQ}(0)}=
	\frac{\lambda}{mnc_p},\quad
	v^2_{{\rm T}Q}=\frac{m\Phi_{QQ}-h^2}{nc_V},\quad
	\gamma=c_p/c_V
\]
$c_p$ and $c_V$ are, correspondingly, thermodynamic values of specific 
heats at the constant pressure and volume, $\lambda$ is a thermal 
conductivity coefficient, $h$ denotes a thermodynamic value of enthalpy, 
$c=\gamma/mnS(k=0)$ denotes an adiabatic sound velocity. In \refp{e6.131}
%
%
\be
\Gamma=\frac{1}{2}(\gamma-1)D_{\rm T}+\frac{1}{2}\eta^{||}\label{e6.133}
\ee
is a sound damping coefficient with
\[
\eta^{||}=\frac{v^2_{p\pi}}{\varphi_{\pi\pi}(0)}=
	\lp\frac{4}{3}\eta+\kappa\rp/nm,
	\qquad v^2_{p\pi}=\frac{mS(0)\Phi_{\pi\pi}(0)-\gamma}{mnS(0)},
\]
where $\eta$ and $\kappa$ are shear and bulk viscosities. This spectrum 
coincides with the results of papers \cite{134,136,137}. However, at fixed 
values of $\bmv{k}$ and $\omega$, the transport kernels $\varphi_{\pi\pi}$, 
$\varphi_{\pi Q}$, $\varphi_{Q\pi}$, $\varphi_{QQ}$ are expressed via the 
generalized transport kernels $D_{\nu\mu}(\bmv{k};z)$ of matrix 
\refp{e6.128}, i.e. via $\varphi_{nn}(\bmv{k},\bmv{p},\bmv{p'};t,t')$, 
$\varphi_{nh}(\bmv{k},\bmv{p};t,t')$, $\varphi_{hn}(\bmv{k},\bmv{p'};t,t')$, 
$\varphi_{hh}(\bmv{k};t,t')$ \refp{e6.50}--\refp{e6.53}, according to the 
definition $D_{\nu\mu}(\bmv{k};z)$ \refp{e6.94}. Here, it is important to 
point out that passing from the system of transport equations of a 
self-consistent description of kinetics and hydrodynamics to the equations
of generalized hydrodynamics, we can connect the generalized transport 
kernels \refp{e6.50}--\refp{e6.53} with the hydrodynamic transport kernels 
$D_{\nu\mu}(\bmv{k};z)$ in \refp{e6.123} or \refp{e6.128}. Therefore, we 
gain the aim of the present section, namely, to connect transport kernels 
of a self-consistent description of kinetics and hydrodynamics with 
hydrodynamic transport kernels for time correlation functions for densities 
of the number of particles $\hat n_{\bmv{k}}$, momentum 
$\hat{\bmv{p}}_{\bmv{k}}$, enthalpy $\hat h_{\bmv{k}}$, which can be 
calculated by using the MD method 
\cite{137,138,139,140,141,142,143,144,145,146,147,154}, as well as by 
experiments on light scattering for different real systems. Evidently, the 
most important problem is model calculations of transport kernels 
\refp{e6.50}--\refp{e6.53}. Such calculations must take into account a type 
of interaction of particles at short and long distances, as well as their 
structural distribution. From this point of view, of significant interest 
is the modelling of transport kernels \refp{e6.50}--\refp{e6.53} 
for a system of charged hard spheres, since for a system of hard spheres a 
transport kernel of the type $\varphi_{nn}(\bmv{k};\bmv{p},\bmv{p'};t,t')$ 
\refp{e6.50} was computed via the Enskog collision integral \cite{115,116}.

%% file: s_04.tex
We have presented one of the approaches for a self-consistent description of
kinetic and hydrodynamic processes in systems of interacting particles,
formulated by the NSO method. It is based on a modification of the
boundary condition to the Liouville equation, which takes into account
a nonequilibrity of the one-particle distribution function, as well as the
local conservation laws which constitute a basis for the hydrodynamic
description. Using such a description, generalized kinetic equations
for dense gases and liquids can be derived. The result obtained can
be extended to quantum systems of interacting particles \cite{15,16,17}.
In particular, it can be applied to nuclear matter and chemically reacting
systems, where kinetic processes play an important role together with
hydrodynamic ones. The method applied can be extended to the investigations
of nonlinear and hydrodynamic fluctuations in polar, ionic and magnetic
liquids and electrolyte solutions.

There is a significant interest in investigations of collective excitations,
time correlation functions and transport coefficients on the basis of
equations \refp{e6.93}, \refp{e6.94}, \refp{e6.123}, \refp{e6.128} at
the presence of transport kernels $\vphi(\bmv{k};z)$ which describe
an interference between kinetic and hydrodynamic processes. In particular,
the transport kernel $\vphi(\bmv{k},\bmv{p},\bmv{p}';t)$ can take into 
account processes of interaction of particles at short and long distances 
(at Coulomb and dipole interactions of particles). To do this, we can apply 
an approach of collective modes 
\cite{137,138,139,140,141,142,143,144,145,146,147}.

%% file: article.bbl
\begin{thebibliography}{999}
\itemsep-1pt
\bibitem{1} Zubarev~D.N., Morozov~V.G. Formulation of boundary conditions
	for the Bogolubov hierarchy with allowance for local conservation
	laws. // Teor. Mat. Fiz., 1984, vol. 60, No.~2, p. 270-279 (in Russian)%
	\footnote{Publications in that journal are ususlly translated into 
	English and published by the Plenum Publishing Corporation in the USA 
	annually. If we know where they appear in a translated version, we 
	shall give this information next to the data of the original.}.
\bibitem{2} Zubarev~D.N., Morozov~V.G. Omelyan~I.P., Tokarchuk~M.V. 
	Unification of kinetics and hydrodynamics in theory of transport 
	phenomena. In: Collection of scientific works of Institute for 
	Theoretical and Applied Mechanics of Siberian 
	Branch of USSR Academy of Sciences. Models of mechanics of 
	continuous media. Novosibirsk, 1989, p. 34-51 (in Russian).
\bibitem{3} Zubarev~D.N., Morozov~V.G., Omelyan~I.P., Tokarchuk~M.V.
	On kinetic equations for dense gases and liquids.
	// Teor. Mat. Fiz., 1991, vol. 87, No.~1, p. 113-129 (in Russian).
\bibitem{4} Zubarev~D.N., Morozov~V.G., Omelyan~I.P., Tokarchuk~M.V. 
	The Enskog-Landau kinetic equation for charged hard spheres. In:
	Problems of atomic science and technique. Series: Nuclear physics 
	investigations (theory and experiment). Kharkov, Kharkov Physico
	Technical Institute, 1992, vol. 3(24), p. 60-65 (in Russian).
\bibitem{5} Zubarev~D.N., Morozov~V.G., Omelyan~I.P., Tokarchuk~M.V.
	The unification of kinetic and hydrodynamic approaches in the
	theory of dense gases and liquids.
	// Teor. Mat. Fiz., 1993, vol. 96, No.~3, p. 325-350 (in Russian).
\bibitem{6} Morozov~V.G., Kobryn~A.E., Tokarchuk~M.V. Modified kinetic theory
	with consideration for slow hydrodynamical processes. // Cond. Matt. 
	Phys., 1994, vol. 4, p. 117-127.
\bibitem{7} Kobryn~A.E., Morozov~V.G., Omelyan~I.P., Tokarchuk~M.V.
	Enskog-Landau kinetic equation. Calculation of the transport 
	coefficients for charged hard spheres. //
	Physica A, 1996, vol. 230, No.~1\&2, p. 189-201.
\bibitem{8} Omelyan~I.P., Tokarchuk~M.V. Kinetic equation for liquids with a 
	multistep potential of interaction. $H$-theorem. // Physica A, 1996, 
	vol. 234, No.~1\&2, p. 89-107.
\bibitem{9} Zubarev~D., Morozov~V., R\"opke~G. Statistical mechanics of 
	nonequilibrium processes. Vol. 1, Basic concepts, kinetic theory. 
	Berlin, Academie Verlag, 1996. 
\bibitem{9a} Zubarev~D., Morozov~V., R\"opke~G. Statistical mechanics of 
	nonequilibrium processes. Vol. 2, Relaxation and hydrodynamical 
	processes. Berlin, Academie Verlag, 1997.
\bibitem{10} Tokarchuk~M.V., Omelyan~I.P., Kobryn~A.E. On the kinetic theory 
	of classical interacting particles by means of nonequilibrium 
	statistical operator method. // J. Phys. Studies, 1997, vol. 1, 
	No.~4, p. 490-512 (in Ukrainian).
\bibitem{11} Kobryn~A.E., Omelyan~I.P., Tokarchuk~M.V. The modified group 
	expansions for constructions of solutions to the BBGKY hierarchy. 
	// J. Stat. Phys., 1998, vol. 92, No.~5/6 p. 973-994.
\bibitem{12} Tokarchuk~M.V. On the statistical theory of a nonequilibrium 
	plasma in its electromagnetic self-field. // Teor. Mat. Fiz., 1993, 
	vol. 97, No.~1, p. 27-43 (in Russian). English transl. appears In: 
	Teoreticheskaya i Matematicheskaya Fizika, 1994, by Plenum Publishing 
	Corporation, p. 1126-1136.
\bibitem{13} Tokarchuk~M.V., Ignatyuk~V.V. Consistent kinetic and 
	hydrodynamic description of plasma in self-electromagnetic field. I. 
	Transport equations. // Ukr. Fiz. Zhurn., 1997, vol. 42, No.~2, 
	p. 153-160 (in Ukrainian).
\bibitem{14} Tokarchuk~M.V., Ignatyuk~V.V. Consistent kinetic and 
	hydrodynamic description of plasma in self-electromagnetic field. II. 
	Time correlation functions and collective modes. // Ukr. Fiz. Zhurn., 
	1997, vol. 42, No.~2, p. 161-167 (in Ukrainian).
\bibitem{15} Morozov~V., R\"opke~G. Quantum kinetic equation for 
	nonequilibrium dense systems. // Physica A, 1995, vol. 221, 
	p. 511-538.
\bibitem{16} Vakarchuk~I.O., Glushak~P.A., Tokarchuk~M.V. A consistent 
	description of kinetics and hydrodynamics of quantum Bose-systems I. 
	Transport equations. // Ukr. Fiz. Zhurn., 1997, vol. 42, No.~9. p. 
	1150-1158 (in Ukrainian).
\bibitem{17} Tokarchuk~M.V., Arimitsu~T., Kobryn~A.E. Thermo field 
	hydrodynamic and kinetic equations of dense quantum nuclear systems. 
	// Cond. Matt. Phys., 1998, vol. 1, No.~3(15), p. 605-642.
\bibitem{18} Klimontovich~Yu.L. On the necessity and possibility of the
	united description of kinetic and hydrodynamic processes.
	// Teor. Mat. Fiz., 1992, vol. 92, No.~2, p. 312-330 (in Russian).
\bibitem{19} Klimontovich~Yu.L. The united description of kinetic and
	hydrodynamic processes in gases and plasmas.
	// Phys. Lett. A, 1992, vol. 170, p. 434.
\bibitem{20} Klimontovich~Yu.L. From the Hamiltonian mechanics to a 
	continuous media. Dissipative structures. Criteria of 
	self-organization. // Cond. Matt. Phys., 1994, vol. 4, p. 89-116.
\bibitem{21} Bogolubov~N.N. Problems of a dynamical theory in statistical
	physics. In: Studies in statistical mechanics, vol. 1 (eds.
	J.~de~Boer and G.E.Uhlenbeck). Amsterdam, North-Holland, 1962.
\bibitem{22} Liboff~R.L. Introduction to the Theory of Kinetic Equations. 
	New York, John Willey and Sons, 1969.
\bibitem{23} Klimontovich~Yu.L. Kinetic Theory of Nonideal Gases and
	Nonideal Plasmas. Oxford, Pergamon Press, 1982.
\bibitem{24} Ferziger~J.H., Kaper~H.G. Mathematical Theory of Transport
	Processes in Gases. Amsterdam, North-Holland, 1972.
\bibitem{25} R\'esibois~P., de~Leener~M. Classical Kinetic Theory of 
	Fluids. New York, John Willey and Sons, 1977.
\bibitem{26} Kawasaki~K., Oppenheim~J. Logarithmic term in the density 
	expansion of transport coefficients. // Phys. Rev., 1965, vol. 139, 
	No.~6, p. 1763-1768.
\bibitem{27} Weistock~J. Nonanalyticity of transport coefficients and the 
	complete density expansion of momentum correlation function. // 
	Phys. Rev., 1971, vol. 140, No.~2, p. 460-465.
\bibitem{28} Enskog~D. Kinetische Theorie der Warmeleitung Reibung und 
	Selbstdiffusion in gevissen verdichtetten Gasen und Flussigkeiten. 
	// Svenska Vetenskapsaked. Handl., 1922, vol. 63, p. 4.
\bibitem{29} Chapman~S., Cowling~T.G., The Mathematical Theory of Nonuniform
	Gases. Cambridge, Cambridge University Press, 1952.
\bibitem{30} Hanley~H., MacCarthy~R., Cohen~F. Analysis of the transport 
	coefficients for simple dense fluids: application of the modified 
	Enskog theory. // Physica, 1972, vol. 60, No.~2, p. 322-356.
\bibitem{31} Davis~H.T., Rise~S.A., Sengers~J.V. On the kinetic theory of 
	dense fluids. The fluid of rigid spheres with a square well 
	attraction. // J. Chem. Phys., 1961, vol. 35, No.~6, p. 2210-2233.
\bibitem{32} Grmela~M., Rosen~R., Garcia-Colin~L.S. Compatibility of the 
	Enskog theory with hydrodynamics. // J. Chem. Phys., 1981, vol. 75, 
	No.~11, p. 5474-5484.
\bibitem{33} Ernst~M.H., van~Beijeren~H. The modified Enskog equation. // 
	Physica, 1973, vol. 68, No.~3, p. 437-456.
\bibitem{34} R\'esibois~P. $H$-theorem for the  (modified) nonlinear Enskog 
	equation. // Phys. Rev. Lett., 1978, vol. 40, No.~22, p. 1409-1411.
\bibitem{35} R\'esibois~P. $H$-theorem for the (modified) nonlinear Enskog 
	equation. // J. Stat. Phys., 1978, vol. 19, No.~6, p. 593-603.
\bibitem{39} Karkheck~J., van~Beijeren~H., de~Schepper~I.M. Kinetic theory 
	and $H$-theorem for a dense square well fluid. // Phys. Rev. A, 1985, 
	vol. 32, No.~4, p. 2517-2520.
\bibitem{36} Karkheck~J., Stell~G. Kinetic mean-field theories. // J. Chem. 
	Phys., 1981, vol. 75, No.~3, p. 1475-1487.
\bibitem{37} Stell~G., Karkheck~J., van~Beijeren~H. Kinetic mean-field 
	theories: rezultz of energy constraint in maximizing entropy. // J. 
	Chem. Phys., 1983, vol. 79, No.~6, p. 3166-3167.
\bibitem{38} Tokarchuk~M.V., Omelyan~I.P. Normal solution of Enskog-Vlasov
	kinetic equation using boundary conditions method. In:
	Proceedings, Contributed papers of conference ``Modern problems of
	statistical physics'', vol. 1. Lviv, February 3-5, 1987,
	p. 245-252 (in Russian).
\bibitem{40} Leegwater~J.A., van~Beijeren~H., Michels~J.P.J. Linear kinetic 
	theory of the square well fluid. // J. Phys: Cond. Matt., 1989, 
	vol. 1, p. 237-255.
\bibitem{41} Rudyak~V.Ya. On the theory of kinetic equations for dense
	gas. // Jurn. Tehn. Fiz.%
	\footnote{Engl. transl. reads ``Journal for Technical Physics''},
	1984, vol. 54, No.~7, p. 1246-1252 (in Russian).
\bibitem{42} Rudyak~V.Ya. About derivation of Enskog-like kinetic equation
	for dense gas. // Teplofiz. Vys. Temp.%
	\footnote{Engl. transl. reads ``Heat Physics of High Temperatures''},
	1985, vol. 23, No.~2, p. 268-272 (in Russian).
\bibitem{43} Rudyak~V.Ya. Statistical Theory of Dissipative Processes in
	Gases and Liquids.  Novosibirsk, Nauka, 1987 (in Russian).
\bibitem{44} Tokarchuk~M.V., Omelyan~I.P. Model kinetic equations for
	dense gases and liquids. // Ukr. Fiz. Zhurn., 1990, vol. 36,
	No.~8, p. 1255-1261 (in Ukrainian).
\bibitem{45} Kobryn~A.E., Omelyan~I.P., Tokarchuk~M.V. Normal solutions and 
	transport coefficients to the Enskog-Landau kinetic equation for a 
	two-component system of charged hard spheres. The Chapman-Enskog 
	method. // Physica A (to be published).
\bibitem{46} Lenard~A. On Bogolubov's kinetic equation for a spatially 
	homogeneous plasma. // Ann. Phys., 1960, vol. 3, p. 390-401.
\bibitem{47} Balescu~R. Irreversible processes in ionized  gases. // Phys. 
	Fluids., 1960, vol. 3, p. 52-71.
\bibitem{48} Shelest~A.V. Bogolubov's Method in Dynamical Theory of Kinetic 
	Equations. Moscow, Nauka, 1990 (in Russian).
\bibitem{49} Zubarev~D.N. Nonequilibrium Statistical Thermodynamics.
	New York, Consultant Bureau, 1974.
\bibitem{50} Zubarev~D.N. Modern methods of statistical theory of 
	nonequilibrium processes. In: Reviews of science and technology.
	Modern problems of mathematics. Moscow, VINITI, 1980, vol. 15, 
	p. 131-220 (in Russian).
\bibitem{51} Zubarev~D.N., Novikov~M.Yu. Generalized formulation of the
	boundary condition for the Liouville equation and for BBGKY hierarchy.
	// Teor. Mat. Fiz. 1972, vol. 13, No.~3, p. 406-420 (in Russian).
\bibitem{52} Zubarev~D.N., Novikov~M.Yu. Die Boltzmangleichung und m\"ogliche 
	Wege zur Entwicklung dynamischer Methoden in der kinetische Theorie. 
	// Fortschritt f\"ur Physik, 1973, vol. 21, No.~12, p. 703-734.
\bibitem{53} Zubarev~D.N., Morozov~V.G., Omelyan~I.P., Tokarchuk~M.V.
	Derivation of kinetic equations for the system of hard spheres
	using nonequilibrium statistical operator method. / Preprint 
	ITP-90-11R, Kiev 1990 (in Russian).
\bibitem{54} Viner~N. Fourier Integral and Some its Applications. Moscow, 
	Fizmatgiz, 1963 (in Russian).
\bibitem{55} Ditkin~V.A., Kuznetsov~P.I. Reference Book on Operating 
	Calculus. Moscow-Le\-nin\-grad, Gostehizdat, 1951 (in Russian).
\bibitem{56} Boltzmann~L. In: Lectures on Gas Theory. Barkeley, University 
	of California, 1964.
\bibitem{57} Zubarev~D.N., Khonkin~A.D. The method for construction of
	normal solutions for kinetic equations with the help of boundary
	conditions. // Teor. Mat. Fiz., 1972, vol. 11, No.~3, p. 403-412
	(in Russian).
\bibitem{58} Yukhnovskii~I.R., Holovko~M.F. Statistical Theory of Classical 
	Equilibrium Systems. Kiev, Naukova Dumka, 1980 (in Russian).
\bibitem{59} Rudyak~V.Ya., Yanenko~N.N. On taking into account
	intermolecular attractive forces at the derivation of kinetic
	equations. // Teor. Mat. Fiz., 1985, vol. 64, No.~2, p. 277-286
	(in Russian).
\bibitem{59a} Rudyak~V.Ya. Kinetic equations of nonideal gas with real 
	interaction potentials. // Jurn. Tehn. Fiz., 1987, vol. 57, No.~8, p. 
	1466-1475 (in Russian).
\bibitem{60} Lifshiz~E.M., Pitaevskii~L.P. Physical Kinetics.
	Oxford, Pergamon Press, 1981.
\bibitem{61} Balescu~R. Statistical Mechanics of Charged Particles.
	New York, Interscience, 1963.
\bibitem{62} Ecker~G. Theory of Fully Ionized Plasmas. New York, London, 
	Academic Press, 1972.
\bibitem{63} Kobryn~A.E., Omelyan~I.P., Tokarchuk~M.V. Normal solution to the 
	Enskog-Landau kinetic equation using boundary conditions method. // 
	Phys. Lett. A, 1996, vol. 223, No.~1\&2, p. 37-44.
\bibitem{64} Kobryn~A.E., Omelyan~I.P., Tokarchuk~M.V. Nonstationary solution 
	to the Enskog-Landau kinetic equation using boundary conditions method. 
	// Cond. Matt. Phys., 1996, No.~8, p. 75-98.
\bibitem{65} Green~H.S. The Molecular Theory of Gases. Amsterdam, North 
	Holland, 1952.
\bibitem{66} Cohen~E.G.D. Cluster expansion and hierarchy. // Physica, 1962, 
	vol. 28, No.~10, p. 1045-1059.
\bibitem{67} Cohen~E.G.D. On the kinetic theory of dense gases. // J. Math. 
	Phys., 1963, vol. 4, No.~2, p. 183-189.
\bibitem{68} Bogolubov~N.N. On the stochastic processes in the dynamical 
	systems. / Preprint JINR-E17-10514, Dubna, 1977 (in English).
	Bogolubov~N.N. On the stochastic processes in the dynamical systems. 
	// Fizika Elementarnyh Chastic i Atomnogo Yadra (Particles \& Nuclei), 
	1978, vol. 9, No.~4, p. 501-579 (in Russian).
\bibitem{69} Gould~H.A., DeWitt~H.E. Convergent kinetic equation for a 
	classical plasma. // Phys. Rev., 1967, vol. 155, No.~1, p. 68-74.
\bibitem{70} Ernst~M.H., Dorfman~J.R. Nonanalitic dispersion relations in 
	classical fluids I. The hard sphere gas. // Physica, 1972, vol. 61, 
	p. 157-181.
\bibitem{71} Scold~K., Rowe~J.M., Ostrowski~G., Randolph~R.D. Coherent and 
	incoherent scattering laws of liquid Argon. // Phys. Rev. A, vol. 6, 
	No.~3, p. 1107-1131.
\bibitem{72} Copley~J.R.D., Rowe~J.M. Density fluctuation in liquid Rubidium. 
	I. Neutron scattering measurements. // Phys. Rev. A, 1974, vol. 9, 
	No.~4, p. 1656-1666.
\bibitem{73} Copley~J.G.D., Lovesey~S.W. The dynamic properties of monoatomic 
	liquids. // Rept. Progr. Phys., 1975, vol. 38, No.~4, p. 461-563.
\bibitem{74} Dynamics of Solids and Liquids by Neurton Scattering (eds. 
	S.W.Lovesey and T.Schringer). Berlin, Springer Verlag, 1977.
\bibitem{75} Boon~J.P., Yip~S. Molecular Hydrodynamics. New York, 1980.
\bibitem{76} Hansen~J.P., McDonald~I.R. Theory of Simple Liquids, 2nd edn.
	London, Academic Press, 1986.
\bibitem{77} Balucani~U., Zoppi~M. Dynamics of the Liquid State. Oxford, 
	Clarendon Press, 1994.
\bibitem{78} Kadanoff~L.P., Martin~P.C. Hydrodynamic equations and 
	correlation functions. // Ann. Phys., 1963, vol. 24, No.~1, p. 419-469.
\bibitem{79} Mountain~R.D. Spectral distribution of scattered ligth in a 
	simple fluid. // Rev. Mod. Phys., 1966, vol. 38, No.~1, p. 205-214.
\bibitem{80} R\'esibois~P. On linearized hydrodynamic modes in statistical 
	physics. // J. Stat. Phys., 1970, vol. 2, No.~1, p. 21-51.
\bibitem{81} Forster~D. Hydrodynamic Fluctuations, Broken Symmetry, and 
	Correlation Functions. London, Amsterdam, W.A.Benjamin, Inc., 1975.
\bibitem{82} March~N.H., Tosi~M.P. Atomic Dynamics in Liquids. Macmillan 
	Press, 1976.
\bibitem{83} Zwanzig~R. Ensemble method in the theory of irreversibility. // 
	J. Chem. Phys., 1960, vol. 33, No.~5, p. 1338-1341.
\bibitem{84} Zwanzig~R. Memory effects in irreversible thermodynamics. // 
	Phys. Rev., 1961, vol. 124, No.~4, p. 983-992.
\bibitem{85} Mori~H. Transport, collective motion and Brownian motion. // 
	Progr. Theor. Phys., 1965, vol. 33, No.~3, p. 423-455.
\bibitem{86} Sergeev~M.V. Generalized transport equations in the theory of 
	irreversible processes. // Teor. Mat. Fiz., 1974, vol. 21, No.~3, p. 
	402-414 (in Russian).
\bibitem{87} Tishchenko~S.V. Construction of generalized hydrodynamics 
	using the nonequilibrium statistical operator method. // Teor. Mat. 
	Fiz., 1976, vol. 26, No.~1, p. 96-110 (in Russian).
\bibitem{88} Tserkovnikov~Yu.A. On a method of solution of infinite systems 
	of equations for temperature Green functions. // Teor. Mat. Fiz., 
	1981, vol. 49, No.~2, p. 219-233 (in Russian).
\bibitem{89} Tserkovnikov~Yu.A. Method of two-time Green functions in 
	molecular hydrodynamics. // Theor. Mat. Fiz., 1985, vol. 63, No.~3, 
	p. 440-457 (in Russian).
\bibitem{90} Zubarev~D.N., Tserkovnikov~Yu.A. Method of two-time 
	temperature Green functions in equilibrium and nonequilibrium 
	statistical mechanics. In: Collection of scientific works of 
	Mathematical Institute of USSR Academy of Sciences. Moscow, Nauka, 
	1986, vol. 175, No.~3, p. 134-177 (in Russian).
\bibitem{91} Akcasu~A.Z., Daniels~E. Fluctuation analysis in simple fluids. 
	// Phys. Rev. A, 1970, vol. 2, No.~3, p. 962-975.
\bibitem{92} Desai~R.C., Kapral~R. Translational hydrodynamics and ligth 
	scattering from molecular fluids. // Phys. Rev. A, 1972, vol. 6, 
	No.~6, p. 2377-2390.
\bibitem{93} G\"otze~W., L\"ucke~M. Dynamical current correlation functions of 
	simple classical liquids for intermediate wave number. // Phys. Rev. 
	A, 1975, vol. 11, No.~6, p. 2173-2190.
\bibitem{94} Bosse~J., G\"otze~W., L\"ucke~M. Mode-coupling theory of simple 
	classical liquids. // Phys. Rev. A, 1978, vol. 17, No.~1, p. 434-445.
\bibitem{95} Bosse~J., G\"otze~W., L\"ucke~M. Current fluctuation spetra of 
	liquid Argon near its triple point. // Phys. Rev. A, 1978, vol. 17, 
	No.~1, p. 447-454.
\bibitem{96} Bosse~J., G\"otze~W., L\"ucke~M. Current-fluctuation spectra of 
	liquid Rubidium. // Phys. Rev. A, 1978, vol. 18, No.~3, p. 1176-1182.
\bibitem{97} Bosse~J., G\"otze~W., Zippelius~A. Velocity-autocorrelation 
	spectrum of simple classical liquids. // Phys. Rev. A, 1978, vol. 18, 
	No.~3, p. 1214-1221.
\bibitem{98} Zippelius~A., G\"otze~W. Kinetic theory for the coherent 
	scattering function of classical liquids. // Phys. Rev. A, 1978, 
	vol. 17, No.~1, p. 414-423.
\bibitem{99} G\"otze W. Aspects of Structural Glass Transitions. In: 
	Liquids, Freezing and Glass Transition (eds. J.-P.Hansen, D.Levesque 
	and J.Zinn-Justin). Amsterdam, North-Holland, 1991.
\bibitem{100} Shurygin~V.Yu, Yulmetyev~R.M. Influence of non-Markovian 
	effects in thermal motion of particles on intensity of non-coherent 
	scattering of slow neutrons in liquids. // Teor. Mat. Fiz., 1990, 
	vol. 83, No.~2, p. 222-235 (in Russian).
\bibitem{101} Shurygin~V.Yu, Yulmetyev~R.M. Calculation of the dynamic 
	structure factor for liquids by the reduced description technique. // 
	Zhurn. Exp. Teor. Fiz., 1989, vol. 96, No.~3(9), p. 938-947 (in 
	Russian).
\bibitem{101a} Shurygin~V.Yu, Yulmetyev~R.M. Space dispersion of structure 
	relaxation in simple liquids. // Zhurn. Exp. Teor. Fiz., 1991, 
	vol. 99, No.~1, p. 144-154 (in Russian).
\bibitem{101b} Shurygin~V.Yu, Yulmetyev~R.M. On spectre of non-Markovianity of 
	relaxation processes in liquids. // Zhurn. Exp. Teor. Fiz., 1992, 
	vol. 102, No.~3(9), p. 852-862 (in Russian).
\bibitem{102} Shurygin~V.Yu, Yulmetyev~R.M. Role of asymmetric time 
	correlation functions in kinetics of relaxation processes in 
	liquids. // Ukr. Fiz. Zhurn., 1990, vol. 35, No.~5, p. 693-695 
	(in Russian).
\bibitem{103} Akcasu~A.Z., Duderstadt~J.J. Derivation of kinetic equation 
	from the generalized Langevin equation. // Phys. Rev., 1969, vol. 188, 
	No.~1, p. 479-486.
\bibitem{104} Forster~D., Martin~P.C. Kinetic theory of a weakly coupled 
	fluid. // Phys. Rev. A, 1970, vol. 2, No.~4, p. 1575-1590.
\bibitem{105} Mazenko~G.F. Microscopic method for calculating memory functions  
	in transport theory. // Phys. Rev. A, 1971, vol. 3, No.~6, p. 2121-2137.
\bibitem{106} Mazenko~G.F. Properties of the low-density memory function. // 
	Phys. Rev. A, 1972, vol. 5, No.~6, p. 2545-2556.
\bibitem{107} Mazenko~G.F., Tomas~Y.S., Sidney~Yip. Thermal fluctuation of 
	hard sphere gas. // Phys. Rev. A, 1972, vol. 6, No.~5, p. 1981-1995.
\bibitem{108} Mazenko~G.F. Fully renormalized kinetic theory I. 
	Self-diffusion. // Phys. Rev. A, 1973, vol. 7, No.~1, p. 209-222.
\bibitem{108a} Mazenko~G.F. Fully renormalized kinetic theory II. Low-density. 
	// Phys. Rev. A, 1973, vol. 7, No.~1, p. 222-232.
\bibitem{109} Mazenko~G.F. Fully renormalized kinetic theory III. Density 
	flutuation. // Phys. Rev. A, 1974, vol. 9, Np~1. p. 360-387.
\bibitem{110} Forster~D. Properties of the kinetic memory function in 
	classical fluids. // Phys. Rev. A, 1974, vol. 9, No.~2, p. 943-956.
\bibitem{111} Boley~C.D., Desai~R.C. Kinetic theory of a dense gas: 
	triple-collision memory function. // Phys. Rev. A, 1973, vol. 7, 
	No.~5, p. 1700-1709.
\bibitem{112} Furtado~P.M., Mazenko~G.F., Sidney~Yip. Hard-sphere kinetic 
	theory analisis of classical simple liquid. // Phys. Rev. A, 1975, 
	vol. 12, No.~4, p. 1653-1661.
\bibitem{113} John~M.S., Forster~D. A kinetic theory of classical simple 
	liquids. // Phys. Rev. A, 1975, vol. 12, No.~1, p. 254-266.
\bibitem{114} Sjodin~S., Sjolander~A.  Kinetic model for classical liquids. 
	// Phys. Rev. A, 1978, vol. 18, No.~4, p. 1723-1736.
\bibitem{115} de~Schepper~I.M., Cohen~E.G.D. Collective modes in fluids and 
	neutron scattering. // Phys. Rev. A, 1980, vol. 22, No.~1, p. 287-289.
\bibitem{116} de~Schepper~I.M., Cohen~E.G.D. Very-short wavelength collective 
	modes in fluids. // J. Stat. Phys., 1982, vol. 27, No.~2, p. 223-281.
\bibitem{117} Cohen~E.G.D., de~Schepper~I.M., Zuilhof~M.J. Kinetic theory of 
	the eigenmodes of classical fluids and neutron scattering. // Physica 
	B, 1984, vol. 127, p. 282-291.
\bibitem{118} de~Schepper~I.M., Verkerk~R., van~Well~A.A., de~Graaf~L.A. 
	Short-wavelength sound modes in liquid Argon. // Phys. Rev. Lett., 
	1983, vol. 50, No.~13, p. 974-977.
\bibitem{119} van~Well~A.A., de~Graaf~L.A. Density fluctuations in liquid 
	Argon II. Coherent dynamic structure factor at large wave numbers. // 
	Phys. Rev. A, 1985, vol. 32, No.~4, p. 2384-2395.
\bibitem{120} van~Well~A.A., de~Graaf~L.A. Density fluctuations in liquid 
	Neon studied by neutron scattering. // Phys. Rev. A, 1985, vol. 32, 
	No.~4, p. 2396-2412.
\bibitem{121} Alley~W.E., Alder~B.J. Generalized transport coefficient for 
	hard spheres. // Phys. Rev. A, 1983, vol. 27, No.~6, p. 3158-3173.
\bibitem{122} Alley~W.E., Alder~B.J., Sidney~Yip. The neutron scattering 
	function for hard spheres. // Phys. Rev. A, 1983, vol. 27, No.~6, 
	p. 3177-3186.
\bibitem{123} de~Schepper~I.M., van~Rijs~J.C., van~Well~A.A., Verkerk~P., 
	de~Graaf~L.A., Bruin~C. Microscopic sound waves in dense Lennard-Jones 
	fluids. // Phys. Rev. A, 1984, vol. 29, No.~3, p. 1602-1605.
\bibitem{124} Sidney~Yip., Alley~W.E., Alder~B.J. Evaluation of time 
	correlation function from a generalized Enskog equation. // J. Stat. 
	Phys., 1982, vol. 27, No.~1, p. 201-217.
\bibitem{125} de~Schepper~I.M., Cohen~E.G.D., Zuilhof~M.J. The width of 
	neutron spectra and heat mode of fluids. // Phys. Lett. A, 1984, 
	vol. 101, No.~8, p. 399-404.
\bibitem{126} de~Schepper~I.M., Verkerk~P., van~Well~A.A., de~Graaf~L.A. 
	Nonanalytic dispersion relations in liquid Argon. // Phys. Lett. A, 
	1984, vol. 104, No.~1, p. 29-32.
\bibitem{127} Bruin~C., Mechels~J.P.J., van~Rijs~J.C., de~Graaf~L.A., 
	de~Schepper~I.M. Extended hydrodynamic modes in a dence hard sphere 
	fluid. // Phys. Lett. A, 1985, vol. 110, No.~1, p. 40-43.
\bibitem{128} Kamgar-Parsi~B., Cohen~E.G.D., de~Schepper~I.M. Dynamical 
	processes in hard sphere fluids. // Phys. Rev. A, 1987, vol. 35, 
	No.~11, p. 4781-4795.
\bibitem{129} Cohen~E.G.D., de~Schepper~I.M. Effective eigenmode description
	of dynamic processes in dense classical fluid mixtures. // Nuovo 
	Cimento~D, 1990, vol.~12, No.~4/5, p.~521--542.
\bibitem{130} de~Schepper~I.M., Cohen~E.G.D., Kamgar-Parsi~B. On the 
	calculation of equilibrium time correlation functions in hard sphere 
	fluids. // J. Stat. Phys., 1989, vol. 54, No.~1/2, p. 273-313.
\bibitem{131} Ernst~M.H., Dorfman~J.R. Nonanalitic dispersion relations for 
	classical fluids II. General fluid. // J. Stat. Phys., 1975, vol. 12, 
	No.~4, p. 311-359.
\bibitem{132} van~Well~A.A., Verkerk~P., de~Graaf~L.A., Suck~J.-B., 
	Copley~J.R.D. Density fluctuation in liquid Argon: coherent dynamic 
	structure factor along the 120 K isoterm obtained by neutron 
	scattering. // Phys. Rev. A, 1985, vol. 31, No.~5, p. 3391-3414.
\bibitem{133} Egelstaff~P.A., Glase~W., Litchensky~D., Schneider~E., 
	Suck~J.-B. Two-body time correlations in dense Krypton gas. // Phys. 
	Rev. A, 1983, vol. 27, No.~2, p. 1106-1115.
\bibitem{134} de~Schepper~I.M., Cohen~E.G.D., Bruin~C., van~Rijs~J.C., 
	Montfrooij~W., de~Gra\-af~L.A. Hydrodynamic time correlation functions 
	for a Lennard-Jones fluid. // Phys. Rev. A, 1988, vol. 38, No.~1, 
	p. 271-287.
\bibitem{135} Zubarev~D.N., Tokarchuk~M.V. Nonequilibrium statistical 
	hydrodynamics of ionic systems. // Teor. Mat. Fiz., 1987, vol. 70, 
	No.~2, p. 234-254 (in Russian).
\bibitem{136} Mryglod~I.M., Tokarchuk~M.V. On statistical hydrodynamics of 
	simple liquds. In: Problems of atomic science and technique. Series: 
	Nuclear physics investigations (theory and experiment). Kharkov, 
	Kharkov Physico Technical Institute, 1992, vol. 3(24), p. 134-139 (in 
	Russian).
\bibitem{137} Mryglod~I.M., Omelyan~I.P. Generalized collective modes of a 
	Lennard-Jones fluid. High-mode approximation. // Cond. Matt. Phys., 
	1994, vol. 4, p. 128-160.
\bibitem{138} Mryglod~I.M., Hachkevych~A.M. On nonequilibrium statistical 
	theory of a fluid: linear relaxation theories with different sets of 
	dynamic variables. // Cond. Matt. Phys., 1995, vol. 5, p. 105-123.
\bibitem{139} Mryglod~I.M., Omelyan~I.P., Tokarchuk~M.V. Generalized 
	collective modes for the Lennard-Jones fluid. // Mol. Phys., 1995, 
	vol. 84, No.~2, p. 235-259. 
\bibitem{140} Mryglod~I.M., Omelyan~I.P. Generalized collective modes for 
	a Lennard-Jones fluid in higher mode approximations. // Phys. 
	Lett. A, 1995, vol. 205, No.~4, p. 401-406. 
\bibitem{141} Mryglod~I.M., Omelyan~I.P. Generalized mode approach: I. 
	Transverse time correlation functions and generalized shear viscosity 
	of a Lennard-Jones fluid. // Mol. Phys., 1997, vol. 90, No.~1, 
	p. 91-99.
\bibitem{142} Mryglod~I.M., Omelyan~I.P. Generalized mode approach: II. 
	Longitudinal time correlation functions of a Lennard-Jones fluid. // 
	Mol. Phys., 1997, vol. 91, No.~6, p. 1005-1015. 
\bibitem{143} Mryglod~I.M., Omelyan~I.P. Generalized mode approach: III. 
	Generalized transport coefficients of a Lennard-Jones fluid. // Mol. 
	Phys., 1997, vol. 92, No.~5, p. 913-927.
\bibitem{144} Mryglod~I.M. Generalized hydrodynamics of liquids. II. Time 
	correlation functions using the formalism of generalized collective 
	modes. // Ukr. Fiz. Zhurn., 1998, vol. 43, No.~2, p. 252-256 
	(in Ukrainian). 
\bibitem{145} Mryglod~I.M., Hachkevych~A.M. Simple iteration schema for 
	calculation of memory function: a ``shoulder'' problem for 
	generalized shear viscosity. // Ukr. Fiz. Zhurn., 1998 
	(submitted).
\bibitem{146} Mryglod~I.M. Generalized statistical hydrodynamics of fluids: 
	approach of generalized collective modes. // Cond. Matt. Phys., 1998 
	(submitted).
\bibitem{147} Mryglod~I.M. Mode-coupling behaviour of the generalized 
	hydrodynamic modes for a Lennard-Jones fluid. // J. Phys. Studies, 
	1998 (submitted).
\bibitem{148} Mryglod~I.M., Ignatyuk~V.V. Generalized hydrodynamics of 
	binary mixtures. // J. Phys. Studies, 1997, vol. 1, No.~2, p. 181-190 
	(in Ukrainian).
\bibitem{149} Bryk~T.M., Mryglod~I.M., Kahl~G. Generalized collective modes 
	in a binary He$_{0.65}$-Ne$_{0.35}$ mixture. // Phys. Rev. E, 1997, 
	vol. 56, No.~3, p. 2903-2915.
\bibitem{150} Mryglod~I.M. Generalized hydrodynamics of multicomponent fluids. 
	// Cond. Matt. Phys., 1997, vol. 10, p. 115-135.
\bibitem{151} Bryk~T.M., Mryglod~I.M. Spectra of transverse excitations in 
	liquid glass-forming metallic alloy Mg$_{70}$Zn$_{30}$: temperature 
	dependence. // Cond. Matt. Phys., 1998 (submitted).
\bibitem{152} Omelyan~I.P. Wavevector- and frequency-dependent dielectric 
	constant of the Stockmayer fluid. // Mol. Phys., 1996, vol. 87, No.~6, 
	p. 1273-1283.
\bibitem{153} Omelyan~I.P. Temperature behaviour of the frequency-dependent 
	dielectric constant for a Stockmayer fluid. // Phys. Lett. A, 1996, 
	vol. 216, No.~1/5, p. 211-216.
\bibitem{154} Omelyan~I.P. Wavevector dependent dielectric constant of the 
	MCY water model. // Phys. Lett. A, 1996, vol. 220, No.~1/3, p. 167-177.
\bibitem{155} Omelyan~I.P. On the reaction field for interaction site models 
	of polar systems. // Phys. Lett. A, 1996, vol. 223, No.~4, p. 295-302.
\bibitem{156} Omelyan~I.P. Generalized collective mode approach in the 
	dielectric theory of dipolar systems. // Physica A, 1997, vol. 247, 
	No.~1/4, p. 121-139.
\bibitem{157} Omelyan~I.P., Zhelem~R.I., Tokarchuk~M.V. Generalized 
	hydrodynamics of polar liquids in external inhomogeneous electric 
	field. Method of nonequilibrium statistical operator. // Ukr. Fiz. 
	Zhurn., 1997, vol. 42, No.~6, p. 684-692 (in Ukrainian).
\bibitem{158} Omelyan~I.P. Longitudinal wavevector- and frequency-dependent 
	dielectric constant of the TIP4P water model. // Mol. Phys., 1998, 
	vol. 93, No.~1, p. 123-135.
\bibitem{159} Omelyan~I.P., Mryglod~I.M., Tokarchuk~M.V. Generalized dipolar 
	modes of a Stockmayer fluid in high-order approximations. // Phys. 
	Rev. E, 1998, vol. 57, No.~6, p. 6667-6676.
\bibitem{160} Omelyan~I.P., Mryglod~I.M., Tokarchuk~M.V. Dielectric 
	relaxation in dipolar fluid. Generalized mode approach. // Cond. Matt. 
	Phys., 1998, vol. 1, No.~1(13), p. 179-200.
\bibitem{161} Peletminskii~S.V., Sokolovskii~A.I. General equations of 
	fluctuation hydrodynamics. // Ukr. Fiz. Zhurn., 1992, vol. 37, 
	No.~10, p. 1521-1528 (in Russian).
\bibitem{162} Peletminskii~S.V., Slusarenko~Yu.V. Kinetic and hydrodynamic 
	theory of long-wave nonequilibrium fluctuations. In: Abstracts and 
	contributed papers of International conference ``Physics in Ukraine''. 
	Kiev, 22-27 June, 1993. Vol.: Statistical physics and phase 
	transitions. Kiev, Institute for Theoretical Physics, 1993, p. 103-106.
\bibitem{163} Akhiezer~A., Peletminskii~S. Methods of Statistical
	Physics. Oxford, Pergamon, 1981.
\bibitem{164} Sokolovskii~A.I. Equations of hydrodynamics near equilibrium 
	at absence of long-range correlations. // Ukr. Fiz. Zhurn., 1993, 
	vol. 37, No.~10, p. 1528-1536 (in Russian).
\bibitem{164a} Peletminskii~S.V., Sokolovskii~A.I. Space dispersion of 
	transport coefficients of a liquid near equilibrium. // Ukr. Fiz. 
	Zhurn., 1992, vol. 37, No.~11, p. 1702-1711 (in Russian).
\bibitem{165} Balabanyan~G.O. Construction of hydrodynamical asymptotics of 
	classical equilibrium correlation Green functions by means of the 
	Boltzmann kinetic equation. // Teor. Mat. Fiz., 1990, vol. 82, No.~1, 
	p. 117-132 (in Russian).
\bibitem{166} Balabanyan~G.O. Classical equilibrium generalized 
	hydrodynamic correlation Green functions. I. // Teor. Mat. Fiz., 
	1990, vol. 82, No.~3, p. 450-465 (in Russian).
\bibitem{166a} Balabanyan~G.O. Classical equilibrium generalized hydrodynamic 
	correlation Green functions. II. Transversal autocorrelation Green 
	function. // Teor. Mat. Fiz., 1990, vol. 83, No.~2, p. 311-319 
	(in Russian).
\bibitem{166b} Balabanyan~G.O. Classical equilibrium generalized hydrodynamic 
	correlation Green functions. III. ``Density-density" correlation 
	Green function. // Teor. Mat. Fiz., 1990, vol. 85, No.~1, p. 102-114 
	(in Russian).
\bibitem{166c} Balabanyan~G.O. Construction of equations for classical 
	equilibrium correlation Green functions by means of kinetic equations. 
	// Teor. Mat. Fiz., 1991, vol. 88, No.~2, p. 225-245 (in Russian).
\bibitem{167} Tserkovnikov~Yu.A. Molecular hydrodynamics of weakly nonideal 
	nondegenerate Bose-gas. I. Green functions of transverse components 
	of density of particle current. // Teor. Mat. Fiz., 1990, vol. 85, 
	No.~1, á. 124-149 (in Russian).
\bibitem{168} Tserkovnikov~Yu.A. Molecular hydrodynamics of weakly nonideal 
	nondegenerate Bose-gas. II. Green functions of longitudinal 
	fluctuations of particles and energy densities. // Teor. Mat. Fiz., 
	1990, vol. 85, No.~2, p. 258-287 (in Russian).
\bibitem{169} Robertson~B. Equations of motion in nonequilibrium statistical 
	mechanics. // Phys. Rev., 1966, vol. 144, No.~1, p. 151-161.
\bibitem{170} Robertson~B. Equations of motion in nonequilibrium statistical 
	mechanics II. Energy transport. // Phys. Rev., 1967, vol. 160, No.~1, 
	p. 175-183.
\bibitem{171} Kawasaki~K., Gunton~J.D. Theory of nonlinear shear viscosity 
	and normal stress effects. // Phys. Rev. A, 1973, vol. 8, No.~4, 
	p. 2048-2064.
\bibitem{172} Gould~H., Mazenko~G.F. Microscopic theory of the self-diffusion 
	in a classical one-component plasma. // Phys. Rev. A, 1977, vol. 15, 
	No.~3, p. 1274-1287.
\end{thebibliography}
